\documentclass[11pt, a4paper]{article} 

\usepackage[utf8]{inputenc}
\usepackage[english]{babel}
\usepackage{comment}

\usepackage{graphicx, subcaption}

\usepackage[left=2cm, right=2cm, top=2cm, bottom=2cm]{geometry}

\usepackage{authblk}

\usepackage[colorlinks=true,allcolors=blue]{hyperref}

\usepackage{amsmath,amsthm,amssymb,mathabx}
\usepackage{bbm,bm}

\usepackage[autostyle]{csquotes}
\usepackage[
    backend=bibtex,
    style=authoryear,
    natbib=true,
    sortlocale=en_US,
    url=false,
    doi=true,
    eprint=true,
    isbn=false,
    firstinits=true,
    useprefix=true,
    maxbibnames=10,
    maxcitenames=2,
    dashed=false,
	date = year
]{biblatex}
\renewbibmacro{in:}{}

\AtEveryBibitem{%
  \clearfield{note}%
  \clearlist{language}%
}
\DeclareFieldFormat*{title}{#1}
\DeclareNameAlias{sortname}{last-first}
\DeclareNameAlias{default}{last-first}
\usepackage{enumerate}
\usepackage{url, nth}

\usepackage{algorithm2e}

\usepackage{multirow}
\usepackage{booktabs,colortbl,xcolor}

\newcommand{\soft}{\textsf}

\newcommand{\N}{\mathbb{N}}

\newcommand{\calG}{\mathcal{G}}

\newcommand{\esp}[1]{\mathbb{E}\left[#1\right]}

\newcommand{\Rvec}{\boldsymbol{R}}

\newcommand{\Yvec}{\boldsymbol{Y}}

\newcommand{\Avec}{\boldsymbol{A}}

\newcommand{\Wvec}{\boldsymbol{W}}

\newcommand{\Cvec}{\boldsymbol{C}}

\newcommand{\Evec}{\boldsymbol{E}}

\newcommand{\wvec}{\boldsymbol{w}}

\newcommand{\uvec}{\boldsymbol{u}}

\newcommand{\thetavec}{\boldsymbol{\theta}}

\newcommand{\Sigmavec}{\boldsymbol{\Sigma}}
\newcommand{\zetavec}{\boldsymbol{\zeta}}

\newcommand{\gammavec}{\boldsymbol{\gamma}}

\newcommand{\Phivec}{\boldsymbol{\Phi}}

\newcommand{\Upsilonvec}{\boldsymbol{\Upsilon}}

\newcommand{\qvec}{\boldsymbol{q}}
\newcommand{\V}{\mathbb{V}}

\newcommand{\systL}{\left\{\begin{array}{ll}}
\newcommand{\systR}{\end{array}\right.}
\newcommand{\matL}{\left(\begin{matrix}}
\newcommand{\matR}{\end{matrix}\right)}
\newcommand{\detL}{\left|\begin{matrix}}
\newcommand{\detR}{\end{matrix}\right|}

\usepackage{dsfont}

\makeatletter
\renewcommand{\@biblabel}[1]{\ }

\makeatother

\title{Impacts of Climate Change on Mortality: An extrapolation of temperature effects based on time series data in France}

\author[1]{Quentin Guibert\footnote{Email: \href{mailto:guibert@ceremade.dauphine.fr}{guibert@ceremade.dauphine.fr}.}}

\author[2]{Gaëlle Pincemin
\footnote{Email: \href{mailto:gaelle_pincemin@gallagherre.com}{gaelle\_pincemin@gallagherre.com}.}
}

\author[3,4]{Frédéric Planchet
\footnote{Email: \href{mailto:frederic.planchet@univ-lyon1.fr}{frederic.planchet@univ-lyon1.fr}.}
}

\affil[1]{CEREMADE, Université Paris-Dauphine, PSL University, CNRS, 75016 Paris, France}
\affil[2]{Gallagher Re, 251 Boulevard Pereire, 75017 Paris, France}
\affil[3]{Univ Lyon, Université Claude Bernard Lyon 1, Institut de Science Financière et d'Assurances (ISFA), Laboratoire SAF EA2429, F-69366, Lyon, France}
\affil[4]{Prim'Act, 42 avenue de la Grande Armée, 75017 Paris, France}

\begin{document}
\date{January 24, 2024}

\maketitle

\begin{abstract}%
Most contemporary mortality models rely on extrapolating trends or past events. However, population dynamics will be significantly impacted by climate change, notably the influence of temperatures on mortality. In this paper, we introduce a novel approach to incorporate temperature effects on projected mortality using a multi-population mortality model. This method combines a stochastic mortality model with a climate epidemiology model, predicting mortality variations due to daily temperature fluctuations, be it excesses or insufficiencies. The significance of this approach lies in its ability to disrupt mortality projections by utilizing temperature forecasts from climate models and to assess the impact of this unaccounted risk factor in conventional mortality models. We illustrate this proposed mortality model using French data stratified by sex and age, focusing on past temperatures and mortality. Utilizing climate model predictions across various IPCC scenarios, we investigate gains and loss in life expectancy linked to temperature and the additional mortality induced by extreme heatwaves, and quantify them by assessing this new risk factor in prediction intervals. Furthermore, we analyze the geographical differences across the Metropolitan France. 


\end{abstract}

\section{Introduction\label{sec:intro}}

The study of past trends since the 19th century has demonstrated unprecedented improvements in human mortality worldwide. In the fields of actuarial science and demography, the analysis of these trends has led to a multitude of models and approaches aimed at studying the dynamics of mortality, see \citet{lee_modeling_1992, brouhns02, barrieu_understanding_2012, plat_stochastic_2009, dowd_cbdx_2020} among others. These dynamics are complex and arise from numerous factors that are more or less shared among different population groups, such as health factors, lifestyles, and the performance of health systems. These elements typically emerge through socioeconomic analyses that reveal mortality disparities within a region, see for example \citet{bonnet_mortality_2023} or \citet{cairns_drivers_2024} for recent results.

Climate change and the disruptions caused by humanity to the environment will have consequences on human health and mortality \citep{ipcc_ar6_wg2_ts}, including European countries \citep{weilnhammer_extreme_2021}. The \citet{who14} identify five main causes contributing to increase mortality in response to climate change: undernutrition, malaria, diarrhoeal disease, dengue, and heat. These causes will affect populations differently in the future depending on their level of education, access to healthcare, and age. For example, undernutrition primarily affects children under 15 years old, while heat mainly impacts mortality of the elderly. Malaria and diarrhoeal diseases are prevalent in children under 5 years, especially in South Asia and Africa. Heatwaves are expected to affect the entire world, with more than 68,000 heat-related deaths in Europe by 2030 without adaptation measures \citep{ballester_heat-related_2023}.
In these regions, the main mortality concerns among those listed by the \citet{who14} are actually heatwaves \citep{pottier_climate_2021} and vector-borne diseases \citep{semenza_vector-borne_2018, paz_climate_2020} with potential impacts on longevity and life expectancy. Recently, a significant number of studies have highlighted the impact of temperature fluctuations resulting from climate change on mortality \citep{guo_gasparrini15, gronlund18, guo_quantifying_2018, vicedo-cabrera_multi-country_2018, lee_projections_2020, martinez-solanas_projections_2021}. With human activity, the rise in temperatures has accelerated since the pre-industrial era, reaching an average global increase of $1.1^{\circ}\text{C}$ compared to the pre-industrial period (1850-1900) \citep{ipcc_ar6_wg2_ts}. These changes vary in magnitude according to latitude. For example, French population experienced exceptionally warm temperatures in 2022 \citep{hautconseilclimat2023} with consequences in terms of excess mortality \citep{ballester_heat-related_2023}. According to IPCC projections, the direct impacts of temperature on human health are expected to intensify regardless of future climate trajectories. This could lead to changes in premature mortality due to temperature and adjustments in mortality trends related to the intra-annual death distribution. Under these conditions, it becomes crucial to study and incorporate the effects of climate change, particularly temperature, into the stochastic mortality models used for population projections.
This is also a significant issue for insurers' and pension funds' models, as well as for supervisory authorities seeking to assess the vulnerability of financial institutions to climate change \citep{acpr_report_2024}. Therefore, the primary aim of this paper is to account for the effect of temperature variations on future projected mortality rates within the framework of multi-population stochastic mortality models.

The proposed model comprises two components, estimated separately. The first component is a conventional stochastic mortality model calibrated on historical trends. The second component is constructed using a climate epidemiology model. This last model is calibrated using daily temperature and mortality data allowing for the attribution of a fraction of deaths to the effects of temperature. By coupling these two models, our approach produces a stochastic mortality model accounting for the trend induced by temperatures. Our two-step approach provides significant flexibility to the mortality model, enabling projections that incorporate various climate scenarios concerning temperature evolution. Additionally, it allows for the measurement of gains or losses in life expectancy due to climate change. More precisely, we use daily temperature projections at different geographical levels and quantify the local and specific impact of temperature on mortality over specific periods of the year, e.g. isolating extreme temperatures. Furthermore, our approach also incorporates the uncertainty associated with various climate models.

The approach implemented in our article falls within the framework of modeling exogenous shocks to mortality rate dynamics. Starting from single-population mortality models such as the \citet{lee_modeling_1992} model, see \citet{basellini_thirty_2022} for a review, a relatively sparse literature has explored the introduction of stochastic shocks. Most of these models have been introduced to incorporate the effects of economic shocks \citep{french_forecasting_2014, niu_trends_2014, boonen_modeling_2017, seklecka_mortality_2019, cupido_space_2024, ma_longevity_2024} 
or jumps in  mortality rate dynamics, such as catastrophes, wars or pandemic shocks \citep{liu_age_2015, robben_assessing_2022, goes_bayesian_2023} into the dynamics of mortality rates. 
In an insurance context, this may have substantial impacts on the amount of capital requirement, see e.g. \citet{planchet_adding_2024} among others, or in pricing applications, see e.g. \citet{chen_modeling_2009,li_pricing_2023}. 
Incorporating jumps whose effects gradually disappear over several years is part of model enhancement. Recently, \citet{robben_catastrophe_2024} develop a multi-population approach comprising a regime-switching model with a high-volatility state where Gaussian shocks on the mortality trend can occur. Their model involves maximum likelihood estimation on improvement rates in several steps and requires preprocessing of data labeled as outlines. Another recent example is provided by \citet{goes_bayesian_2023} who use a Bayesian formulation and consider the impact of shocks that gradually diffuse over several years.
So far, very few mortality models incorporating a temperature-related effect have been developed, despite the abundance of epidemiological literature on the effects of heat, cold, or heatwaves on mortality, as discussed in Section~\ref{sec:back}.
\citet{seklecka_mortality_2017} extend the framework of the \citet{lee_modeling_1992} and \citet{plat_stochastic_2009} models by adding a component related to temperature-related mortality. Their model is applied to data from the United Kingdom. \citet{li_joint_2022} implement an approach based on extreme value theory on US monthly data to model the dependence between extreme temperatures, both hot and cold, and excess mortality.
For these different approaches, the generated mortality shocks are estimated from outliers typically observed in annual mortality rates. These outliers are generated by catastrophic events, such as pandemics, wars or other major events. In contrast, temperatures are responsible for lower-intensity mortality fluctuations that vary spatially, incorporate lagged effects, and are observed over specific periods of the year. Additionally, compensations between the effects of hot and cold temperatures can be observed over a calendar year. For instance, heatwave-type shocks are acute and can be partly or totally compensate throughout the year due to harvest effects, i.e. the decrease in mortality observed after a peak of heat-related mortality. This effect has been observed notably in the French population \citep{pascal_heat_2018}, but is not systematically found in all studies. For these reasons, approaches based on annual mortality data may encounter difficulties in being extended to model the effect of
future temperatures on mortality.  Another important aspect to consider is the increasing impact of climate change, particularly the consequences of rising temperatures and extreme heat events. This trend tends to limit the relevance of models in which shocks are estimated based on previous annual observations and suggests studying the temperature-mortality relationship.

It is almost universally observed that the mortality response to hot and cold temperatures follows a U- or V-shaped curve, depending on the population \citep{kunst_outdoor_1993, guo_gasparrini14}. These effects are generally modeled using daily data through time-series regression models that capture exposure–response associations.
From the aggregated temperature-mortality time series data, two traditional models are explored to capture the nonlinear correlation between temperature and mortality.
\citet{muggeo_modeling_2008} introduces the constrained segmented distributed lag (CSDL) model. This model assesses the effects of heat and cold using a fixed threshold within a Poisson regression framework. Recently, \citet{pincemin_risques_2021} implement it to measure the effect of temperatures on projected life expectancy of the French population. Alternatively, the distributed lag non-linear generalized model (DLNM)\citep{gasparrini_distributed_2010, gasparrini_modeling_2014} is a reference in epidemiology and has been used in numerous studies, see among others \citet{guo_gasparrini15, gasparrini_projections_2017, vicedo-cabrera_burden_2021}. It estimates the number of deaths attributable to temperature effects over specific periods, accounting for different lag effects depending on whether the temperatures are hot or cold. In a prospective approach, these temperature variations should be studied simultaneously, as the number of deaths induced by cold temperatures is expected to decrease while that induced by hot temperatures is expected to increase in Europe due to the climate change \citep{martinez-solanas_projections_2021}.

The numerical application developed in this paper focuses on the use of a mortality model for two populations using the augmented common factor approach introduced by \citet{li_coherent_2005}. Many multi-population mortality models have been developed in the literature in recent years. Most of these models are based on the framework proposed by \citet{lee_modeling_1992} or the CBD models \citep{cairns_two-factor_2006}, as reviewed by \citet{villegas_comparive_2017}. These models typically assume coherence or semi-coherence \citep{li_semicoherent_2017} among several sub-populations. In our paper, we impose a coherence assumption between women and men in France regarding mortality not affected by temperatures. Although the long-term convergence hypothesis is challenging to demonstrate for women and men, this modeling illustrates the possibility of coupling the DLNM climate epidemiology model with a standard multi-population mortality model. This issue is indeed essential given the importance of multi-population mortality models \citep{antonio_producing_2017} and the need to account for geographical factors when modeling temperature dynamics.

The contribution of our paper is fourfold. First, we develop a new specification of a multi-population mortality model that incorporates the effect of temperature variations on central mortality rates. These effects require a preliminary phase of estimating temperature-attributable deaths using a DLNM model. By subsequently modeling deaths not attributable to temperatures through a multi-population model, our approach associates temperature-related and non-temperature-related deaths while preserving the Poisson assumption of the number of deaths. To the best of our knowledge, this is a novel approach in the literature enabling an easy coupling of most single-population or multi-population mortality models with temperature. We illustrate it using the \citet{li_coherent_2005} model on French mortality data. Once calibrated, our modeling framework can project future mortality by incorporating various temperature-related effects. Second, the temperature effects are calibrated by sex and age groups, allowing us to integrate their specific sensitivities to temperature into the projections. Third, our multi-population mortality model is associated to temperature projections from 12 climate models across three Representative Concentration Pathways (RCP) scenarios for the period 2020-2100. Then, we evaluate the loss or gain in life expectancy due to future temperature trends in these different climate scenarios in France, highlighting the specific impact of extreme heatwaves. Furthermore, this approach allows us to measure the impact of uncertainties caused by different climate models. Fourth, we examine the sensitivity of mortality projections to the local temperature dynamics of various French cities. This reveals divergences and a potentially new source of heterogeneity in mortality projections.

This paper is organized as follows. Section~\ref{sec:back} presents a brief review of the literature on the relationship between mortality and temperature. Section~\ref{sec:method} introduces the notations and the modeling framework used. In particular, we describe our multi-population mortality model for all causes of death excluding temperature effects and explain how temperature-attributable mortality is incorporated through the DLNM model. Section~\ref{sec:estimation} details the various estimation procedures adopted, starting with the DLNM model. We then explain how to calculate the total temperature-attributable fractions, followed by the estimation of the mortality model. Section~\ref{sec:forecast_method} describes the calibration of time series models used to project mortality trends. In addition, we present the projection of the attributable fractions based on climate scenarios and the life expectancy loss due to temperature. In Section~\ref{sec:application}, we estimate and project the series of central death rates by adding the effect of future temperatures. We then quantify the gains and losses in life expectancy induced by temperatures up to 2100 in Metropolitan France under different RCP scenarios and for different locations. Finally, Section~\ref{sec:conclusion} concludes the paper. 

\section{Background on temperature-mortality effects\label{sec:back}}

This section presents a brief overview of the influence of temperatures on human mortality. These changes can have diverse consequences, directly impacting the body's function during extreme cold or heat situations, or indirectly affecting our environment-leading to issues such as disease outbreaks, food shortages, and deteriorating living conditions. Our specific focus lies in examining the effects of temperature fluctuations on mortality rates. For a comprehensive summary of other consequences of climate change on mortality, including aspects like vector-borne diseases, under-nutrition, and related factors, we refer to the sixth assessment report of the IPCC \citep[][TS.C.6]{ipcc_ar6_wg2_ts}.

\subsection{The effects of heat and cold on human body\label{subsec:heat}}

Human beings are homeotherms, equipped with various protective mechanisms to maintain stable internal temperatures. Prolonged exposure to high external temperatures can elevate internal heat, while cold temperatures can lower it \citep{beker18}. During periods of high heat, the body experiences several effects, including increased sweating, elevated heart rate, lower blood pressure, confusion, heat exhaustion, dehydration, heatstroke, heat rash, fainting, and heat edema, among other symptoms \citep{who18}. 
To counter high temperatures, the body employs external mechanisms such as sweating and internal mechanisms that involve the production of specific proteins to safeguard internal cells \citep{bouchama02}. However, these mechanisms have limitations and become inadequate with prolonged exposure. This can lead to more severe conditions like hyperthermia, identifiable through symptoms like rapid pulse, headaches, nausea, and loss of consciousness \citep{bouchama02}. Severe dehydration is another consequence, carrying significant cardiovascular and pulmonary risks, potentially resulting in a rapid increase in body temperature and a heightened risk of fatality \citep{whoEU04}. The impact of heatstroke and dehydration during heatwaves can persist for up to three days post-exposure \citep{alsaiqali22}. The risks escalate when high temperatures endure. Heatwaves pose greater threats when protective measures for vulnerable populations are lacking. Elderly individuals, those with existing health conditions, pregnant women, and outdoor workers face higher vulnerability. Consequently, targeted protection plans are implemented for these groups.

As with hot temperatures, cold temperatures elicit various physiological responses in individuals, see \citet{ryti_global_2016} for a review. Upon exposure to cold, initial symptoms include reduced blood circulation in extremities, arrhythmia, breathing challenges, muscle discomfort, and restricted mobility.
Prolonged exposure to cold can increase levels of hypothermia. As the body temperature drops, the heart rate decreases, and severe bradycardia can occur when the internal temperature falls below $32^{\circ}\text{C}$, potentially resulting in fatality \citep{beker18}.
Cold weather directly impacts heart and pulmonary health, notably affecting cardiovascular diseases (CVD). Particularly, mortality factors associated with cold weather include coronary heart disease, stroke and respiratory disorders. These health risks are notably prevalent in socioeconomically disadvantaged groups, and increase when temperature falls below about $0^{\circ}\text{C}$. Paradoxically, mortality rates also rise sharply when ambient temperature falls in regions with milder winters \citep{mercer03}.

\subsection{Seasonal variations in mortality\label{subsec:season}}

The climate and seasonal variations significantly influence human mortality. 
Episodes of excess mortality can be observed during periods of heat, cold, or substantial temperature fluctuations, with characteristics varying by region. These differences in mortality patterns can be considerable across studied areas, yet the temperature-mortality relationship is generally assessed as having a U or V--shaped-relationship \citep{kunst_outdoor_1993, guo_gasparrini14}. Both heat and cold increase mortality risk: cold effects can persist over several weeks, while heat impacts are immediate but shorter-lived.

Each region exhibits an optimal temperature, known as the Minimum Mortality Temperature (MMT), see \citet{yin_mapping_2019} for a current and projected overview across the planet. This temperature corresponds to the point at which the death rate is the lowest, reflecting the population's acclimatization to its surroundings. The MMT varies significantly across countries, influenced by latitude, climate type, and average annual temperature, ranging from 14.2 to 31.1°C \citep{tobias21}. The observed MMT depends on the timeframe, serving as a crucial indicator of population adaptation. For instance, in France, the MMT is reported to have increased from 17.5°C (1968-1981) to 18.2°C (1996-2009) \citep{todd15}. Globally, for every 1°C rise in average annual temperatures, the MMT increased by 0.8°C \citep{tobias21}. Of course, socioeconomic factors also influence these thresholds \citep{krummenauer19}.

Sensitivity to ambient temperatures varies by age, sex, health status, and social category \citep{who11}. For example, the elderly (over 65 years) and young children (0-5 years) in tropical areas are particularly vulnerable to temperature fluctuations \citep{egondi12}. Among the elderly, women have a higher risk of heat-related mortality during heatwaves \citep{maridellolmo19,yu10}. Geographic regions differ in their exposure to temperature variations and their capacity to respond to climate change, reflecting population acclimatization. Urban density within the same geographical area amplifies the risk of heat-related mortality, positively correlated with population density, air quality (e.g., fine particle concentration), or social inequality \citep{sera19}. 

In countries with temperate and continental climates, mortality follows a distinct seasonal pattern \citep{madaniyazi22}. Mortality rates are higher in cold seasons compared to warm seasons, a trend observed globally but with varying seasonal amplitudes between regions. Temperature substantially explains this pattern, as stated in the literature \citep{ipcc_ar6_wg2_ts}, although the relation between mortality and temperatures can involve complex physiological aspects or indirect environmental and social aspects.

In tropical regions, besides temperature, rainfall plays also a crucial role. During monsoon periods, diseases like malaria, diarrheal diseases, or vector-borne illnesses spread more rapidly, significantly affecting women \citep{egondi12}. These diseases, identified by the World Health Organization (WHO) as major concerns related to climate change, contribute significantly to mortality \citep{who14}.

Additionally, the human body's cooling mechanisms, primarily reliant on perspiration, can be compromised if ambient humidity levels are either too low or too high. From a physiological standpoint, scientific studies have highlighted the potential health hazards associated with extreme temperatures in highly humid environments \citep{davis_humidity_2016}. However, epidemiological evidence regarding the consequences of wet-hot events seems remain less conclusive \citep{armstrong_role_2019}, while \citet{fang_joint_2023} indicating an increased risk of mortality in the presence of dry-hot events.

\subsection{Temperature related deaths : past and expectation\label{subsec:death}}

Over the period 1998-2017, the WHO estimates the number of heatwave induced deaths at over 166,000 worldwide, including 70,000 deaths in Europe in the summer of 2003 alone \citep{who17}. The number of people exposed to heatwaves is also expected to continue to rise. In Europe, it is 9.3 million and would reach almost 300 million in 2100 in a +3°C scenario \citep{europe20}. The annual number of deaths linked to heatwaves is currently estimated at more than 61,000 in Europe during the summer in 2022 and would be more than 120,000 in 2050 \citep{ballester_heat-related_2023}. However, it is important to note that its results strongly depend on the definition given for a heatwave, see~Appendix~\ref{sec:app_hist_france}. In European countries and Japan, where the proportion of older people in the population is increasing, the damage from heat could be even more severe in the long term. Indeed, the average temperature of the planet has already increased by +1.1°C \citep[][TS.C.6]{ipcc_ar6_wg2_ts} since the pre-industrial era and is expected to continue to rise under any scenario. Thus, extreme weather events such as heatwaves are expected to become more frequent while at the same time the probability of cold waves is expected to decrease \citep{ipcc_ar6_wg2_ts_chap16}. The seasonality of deaths observed today in temperate climates could strongly evolved. Climate change will reduce extreme cold spells and winter seasons period \citep{ipcc_ar6_wg2_ts_chap16}. This would conduct to reduce the cold exposure and cold fatalities \citep{madaniyazi21,europe20} when most of the temperature related deaths was caused by cold period \citep{guo_gasparrini15}. 
In some parts of Europe, heat-associated health effects may even outweigh the effects of cold \citep{gronlund18, martinez-solanas_projections_2021}.

So far, no real mitigation or change in the distribution of annual deaths has already been observed in temperate countries \citep{cheng15}, which might explain a lesser focus in demographic or actuarial literature. 

\section{A multi-population mortality model with temperature effects\label{sec:method}}

In this section, we explain our approach to incorporating temperature effects into our mortality model. After introducing the data and notations in Section~\ref{subsec:data_notations}, Section~\ref{subsec:LL_model} presents the proposed dynamics for central mortality rates excluding temperature effects, followed by the Poisson formulation used to estimate the model parameters. Finally, Section~\ref{subsec:daily_death} describes the DLNM model used to estimate temperature-attributable deaths.

\subsection{Notation, death counts and temperature-attributable deaths\label{subsec:data_notations}}
 
This paper considers two distinct time scales, daily and annual, as well as two human populations (women and men) in Metropolitan France. In the following, some notations are introduced for modeling mortality. For clarity, bold notations are used for vectors and matrices.

Let us introduce the notations for an annual mortality model that captures long-term trends by age and period in mortality rates. This model is estimated using yearly data observed during a calibration period.
More precisely, we use annual mortality data derived from the series of observed numbers of deaths and observed exposure to risk, both available from the \citet{human_mortality_database_university_2024}. Further details on these data sources are provided in Section~\ref{subsec:source}. 

Let $\mu_{x,t}^{(g)}$, $E_{x,t}^{(g)}$, and $D_{x,t}^{(g)}$ represent, respectively, the force of mortality, observed exposure to risk, and observed number of deaths for sex $g \in \mathcal{G} = \{\text{female}, \text{male}\}$ within the age range $[x,x+1)$ and the calendar year $[t,t+1)$, where $x \in \mathcal{X}= \{x_{\min}, \ldots, x_{\max}\}$ denotes the set of integer ages, and $t \in \mathcal{T}_y = \{ y_{\min}, \ldots, y_{\max} \}$ represents the set of calendar years considered as the calibrated period for mortality models.
Common stochastic mortality models such as the Lee-Carter model \citep{lee_modeling_1992} rely on the crude central death rate of mortality $\widehat{m}_{x,t}^{(g)} = D_{x,t}^{(g)}/E_{x,t}^{(g)}$. It is equivalent to the estimated force of mortality $\widehat{\mu}_{x,t}^{(g)}$ assuming a piecewise constant force of mortality. Traditional mortality models may also use the mortality rate $\widehat{q}_{x,t}^{(g)} \simeq 1- \exp{(- \widehat{\mu}_{x,t}^{(g)})}$ under the classical piecewise constant force of mortality assumption.

In the following sections, our focus is on the dynamics of the annual number of deaths, denoted as $D_{x,t}^{(g)}$. We assume that the annual number of deaths can be decomposed into two components
\begin{equation}\label{eq:decom_nb_deaths}
D_{x,t}^{(g)} = \widetilde{D}_{x,t}^{(g)} + \widebar{D}_{x,t}^{(g)},
\end{equation}
where $\widetilde{D}_{x,t}^{(g)}$ represents the number of deaths at age $x$ for sex $g$ and for year $t$ not attributable to temperature effects, i.e. death counts for all causes of death excluding those attributed to temperature.  The term $\widebar{D}_{x,t}^{(g)}$ corresponds to the number of deaths attributable to temperature variations during the year. 
Additionally by dividing by $E_{x,t}^{(g)}$, we denote $\widebar{m}_{x,t}^{(g)}$ and $\widetilde{m}_{x,t}^{(g)}$  as the components of the crude central death rate attributable and not attributable to temperature, respectively, such as
\begin{equation}\label{eq:decom_mxt}
\widehat{m}_{x,t}^{(g)} = \widetilde{m}_{x,t}^{(g)} + \widebar{m}_{x,t}^{(g)}.
\end{equation}
This relationship reveals two causes of mortality. The central death rate $\widebar{m}_{x,t}^{(g)}$ associated with the temperature component of mortality captures a portion of the observed seasonal variations in daily deaths.
We can also express the all-cause central death rate in a multiplicative form
\begin{equation}\label{eq:multi_deaths_rates}
\widehat{m}_{x,t}^{(g)} = \widetilde{m}_{x,t}^{(g)}T_{x,t}^{(g)},
\end{equation}
where $T_{x,t}^ {(g)}  = \left( 1 - \text{AF}^{(g)}_{x,t}\right)^{-1}$,
and $\text{AF}_{x,t}^{(g)}$ is the total attributable fraction related to temperature in year $t$ at age $x$ for sex $g$
\begin{equation}\label{eq:attributable_fract}
\text{AF}_{x,t}^{(g)}= \dfrac{\widebar{D}_{x,t}^{(g)}}{D_{x,t}^{(g)}}.
\end{equation}

To use the decomposition~\eqref{eq:decom_nb_deaths} and then calculation of the total attributable fraction~\eqref{eq:attributable_fract}, it is necessary to establish a count of deaths attributable to temperature (heat, cold, or both). However, to the best of our knowledge, this categorization of deaths is not explicitly recorded. Two approaches are commonly employed in the literature to estimate temperature-attributable deaths. The first approach involves a detailed examination of causes of death, based notably on the International Classification of Diseases (ICD) established by the World Health Organization (WHO). This classification contains codes corresponding to diseases, symptoms, signs, or medical observations. The temperature-attributable fraction can then be estimated for different causes of death, following methodologies developed by the Global Burden of Disease Study, see e.g. \citet{burkart_estimating_2021} or \citet{song_ambient_2021}. The second approach considers all-cause mortality and estimates the component attributable to temperature using a time-series regression model \citep{bhaskaran_time_2013}. In epidemiology, this method explains short-term variations in an outcome as a function of variations in other factors. It is particularly applied to analyze the relationship between temperature and mortality, based on the approach developed by \citet{gasparrini_distributed_2010}. Their model serves as a reference methodological framework for numerous studies investigating the seasonality of deaths and the mortality response to external factors, as discussed in Section~\ref{sec:back}. It is important to note that in both cases, the number of deaths attributable to temperature remains an estimated quantity and not an observed one. The number of deaths $\widetilde{D}_{x,t}^{(g)}$, which is also unobserved, can either be estimated with a model or calculated by the difference between observed all-cause death counts and the estimate of $\widebar{D}_{x,t}^{(g)}$.

To precisely estimate the annual death counts $\widebar{D}_{x,t}^{(g)}$, we focus on daily mortality and the impact of temperature on intra-annual seasonal variations in mortality. The daily number of temperature-attributable deaths is fitted within the daily calibration period $\mathcal{T}_d = \{ d_{\min}, \ldots, d_{\max} \}$, which includes all the days encompassed in the set $\mathcal{D}^\star = \left\lbrace 1,2,\ldots, 365, (366)\right\rbrace$ over the entire calibration period $\mathcal{T}_y$, as detailed in Section~\ref{sec:estimation}. Specifically, to evaluate the impact of temperature, we gather daily records of the death counts over the calibration period which are sourced from the \textit{Institut national de la statistique et des études économiques} \citep{insee_data}. Hereto, we denote the number of deaths in day $d \in \mathcal{D}^\star$ of year $t \in \mathcal{T}_y$ for sex $g \in \mathcal{G}$ as $D_{x,t,d}^{(g)}$. It can be decomposed between attributable and non-attributable death counts, respectively $\widebar{D}_{x,t,d}^{(g)}$ and $\widetilde{D}_{x,t,d}^{(g)}$, as in Equation~\eqref{eq:decom_nb_deaths}. It is noteworthy that the data series for observed exposure to risk are typically not available on a daily basis.

			\subsection{Incorporating the effect of temperature into the Li-Lee model} 
 			\label{subsec:LL_model}
 			
Our approach involves introducing first a baseline stochastic mortality model for crude central death rates not attributable to temperature $\widetilde{m}_{x,t}^{(g)}$. A wide range of mortality models is conceivable for a single population or multiple populations. However, these models primarily focus on all-cause mortality and do not differentiate temperature-related effects. Our approach assumes that long term mortality trends are consistently observed in central death rates not attributable to temperature effects. Among the different mortality models, we opt for the \citet{li_coherent_2005} model, also known as the Augmented Common Factor (ACF) model. Here, the dynamics of mortality is defined for one country and two sexes.

Several reasons motivate this choice. Primarily, a model employing a biological coherence principle is favored here for projecting populations by sex. This choice allows us to observe whether the addition of a temperature component at a later stage diverges the mortality between men and women. In the context of multi-population mortality models, other authors emphasize a shared coherence assumption across multiple countries, see \citet{dowd_gravity_2011, kleinow_common_2015, li_coherent_2017, bergeron-boucher_coherent_2017} among others. The selection of the ACF model is justified by its status as a benchmark model in multi-population projections, building upon the Lee-Carter model. It's indeed one of the most commonly used mortality models across the literature, as highlighted by \citet{antonio_producing_2017} for mortality projections in the Netherlands and Belgium. It is suitable for modeling mortality at all ages, whereas models like CBDX type models \citep{dowd_cbdx_2020} are generally better suited for retired populations. In addition, it should be noted that the approach subsequently developed to integrate the effect of temperatures is not dictated by the choice of the mortality model; other specifications based on forces of mortality or mortality rates are indeed entirely possible.

We introduce our multi-population mortality model based on the \citet{li_coherent_2005} model for the crude central death rates not attributable to temperature effects as
\begin{equation}\label{eq:2_LL}
\ln \left(\widetilde{m}_{x,t}^{\left(g\right)} \right)= A_x + B_{x}K_{t} + \alpha_{x}^{\left( g \right)} + \beta_{x}^{\left( g \right)} \kappa_{t}^{\left( g \right)}.
\end{equation} 
This model imposes a common unisex trend $K_{t}$ in the logarithm of crude central death rates over time and considers specific dynamics by sex $\kappa_{t}^{\left( g \right)}$. The parameters $A_{x} $ and $\alpha_{x}^{\left( g \right)}$ represent level parameters at age $x$ for mortality, while the parameters $B_x$ and $\beta_x^{(g)}$ modulate the trend of the logarithm of death rates by age and by sex respectively.

The assumption of long-term coherence is ensured by the dynamics given to the series $K_t$ and $\kappa_t^{(g)}$
\begin{align}  \label{eq:LL_time_spe-1}
K_{t} &= \delta + K_{t-1} + e_{t},
\end{align}
\begin{align}  \label{eq:LL_time_spe-2}
\kappa_{t}^{\left( g \right)} &= c^{\left( g \right)} + \phi^{\left( g \right)}  \kappa_{t-1}^{\left( g \right)} + r_{t}^{\left( g \right)}.
\end{align}
Here, we consider a first-order autoregressive (AR) model with drift for $\kappa_t^{(g)}$ with estimated coefficients $c^{\left( g \right)}$ and $\phi^{\left( g \right)}$, and a random walk with drift (RWD) for $K_t$ with a drift parameter $\delta$. As proposed by \citet{antonio_producing_2017, antonio_iabe_2020}, the innovation errors $\left(e_{t}, r_{t}^{\left( f \right)}, r_{t}^{\left( m \right)}\right)$ is assumed to be a vector of Gaussian white noises with a mean of zero and a variance-covariance matrix $\Sigmavec$. The parameters $c^{(g)}$ and $\phi^{\left( g \right)}$ are specific to sex $g$, and are estimated as described in Section~\ref{subsec:baseline_time_series}. This specification ensures that the mortality of women and men does not diverge in the long run.

For calibrating this model, we apply the classical identifiability constraints for ensuring parameter uniqueness and efficient computations

\begin{equation}\label{eq:contraint_li_lee_common}
\sum\limits_{t \in \mathcal{T}_y} K_{t} = 0 \text{ and } \sum\limits_{x \in \mathcal{X}} B_{x}^2 = 1.
\end{equation}
\begin{equation}\label{eq:contraint_li_lee}
\sum\limits_{t \in \mathcal{T}_y} \kappa_{t}^{\left( g \right)} = 0 \text{ and } \sum\limits_{x \in \mathcal{X}} (\beta_{x}^{\left( g \right)})^2 = 1, \text{ for } g \in \mathcal{G}.
\end{equation}

Finally, building on the specification from \citet{brouhns02}, we assume that the number of deaths not attributable to temperature at age $x$ and in year $t$ for sex $g$ follows a Poisson distribution
\begin{equation}\label{eq:poisson_brouhns}
\widetilde{D}_{x,t}^{(g)} \sim \text{Pois}\left( E_{x,t}^{(g)}  \widetilde{m}_{x,t}^{(g)}\right).
\end{equation}
This model cannot be estimated directly because these death counts are not observed and depend on how the mortality-related temperature component is defined. Therefore, we reformulate the relation~\eqref{eq:poisson_brouhns} using Equation~\eqref{eq:attributable_fract} to reintroduce the (observable) all-cause death counts as  $D_{x,t}^{(g)} = T_{x,t}^ {(g)}  \widetilde{D}_{x,t}^{(g)}$. We deduce again a Poisson formulation with log-link function for ${D}_{x,t}^{(g)}$ as 
\begin{equation}\label{eq:poisson_adj_expo}
{D}_{x,t}^{(g)} \sim \text{Pois}\left( E_{x,t}^{(g)} T_{x,t}^{(g)}  \widetilde{m}_{x,t}^{(g)}\right).
\end{equation}

As we see in Section \ref{subsec:excess_mortality}, the prior calculation of total attributable fractions requires processing daily data to estimate the effects of seasonal temperature fluctuations on mortality, and then summing these effects on an annual basis to obtain the terms $\text{AF}^{(g)}_{x,t}$ and $T_{x,t}^{(g)}$. 
From Equation~\eqref{eq:poisson_adj_expo}, the model is estimated as a Poisson generalized linear model (GLM) through maximum likelihood estimation, as described in Section \ref{subsec:baseline_estimation}. We consider $\ln{\left(E_{x,t}^{(g)} T_{x,t}^{(g)}\right)}$ as an offset term when we compute the predicted number of deaths based on the GLM structure
$$
\ln{\left(\esp{{D}_{x,t}^{(g)}}\right)} = \ln{\left(E_{x,t}^{(g)} \right)} + \ln{\left(T_{x,t}^{(g)}\right)} + \ln{\left(\widetilde{m}_{x,t}^{(g)}\right)},
$$
for each age $x$, sex $g$ and year $t$ based on the fitted model from~Equation~\eqref{eq:2_LL}.

\subsection{Modeling daily deaths with the distributed lag non-linear model 
\label{subsec:daily_death}}

We now aim to model the influence of temperature on the daily number of deaths using the distributed lag non-linear model. We begin by describing the general framework of this model in Section~\ref{subsec:general_framework}, followed in Section~\ref{subsec:dlnm_mortality} by the precise specification adopted in our study.

\subsubsection{General framework}\label{subsec:general_framework}

The so-called distributed lag models (DLMs) \citep{almon_distributed_1965} are commonly employed by econometricians and epidemiologists to address the impact of delayed environmental factors on a response variable. 
\citet{armstrong_models_2006, gasparrini_distributed_2010} proposed a more versatile approach known as distributed lag non-linear generalized model (DLNM). This model, based on a time series regression framework, has the ability to capture the short-term and nonlinear effects of explanatory variables on a response variable, as well as the seasonal and long-term trend effects. It has found extensive use in the epidemiological literature related to mortality and temperature, see \citet{gronlund18,guo_quantifying_2018, vicedo-cabrera_multi-country_2018,lee_projections_2020,vicedo-cabrera_burden_2021} among others. 

The model employs an exposure–lag–response function that captures the lagged impacts of exposure variables, which may persist several days after exposure. Indeed, immediate and delayed effects on a daily basis are critical to consider in the temperature-mortality relationship. Specifically, during a heatwave, the most vulnerable populations, particularly the elderly, are the first to be affected. This group experiences an immediate increase in mortality, possibly followed by a period of reduced mortality called the harvesting effect. Conversely, the effects of cold temperatures usually manifest after a delay of several days.

The DLNM is designed to examine the impact of exposure variables, such as temperature, humidity or air pollution, on univariate time series data of the outcome, e.g. the number of deaths, considering delayed effects and nonlinear relationships. This model describes the relationship between a time series $(Y_d, d \in \mathcal{T})$ for a set of indices $\mathcal{T}$ as follows
\begin{equation}\label{eq:dlnm_general}
\rho(\esp{Y_d}) = \eta + \sum_{j=1}^{J}{s_j(x_{d,j},L; \thetavec_j)}
+ \sum_{m=1}^{M}{r_m(u_{d,m};\gammavec_m)},
+ \sum_{p=1}^{P}{h_p( z_{d,p} ; \zetavec_p )},
\end{equation}
where $\rho(\cdot)$ is a monotonic link function and the outcome variable $Y$ is assumed to belong to the exponential family of distributions \citep{mccullagh_generalized_1989}. The mean $\esp{Y_d}$ 
is modeled based on different explanatory variables $\left( x_{d,1}, \ldots, x_{d,J}\right)$, $\left( u_{d,1}, \ldots, u_{d,M}\right)$, and $\left( z_{d,1}, \ldots, z_{d,P}\right)$ for $J, M, P \in \N^\star$. The terms $\eta$, $\thetavec_{j}$, $\gammavec_{m}$ and $\zetavec_{p}$ represent the unknown parameters to be estimated.

The functions $s_j(\cdot,\cdot)$, $j \in \{ 1, \ldots, J \}$, represent smooth bi-dimensional functions of the parameters $\thetavec_j$ that capture delayed effects of past exposures $\qvec_{x_j,d} = \left(x_{d,j}, x_{d-1,j}\ldots, x_{d-L,j}\right)^\top$ with a lag $L \in \N$. \citet{gasparrini_distributed_2010, gasparrini_modeling_2014} specify these smooth functions using the so-called \textit{basis} transformation \citep{wood_generalized_2006}, which refers to a space of known functions, e.g. polynomial functions, splines or B-splines. The transformation of a variable $x_{d,j}$ using a basis function is called a \textit{basis variable}. Following this methodology and omitting the index $j$ to simplify the notation, each function $s_j(\cdot,\cdot)$ can be expressed as a generic function $s(\cdot, \cdot)$ 
\begin{equation}\label{eq:bidim_sline}
s(x_{d} , L; \thetavec ) = \int_{0}^{L}{f\cdot w(x_{d-l},l;\thetavec)dl} \approx \sum_{l=0}^{L}{f\cdot w(x_{d-l},l;\thetavec)},
\end{equation}
where $f\cdot w \left( \cdot, \cdot \right)$ is a bi-dimensional integrable function, called the \textit{exposure–lag–response function}. This function is constructed using the concept of \textit{cross-basis function}, which corresponds to the bi-dimensional function space obtained by combining two independent sets of basis functions, $f(\cdot)$ and $w(\cdot)$, with respective dimensions $v_{x} \in \N^\star$ and $v_{l}\in \N^\star$. Formally, the function $s(\cdot, \cdot)$ can also be expressed as a tensor product
\begin{equation}\label{eq:bidim_tensor}
s(x_{d}, L ; \thetavec ) \approx \wvec_{x,d}^\top \thetavec
= \left( \textbf{1}_{v_{x} \cdot v_l}^\top \Avec_{x, d}\right) \thetavec, 
\end{equation}
where $\textbf{1}_{v}$ is a $v$-dimensional vector of 1's, and $v_{x} \cdot v_l$ represents the dimension of the cross-basis function $f\cdot w$. It is important to note that an intercept term should not be included in $f(\cdot)$ to ensure identifiability. The matrix $\Avec_{x, d}$ is defined as
$$
\Avec_{x, d} = \left( \textbf{1}_{v_l}^\top \otimes \Rvec_{x,d} \right)
\odot \left( \Cvec  \otimes  \textbf{1}_{v_{x}}^\top\right),
$$
where $\Cvec$ is a $(L+1)\times v_l$ matrix of basis variables from $w(\cdot)$, i.e. the transformation of the lag vector $\left( 0, \ldots, L\right)^\top$ using $w(\cdot)$, and  
$\Rvec_{x,d}$ is a $(L+1) \times v_x $ matrix obtained by applying $f \cdot w(\cdot,\cdot)$ to the vector of exposures $\qvec_{x,d}$.  The symbols $\otimes$ and $\odot$ respectively denote the Kronecker product and the Hadamard product. We introduce the matrix of cross-basis functions $\Wvec$ of dimension $\vert \mathcal{T} \vert \times (v_x \times v_l)$ composed of the terms $\wvec_{x,d}$ from Equation~\eqref{eq:bidim_tensor}.

The functions $r_m\left( \cdot \right)$, $m \in \{ 1, \ldots, M \}$, in Equation~\eqref{eq:dlnm_general} are smooth univariate functions that capture the effects of confounding daily variables $\uvec_{d,m}$. Linear effects can also be incorporated. For instance, \citet{gasparrini_distributed_2010} note that air pollution is usually modeled using a linear relationship in studies focusing on the mortality-temperature association. 

The outcome variable is likely influenced by exposure variables and daily confounding variables, but it may also experience short-term residual seasonal fluctuations and long-term effects. In the context of time series regression, it is therefore crucial to finely control for the effects of seasonal phenomena, such as work and holiday periods, weekends, and long-term patterns, in order to properly isolate the short-term exposure-outcome association of interest, as emphasized by \citet{bhaskaran_time_2013}. Various approaches can be adopted to control for these effects. The authors notably highlight the advantages of using spline functions of time instead of time stratified model or Fourier terms since they often fail to adequately capture non-seasonal components. Thus, we introduce functions $h_p(\cdot)$, $p \in \{ 1, \ldots, P \}$, as smooth univariate functions of categorical time variables $z_{d,p}$, e.g. day, day of the week, month or year. They capture residual seasonal effects, demographic shifts, or other long-term trends not accounted for by the other covariates.

\subsubsection{The DLNM model for daily mortality}\label{subsec:dlnm_mortality}

We now examine the modeling of daily deaths using the DLNM framework presented above. In our study, we use daily average temperature as the environmental factor, along with its delayed effects.

The DLNM is not explicitly designed to incorporate both age and time dimensions simultaneously.
Hence, to address this limitation, we adopt an approach proposed by \citet{vicedo-cabrera_hands-tutorial_2019} and introduce age and sex effects by fitting a model for age buckets and sex.
To achieve this, we partition the age range $\mathcal{X}$ into $K \in \N^\star$ distinct strata $\mathcal{X}_k=[x_{k-1}, \mathcal{X}_k), k \in \{1, \ldots, K\}$. We aggregate the observed number of deaths for sex $g$ within each stratum $\mathcal{X}_k$, and denote it as $D_{k,t,d}^{(g)}$ for $k \in \{1, \ldots, K\}$ on day $d \in \mathcal{D}^\star$ of year $t \in \mathcal{T}_y$
$$
D_{k,t,d}^{(g)} = \sum_{x \in \mathcal{X}_k}{D_{x,t,d}^{(g)}}.
$$

Let $\lambda_{k,t,d}^{(g)} = \esp{D_{k,t,d}^{(g)}} $ represent the expected number of deaths for day $d$, year $t$, sex $g$ and sub-group $k$. $\vartheta_{d,t}$ denotes the daily average temperature of day $d$ for year $t$.
We assume that $D_{k,t,d}^{(g)}$ follows a Poisson distribution with overdispersion and a canonical log-link.
Using the general formulation of the DLNM model in Equation~\eqref{eq:dlnm_general}, we specify the following quasi-Poisson regression model within each subgroup $k \in \{1, \ldots, K\}$ and sex $g \in \mathcal{G}$ as follows
\begin{equation}\label{eq:dlnm_myspec}
\ln(\lambda_{k,t,d}^{(g)})=\eta_k^{(g)} + s(\vartheta_{d,t}, L; \thetavec_k^{(g)} ) 
+ \sum_{p=1}^{P}{h_p( z_{d,p} ; \zetavec_{k,p}^{(g)})},
\end{equation}
where $\eta_k^{(g)}$, $\thetavec_{k}^{(g)}$ and $\zetavec_{k,p}^{(g)}$ represent the parameters to be estimated.

The function $s (\vartheta_{d,t} , L ; \thetavec_{k}^{(g)})$ denotes a cross-basis non-linear function that simultaneously specifies the association between $\vartheta_{d,t}$ and its lag structure over a maximum of $L$ days. This function captures the short-term cumulative association between mortality and temperature over the period $L$. Based on the work of \citet{gasparrini_projections_2017, lee_projections_2020, martinez-solanas_projections_2021}, we consider for the exposure–response curve $f(\cdot)$ a natural cubic B-spline with three internal knots that are placed at the 10th, 75th, and 90th percentiles of the daily average temperature distribution within the observational period. This choice of a cubic spline and knots allows for easier extrapolation of results to extreme temperatures. We also use a natural cubic B-spline for the lag-response curve $w(\cdot)$ with an intercept and three internal knots placed at evenly spaced intervals on the log scale. A maximum of $L=21$ days is considered to account for the short-term delayed effects of both cold and hot temperatures.
  
In studies focusing on daily mortality, seasonal variations in mortality are observed and well-documented in the literature \citep{madaniyazi22}. This seasonality is only partially explained by temperature effects. Moreover, it is also crucial to consider trends in improving life expectancy and changes in exposure to risk when fitting the model using data observed on many years.
To control these effects, we consider the day of the week indicator and the day number within the year as the time categorical variables $z_{d,p}$. In our specification, we define the functions $h_p(\cdot)$ following the approaches of \citet{guo_gasparrini15, gasparrini_projections_2017, vicedo-cabrera_burden_2021}. Specifically, we assume a linear relationship for the day of the week variable and employ a natural cubic B-splines for the day of the year variable with 8 degrees of freedom per year in the calibration period. Thus, we can reformulate the component related to the residual seasonal and long-term effects in Equation~\eqref{eq:dlnm_myspec} as
$$
\sum_{p=1}^{P}{h_p( z_{d,p} ; \zetavec_{k,p}^{(g)})} = z_{d,1} \zeta_{k,1}^{(g)} + \sum_{t \in \mathcal{T}_y}{h_t( z_{d,t} ; \zetavec_{k,t}^{(g)})},
$$
where $z_{1,d}$ is the day of the week indicator related to day $d$, and $z_{d,t}$ is the day number within year $t$. This specification captures the effects of certain days of the week, such as weekends, as well as the residual seasonal mortality patterns observed in each year. Since our regression model~\eqref{eq:dlnm_myspec} does not include confounding variables, their effects are implicitly captured within the component related to seasonal and long-term effects. Moreover, it does not explicitly incorporate the exposure to risk as an offset factor within the usual Poisson regression model as in \citet{brouhns02}. Hence, the effects of variations in exposure to risk are also captured through $\eta_k^{(g)}$ and $h_p(\cdot)$ functions.

\section{Estimation procedure \label{sec:estimation}}

In this section, we employ a four-step methodology to integrate temperature variations into our estimation procedure. The initial stage of our approach draws from the established methodological framework commonly employed in environmental epidemiology, as seen in studies such as \citet{gasparrini_projections_2017} or \citet{vicedo-cabrera_hands-tutorial_2019}. This step relies on a statistical approach that enables the association between temperature fluctuations and mortality from daily time series data.
The subsequent step involves quantifying the excess mortality and death counts attributable to temperature variations for past periods. These deaths are then considered for calculating a total attributable fraction that we use to calibrate the mortality model in the third step. Finally, a time series model is introduced.
 
\subsection{Estimating the DLNM model
\label{subsec:estim_dlnm}}

We estimate the model given in Equation~\eqref{eq:dlnm_myspec} using the methodology described by \citet{gasparrini_modeling_2014}. We first note that the matrix $\Wvec$ can be integrated into the design matrix of the regression model specified in Equation~\eqref{eq:dlnm_myspec}. Consequently, the estimation of parameters $\thetavec_k^{(g)}$ for $k \in \{1, \ldots, K \}$ and $g \in \mathcal{G}$, along with parameters $\eta_k^{(g)}$, $\zeta_{k,1}^{(g)}$, and $\zetavec_{k,t}^{(g)}$ for $t \in \mathcal{T}_y$, can be performed using a maximum likelihood method \citep{wood_generalized_2006}. It is important to note that this procedure does not impose constraints on the function $s(\cdot, \cdot)$. As a result, this smooth function may exhibit negative values over limited time intervals, which could lead to the prediction of negative death counts attributable to temperature. A potential solution could be to constrain the model estimation by imposing non-negative values. However, this solution is complex to implement when the model is non-linear and is outside the scope of this paper  \citep{gasparrini_modeling_2014}. It is therefore preferable to adopt a specification for the function $s(\cdot, \cdot)$ that minimizes the occurrence of negative values and to verify that the predicted number of negative values remains negligible once the model is estimated.

The uncertainty linked to the estimation of parameters $\thetavec_k^{(g)}$, $k \in \{1, \ldots, K \}$, $g \in \mathcal{G}$, in the smooth function $s(\cdot, \cdot)$ can be measured through its variance-covariance matrix $\V[\thetavec_k^{(g)}]$. Subsequently, this estimated uncertainty is incorporated into our projections using a parametric bootstrap technique based on the assumption of a multivariate normal distribution for the estimated coefficients $\thetavec_k^{(g)}$ \citep{vicedo-cabrera_hands-tutorial_2019}.

Appendix ~\ref{subsec:app_fit_dlnm_model} includes an analysis of the model's goodness of fit. As described in Section~\ref{subsec:dlnm_mortality}, we refer to recent literature in environmental epidemiology to measure the temperature-mortality relationship, set some hyperparameters and calibrate this model. More precisely, our choice of lag $L$, the function $s(\cdot, \cdot)$, and $h_p(\cdot)$ is made consistently with the existing literature to ensure comparability, rather than optimizing hyperparameters specifically for our dataset. In particular, the lag parameter $L$ modulates the short-term seasonality attributable to temperatures and is challenging to determine. If the lag is too short, the temperature effects may fail to capture the harvesting effect resulting from certain intense heatwaves, as well as the effects of cold waves. Conversely, if the lag is too long, the function $s(\cdot, \cdot)$ may capture other seasonal effects unrelated to the short-term impact of temperatures. Therefore, this hyperparameter is conventionally fixed based on references from the literature to ensure a comparable definition of temperature-attributable mortality. Additional analyses to verify the relevance of hyperparameter choices and check the robustness of our results are provided in Appendix~\ref{subsec:app_dlnm_sensi}.

\subsection{Death counts, excess of mortality and attributable risk\label{subsec:excess_mortality}}

Based on the previously estimated model given by Equation~\eqref{eq:dlnm_myspec} using daily average temperature and daily death series on period $\mathcal{T}_d$, the estimated daily death counts for each day $d \in \mathcal{D}^\star$ of year $t \in \mathcal{T}_y$, and for each age $x \in \mathcal{X}_k$, $k \in \{1, \ldots, K\}$ and $g \in \mathcal{G}$ is given by
\begin{align}\label{eq:pred_all_cause_death}
\widehat{D}_{k,t,d}^{(g)} &= \exp{\left(s(\vartheta_{d,t}, L; \widehat{\thetavec}_k^{(g)} )\right)}
\exp{\left(\widehat{\eta}_k^{(g)} + z_{d,1} \widehat{\zeta}_{k,1}^{(g)} + \sum_{t \in \mathcal{T}_y}{h_t( z_{t,p} ; \widehat{\zetavec}_{k,t}^{(g)})}\right)}\\
&= \exp{\left(s(\vartheta_{d,t}, L; \widehat{\thetavec}_k^{(g)} )\right)} \widehat{\widetilde{D}}_{k,t,d}^{(g)},
\end{align}
where  $\widehat{\widetilde{D}}_{k,t,d}^{(g)}$ corresponds to the estimated number of deaths not attributable to temperature, and $\widehat{\thetavec}_k^{(g)}$, $\widehat{\zeta}_{k,1}^{(g)}$, $ \widehat{\zetavec}_{k,t}^{(g)}$ and $\widehat{\eta}_k^{(g)}$ are estimated as described in Section~\ref{subsec:estim_dlnm}. The term $\exp{\left(s(\vartheta_{d,t}, L; \widehat{\thetavec}_k^{(g)} )\right)}$ corresponds to the classical concept of \textit{Relative Risk} (RR) and can be calculated using Equation~\eqref{eq:bidim_sline}. We deduce the estimated attributable fraction for day $d$ as
$$
1-\exp{\left(- 
\sum_{l=0}^{L}{f\cdot w(\vartheta_{d-l,t},l;\widehat{\thetavec}_k^{(g)})}
\right)},
$$
since the attributable fraction satisfies the classical relationship $AF = (RR-1)/RR$. It's important to note that the temperature effect is shared across all ages within the stratum $\mathcal{X}_k$. 

This formulation corresponds to an estimate of attributable risk from a \textit{backward} perspective, i.e. assessing the current risk based on exposure to temperatures on past dates. Following \citet{gasparrini_attributable_2014}, we adopt an alternative estimator of this attributable fraction, based on a \textit{forward} perspective. It is defined for day $d$ as 
\begin{equation}\label{eq:pred_af}
\widehat{\text{AF}}^{(g)}_{k,t,d} = 1-\exp{\left(- 
\sum_{l=0}^{L}{f\cdot w(\vartheta_{d,t},l ;\widehat{\thetavec}_k^{(g)})}
\right)},
\end{equation}
and can be interpreted on a given date $d$ as the cumulative future effects resulting from a temperature variation on that date. With this alternative, the death counts attributable to temperature effects is estimated based on the mean of daily observed death counts $D_{x,t,d}^{(g)}$ for future dates in period $t, \ldots, t+L$
\begin{equation}\label{eq:pred_death_temp}
\widehat{\widebar{D}}_{x,t,d}^{(g)}  
= \widehat{\text{AF}}^{(g)}_{k,t,d} \times \sum_{l=0}^{L} \frac{D_{x,t,d+l}^{(g)}}{L+1}.
\end{equation}
This forward definition offers several advantages. It provides a simpler interpretation of the results and significantly reduces the occurrence of daily negative deaths attributable to temperature \footnote{It is worth noting that negative deaths attributable to temperature are not necessarily anomalous. As explained by \citet{gasparrini_attributable_2014}, the prediction of negative deaths may occur in response to certain events, such as heatwaves, and can be interpreted as a "harvesting paradox."}. Moreover, it enables an additive representation of the risk attributable to different ranges of temperatures. However, this specification introduces a slight negative bias in the estimation of attributable fractions.

Our model's setup allows for a range of predictions that precisely gauge the influence of heat, cold, or specific events occurring on each day of subset $\mathcal{D}_t \subseteq \mathcal{D}^\star $ of year $t$. To determine the total attributable number of deaths to temperatures for a calendar year $t \in \mathcal{T}_y$, we sum of the contributions from each day of a subperiod $\mathcal{D}_t \subseteq \mathcal{D}^\star$ 
\[
\widehat{\widebar{D}}_{x,t}^{(g)} = \sum_{d \in \mathcal{D}_t}{\widehat{\widebar{D}}_{x,t,d}^{(g)}\mathds{1}_{\left\lbrace d \in \mathcal{D}_t\right\rbrace }}.
\]
Finally, we estimate the total attributable fraction as defined in Equation~\eqref{eq:attributable_fract} for each age $x \in \mathcal{X}_k$ and year $t \in \mathcal{T}_y$ as 
\begin{equation}\label{eq:estim_attrib_factor}
\widehat{\text{AF}}_{x,t}^{(g)} (\mathcal{D}_t)= \dfrac{\widehat{\widebar{D}}_{x,t}^{(g)}}{\sum_{d \in \mathcal{D}_t} D_{x,t,d}^{(g)}},
\end{equation}
as well as the temperature adjustment 
$$
\widehat{T}_{x,t}^{(g)} (\mathcal{D}_t) = \left(1 - \widehat{\text{AF}}_{x,t}^{(g)}(\mathcal{D}_t) \right)^{-1}.
$$
This adjustment can be interpreted as a cumulative RR over period $\mathcal{D}_t$.

In practice, the period $\mathcal{D}_t$ considered could correspond to the entire year, certain targeted months, heatwaves, or days above or below a certain temperature threshold. We can also assess heatwave effects by considering only the four hottest months of the year or on specific heatwave dates. To estimate the stochastic mortality model in Section~\ref{subsec:baseline_estimation}, we consider $\mathcal{D}_t = \mathcal{D}^\star$ so that $\widetilde{m}_{x,t}^{(g)}$ corresponds to the central death rate not attributable to temperature effects (both hot and cold). Conversely, if $\mathcal{D}_t$ is strictly included in $\mathcal{D}^\star$, the central death rate would partially capture some of the temperature effects. 

In Equation~\eqref{eq:estim_attrib_factor}, it is worth noting that all quantities are derived from daily data. While annual death counts can be obtained by summing daily deaths, daily exposure to risk is generally unavailable. Thus, an additional annual dataset containing exposures to risk and death counts is required to calculate annual crude central death rates. Before estimating the stochastic mortality model, it is essential to verify that the annual and daily data sources align and provide consistent death counts by age, sex, and year.

\subsection{Estimating the stochastic mortality model 
\label{subsec:baseline_estimation}}

We now detail our estimation approach for the stochastic mortality model under constraints~\eqref{eq:contraint_li_lee_common} and~\eqref{eq:contraint_li_lee} as presented in Section~\ref{subsec:LL_model}. The dynamics of the Li-Lee model~\eqref{eq:2_LL} rely on unobserved central death rates, which are fitted, according to Equation~\eqref{eq:poisson_adj_expo}, using the total attributable fraction~\eqref{eq:estim_attrib_factor}. After estimating these attributable fractions, we calibrate our model using sets of exposure to risk adjusted for temperature effects and death counts from annual data
$$
\mathcal{E}= \left\lbrace {E}_{x,t}^{(g)} \widehat{T}_{x,t}^{(g)}(\mathcal{D}^\star), x \in \mathcal{X}, t\in \mathcal{T}_y, g \in \mathcal{G} \right\rbrace, 
$$
$$
{\mathcal{C}}= \left\lbrace {D}_{x,t}^{(g)}, x \in \mathcal{X}, t\in \mathcal{T}_y, g \in \mathcal{G} \right\rbrace. 
$$
The estimation of the chosen model involves maximizing the following expression derived from the Poisson log-likelihood
$$
\max_{A_x, B_x, K_t, \alpha_x^{(g)}, \beta_x^{(g)}, \kappa_t^{(g)}}\sum_{x \in \mathcal{X}}{\sum_{t\in \mathcal{T}_y}{\sum_{g \in \{f,m\}}{\left( {{D}_{x,t}^{(g)} \ln( \widetilde{m}_{x,t}^{(g)}) - E_{x,t}^{(g)} \widehat{T}_{x,t}^{(g)}(\mathcal{D}^\star) \widetilde{m}_{x,t}^{(g)}}\right)}}},
$$
where $\widetilde{m}_{x,t}^{(g)} = \exp{\left( A_{x} + B_{x}K_{t} +  \alpha_{x}^{\left( g \right)} +\beta_{x}^{\left( g \right)} \kappa_{t}^{\left( g \right)}\right)}$.
The parameter estimation in the model through maximum likelihood can be calculated following an iterative procedure described, for instance, in \citet{li_poisson_2013} or \citet{chen_sex-specific_2018}. In practice, we employ a simpler alternative suggested by \citet{brouhns02} based on a the conditional maximum-likelihood approach, see e.g. \citet{li_poisson_2013} or \citet{robben_assessing_2022}. Hence, the parameters are determined in two steps. Firstly, we estimate under constraints~\eqref{eq:contraint_li_lee_common}
$$
\max_{A_x, B_x, K_t }\sum_{x \in \mathcal{X}}{\sum_{t\in \mathcal{T}_y}{\left( {{D}_{x,t}^{\text{agg}} \ln(\widetilde{m}_{x,t}^{\text{agg}}) - E_{x,t}^{\text{agg}} \widetilde{m}_{x,t}^{\text{agg}}}\right)}},
$$
where ${D}_{x,t}^{\text{agg}} = {D}_{x,t}^{(f)} + {D}_{x,t}^{(m)}$, ${E}_{x,t}^{\text{agg}} = {E}_{x,t}^{(f)}\widehat{T}_{x,t}^{(f)}(\mathcal{D}^\star) + {E}_{x,t}^{(m)}\widehat{T}_{x,t}^{(m)}(\mathcal{D}^\star)$ and $\widetilde{m}_{x,t}^{\text{agg}} = \exp{\left( A_{x} + B_{x}K_{t} \right)}$.

Once these parameters are estimated and using the predicted values of $\widetilde{m}_{x,t}^{\text{agg}}$, we secondly estimate the sex-specific parameters again by maximizing a Poisson log-likelihood under the constraints~\eqref{eq:contraint_li_lee}
$$
 \forall g \in \mathcal{G}, \max_{\alpha_x^{(g)}, \beta_x^{(g)}, \kappa_t^{(g)}}\sum_{x \in \mathcal{X}}{\sum_{t\in \mathcal{T}_y}{\left( {{D}_{x,t}^{(g)} \ln(\widetilde{m}_{x,t}^{(g)}) - E_{x,t}^{(g)}\widehat{T}_{x,t}^{(g)}(\mathcal{D}^\star) \widetilde{m}_{x,t}^{(g)}}\right)}}.
$$

	As presented in Section~\ref{subsec:calibrating_li_lee}, the parameters estimated for the mortality model described above on temperature-reprocessed data are quite similar to those estimated on non-adjusted exposure to risk.

\subsection{Estimating the time series model}
\label{subsec:baseline_time_series}

The dynamics of our model comprise two components: temperature dynamics and a time series model for forecasting the vector $(K_t, \kappa_t^{(f)}, \kappa_t^{(m)})$. In this article, we assume that temperature dynamics are exogenous information, with projected temperature series provided by climate models presented in Section~\ref{subsec:source}. Here, we focus on calibrating the time series model described in Equations~\eqref{eq:LL_time_spe-1} and~\eqref{eq:LL_time_spe-2}, rewritten in matrix form for all $t \in \mathcal{T}_y$ as
\begin{equation}\label{eq:time_series_matrix}
\Yvec_{t} = \Upsilonvec + \Phivec  \Yvec_{t-1} + \Evec_t,
\end{equation}
where
$$
\Yvec_{t} = \begin{pmatrix} K_t \\ \kappa_t^{(f)} \\ \kappa_t^{(m)} \end{pmatrix}, \Upsilonvec  = \begin{pmatrix} \delta \\ c^{(f)} \\ c^{(m)} \end{pmatrix},
\Phivec = \begin{pmatrix}
1 & 0 & 0\\
0 & \phi^{(f)}  & 0\\
0 & 0  & \phi^{(m)}\\
\end{pmatrix} \text{ and }
\Evec_t = \begin{pmatrix} e_t \\ r_t^{(f)} \\ r_t^{(m)} \end{pmatrix}.
$$
Drawing from \citet{antonio_producing_2017}, the parameters $\Upsilonvec $, $\Phivec$ and $\Sigmavec$ are estimated through maximum likelihood since $\Evec_t \sim \mathcal{N}_3(0, \Sigmavec)$ . This estimation can be performed using the \texttt{R}-package \textbf{MultiMoMo} \citep{antonio_multimomo_2022} by maximizing the three-dimensional Gaussian log-likelihood related to $(\Yvec_{t})_{t \in \mathcal{T}_y}$.

\section{Mortality forecast with temperature effects \label{sec:forecast_method}}

In this section, we now aim to incorporate the effects of projected temperatures on mortality. 
Our framework relies on coupling the estimates of the DLNM model with the Li-Lee model. Although the DLNM model can capture both long-term and seasonal effects, it encounters challenges when used to project mortality. Indeed, the functions $h_p(\cdot)$ are B-spline functions with numerous parameters, making it challenging to extract clear trends for projection purposes. Additionally, unlike the Li-Lee model, this approach is purely data-driven. Thus, it is complex to enforce consistency across the population groups, whether for the $K$ age groups or by sex.

These challenges explain why we do not rely solely on a model based on daily data for mortality projections. Instead, we select a Li-Lee-type model applied to annual data for mortality projections. By adjusting for exposure to risk, we remove the total annual effect attributable to temperature when we train the Li-Lee model. As a result, the terms $K_t$ and $\kappa_t^{(g)}$ estimated in Section~\ref{subsec:baseline_estimation} are no longer influenced by the short-term seasonality of temperature. Finally, the future effects of temperature are projected separately using the DLNM model and subsequently added to the projected central mortality rates. For that, we use daily temperature projections from climate models, which reflect changes in the intensity of both cold and hot periods, as well as variations in the severity and duration of heatwaves.

We consider the projection period on annual basis $\mathcal{T}_y^{\text{for}}=\left\lbrace y_{\max}+1, \ldots ,y_{\max}+H\right\rbrace $ where $y_{\max}$ is the last year of the calibration set $\mathcal{T}_y$ and $H \in \N^\star$ is the projection horizon. 

\subsection{Forecasting central death rates not attributable to temperatures \label{subsec:time_series}}

Firstly, we describe the approach adopted to simulate a sample of central death rate trajectories $\widetilde{m}_{x,t}^{(g)}$ for $x \in \mathcal{X}$, $g \in \calG$ and $t \in \mathcal{T}_y^{\text{for}}$. For this purpose, we use Equation~\eqref{eq:2_LL} with the estimated parameters $\widehat{A}_x$, $\widehat{B}_x$, $\widehat{K}_t$,  $\widehat{\alpha}_x^{(g)}$, $\widehat{\beta}_x^{(g)}$ and  $\widehat{\kappa}_t^{(g)}$ obtained as described in Section~\ref{subsec:baseline_estimation}.

The mortality model for central death rates not attributable to temperature effects also involves the projection of the vector $\Yvec_t$, derived from Equation~\eqref{eq:time_series_matrix} using the estimated parameters $\widehat{\Upsilonvec}$, $\widehat{\Phivec}$ and $\widehat{\Sigmavec}$, see Section~\ref{subsec:baseline_time_series}. The dynamics of $\Yvec_t$ are iteratively deduced by simulating the innovation errors $\Evec_t$ from a multivariate Gaussian vector with zero mean and covariance matrix $\widehat{\Sigmavec}$. Thus, we can obtain a Monte Carlo sample of trajectories $\widetilde{m}_{x,t}^{(g)}$, allowing us to derive prediction intervals for any date $t \in \mathcal{T}_y^{\text{for}}$.

\subsection{Forecasting central death rates with temperature effects \label{subsec:add_deaths_temp}}

Secondly, we focus on adding shocks to mortality resulting from daily temperature variations, based on scenarios of temperatures available on a daily basis, see Section~\ref{subsec:source}. Using the usual methodology in ecological epidemiology \citep{vicedo-cabrera_hands-tutorial_2019}, our approach involves projecting the annual central mortality rates not attributable to temperatures, and then adjusting them for the impact of future temperatures.  This effect can be assessed on a period $\mathcal{D}_t$ different from $\mathcal{D}^\star$ to specifically measure the projected impacts of hot and cold temperatures on mortality, distinguishing the compensations that occur throughout the year. 

To do this, we first establish a relationship between the annual central mortality rates and the daily death counts. Indeed, the relationship in Equation~\eqref{eq:attributable_fract} and the estimator in Equation~\eqref{eq:estim_attrib_factor} cannot be used for projection purposes, as future deaths are not observed. We denote $\text{AF}_{x,d,t}^{(g)} = \widebar{D}_{x,t,d}^{(g)} / {D}_{x,t,d}^{(g)}$ for age $x \in \mathcal{X}_k$, $k \in \{1, \ldots, K\}$, sex $g \in \mathcal{G}$ and for any day $d \in \mathcal{D}_t$ of the year $t \in \mathcal{T}^{\text{for}}_y$. Then, we obtain
\begin{align}\label{eq:forecast_all}
\widehat{m}_{x,t}^{(g)} &= {\widetilde{m}}_{x,t}^{(g)} + {\widebar{m}}_{x,t}^{(g)} \nonumber \\
& = {\widetilde{m}}_{x,t}^{(g)} + \frac{\widetilde{D}_{x,t}^{(g)}}{E_{x,t}^{(g)}} 
\sum_{d \in \mathcal{D}_t}{\frac{\widebar{D}_{x,t,d}^{(g)}}{\widetilde{D}_{x,t}^{(g)}}\mathds{1}_{\left\lbrace d \in \mathcal{D}_t \right\rbrace }}\nonumber \\
& = {\widetilde{m}}_{x,t}^{(g)}  \left[ 1 +  
\sum_{d \in \mathcal{D}_t}{\frac{\widetilde{D}_{x,t,d}^{(g)} \text{AF}_{x,d,t}^{(g)} (1- \text{AF}_{x,d,t}^{(g)})^{-1}}{\widetilde{D}_{x,t}^{(g)}}\mathds{1}_{\left\lbrace d \in \mathcal{D}_t \right\rbrace }}\right] \nonumber \\
& = {\widetilde{m}}_{x,t}^{(g)}  \left[ 1 +  
\sum_{d \in \mathcal{D}_t}{ \omega_{x,t,d}^{(g)}  \text{AF}_{x,d,t}^{(g)} (1- \text{AF}_{x,d,t}^{(g)})^{-1}\mathds{1}_{\left\lbrace d \in \mathcal{D}_t \right\rbrace }}\right].
\end{align}

In Equation~\eqref{eq:forecast_all}, the term $\omega_{x,t,d}^{(g)} =\widetilde{D}_{x,t,d}^{(g)}/\widetilde{D}_{x,t}^{(g)}$ corresponds to the weight associated with the distribution of death counts not attributable top temperature effects over the period $\mathcal{D}_t$ of year $t$. These weights are unknown for future periods, but can be estimated by introducing a distribution assumption of deaths throughout the year. If the seasonality in death counts is primarily driven by temperature effects, it is reasonable to assume that $\widetilde{D}_{x,t,d}^{(g)}$ follows a uniform distribution. Under this common assumption in actuarial literature, the hazard rate function remains constant throughout the year, and the weight can be fixed as $\omega_{x,t,d}^{(g)} = 1/\vert \mathcal{D}^\star \vert$. However, if residual seasonality in death counts, i.e. excluding temperature effects, is more significant, alternative assumptions about the distribution of deaths could be considered. For instance, the daily distribution of deaths not attributable to temperature effects estimated over a previous period can be used. However, this implies that the residual seasonality is assumed to remain constant over time \citep{madaniyazi_seasonality_2024}.

The effect of temperatures along a given trajectory of projected temperatures $(\widehat{\vartheta}_{d,t})$ from an external climate scenario is determined by applying Equation~\eqref{eq:pred_af}. Then, using the central death rates not attributable to temperatures, $ \widehat{\widetilde{m}}_{x,t}^{(g)}$ projected as described in Section~\ref{subsec:time_series}, we derive the projected central death rates as
\begin{equation}\label{eq:forecast_all2}
\widehat{\widehat{m}}_{x,t}^{(g)} =
{\widehat{\widetilde{m}}}_{x,t}^{(g)}  \left[ 1 +  
\sum_{d \in \mathcal{D}_t}{
 \omega_{x,t,d}^{(g)}
 \widehat{\text{AF}}_{x,d,t}^{(g)} (1- \widehat{\text{AF}}_{x,d,t}^{(g)})^{-1}
 \mathds{1}_{\left\lbrace d \in \mathcal{D}_t \right\rbrace }}\right].
\end{equation}

Furthermore, we recognize in Equation~\eqref{eq:forecast_all2} a multiplicative form similar to that of Equation~\eqref{eq:multi_deaths_rates}, which implies that the projected total attributable fraction to temperature for the period $\mathcal{D}_t$ can be calculated as
\begin{equation}\label{eq:simul_attrib_fraction}
\widehat{\text{AF}}_{x,t}^{(g)} (\mathcal{D}_t)= 
\frac{\sum_{d \in  \mathcal{D}_t}{\omega_{x,t,d}^{(g)} \widehat{\text{AF}}_{x,d,t}^{(g)} (1- \widehat{\text{AF}}_{x,d,t}^{(g)})^{-1} \mathds{1}_{\left\lbrace d \in \mathcal{D}_t\right\rbrace }}}
{1 + \sum_{d  \mathcal{D}_t}{\omega_{x,t,d}^{(g)} \widehat{\text{AF}}_{x,d,t}^{(g)} (1- \widehat{\text{AF}}_{x,d,t}^{(g)})^{-1} \mathds{1}_{\left\lbrace d \in \mathcal{D}_t \right\rbrace }}}.
\end{equation}

It is worth noting that this quantity is not necessarily positive. Indeed, the reduction in cold-related mortality in a context of overall temperature increase can lead to a decrease in mortality during certain periods.

Finally, using Equation~\eqref{eq:forecast_all2}, we easily obtain an expression for the projected all-cause mortality rates and the projected mortality rates not attributable to temperatures
$$
\widehat{q}_{x,t}^{(g)} = 1 - \exp{\left(-\widehat{\widehat{m}}_{x,t}^{(g)}\right)}, \quad
\widehat{\widetilde{q}}_{x,t}^{(g)} = 1 - \exp{\left(\widehat{\widetilde{m}}_{x,t}^{(g)}\right)}, 
$$
for all  $x \in \mathcal{X}_k$, $k \in \{1, \ldots, K\}$, sex $g \in \mathcal{G}$ and $t \in \mathcal{T}^{\text{for}}_y$.
Additionally, the periodic life expectancy at age $x$ truncated at age $t_{\max}$, is estimated based on
$$
\widehat{e}_{x,t}^{(g)} = \sum_{k=1}^{t_{\max}}{\prod_{j=0}^{k-1}{\left(1-\widehat{q}_{x,j}^{(g)}\right)}}.
$$
In the presence of two causes of mortality, we examine the number of years of life expectancy lost (or gained) due to temperatures for a person of age $x$ at date $t$ due to the temperature effect
\begin{equation}\label{eq:loss_le}
\Delta \widehat{e}_{x,t}^{(g)} = 
 \sum_{k=1}^{t_{\max}}{\left[
\prod_{j=0}^{k-1}{\left(1-\widehat{\widetilde{q}}_{x,j}^{(g)}\right)}
- \prod_{j=0}^{k-1}{\left(1-\widehat{q}_{x,j}^{(g)}\right)}
\right]}.
\end{equation}

\section{Assessing the impact of future temperatures on French mortality projections \label{sec:application}}

Numerical illustrations presented in this case study have been carried out in the \soft{R} statistical software \citep{rsoft22}. We use the DLNM model from package \textbf{dlnm} \citep{dlnm14}. This package proposes a set of functions allowing the construction of the crossbasis transformation presented in Section~\ref{subsec:daily_death}, as well as many basis plot commands. The mortality projection is performed by adapting the \soft{R} code from the \textbf{MultiMoMo} package \citep{antonio_multimomo_2022}.

\subsection{Data sources\label{subsec:source}}

In this section, we provide an illustration of our model using French mortality data for both women and men, both on a daily and annual basis. Daily mortality data are compared to past data on average national temperatures. Additionally, to project the future effect of temperatures, we utilize temperature scenarios from various climate models.

The calibration period chosen for this study corresponds to the range $\mathcal{T}_y = \{1980, \ldots, 2019\}$. The period $\mathcal{T}_d$ includes all the days corresponding to the years in $\mathcal{T}_y$. This selection allows us to focus on relatively recent temperature event. This calibration period ensures adaptability conditions comparable to those experienced by populations at the beginning of the forecasting period. Additionally, we choose to use temperature and mortality data up to the end of the year 2019 to avoid any undesirable effects of the Covid-19 pandemic on the mortality results. We concentrate our analysis on the age group $\mathcal{X} = \{0, \ldots, 105\}$ available in the HMD. Older age groups are excluded due to the lower volumes of available data, which could compromise the stability of the results.

\subsubsection{Temperature dataset for past periods\label{subsubsec:temperature}}

The temperature records come from the GHCN database \citep{GHCN18}.
This database gathers more than 20 different sources and contains the main climatic information that can be observed at the surface of the globe, station by station. It undergoes several testing phases and validation processes to ensure the accuracy of its data.

The oldest data in the GHCN database can be up to 175 years old, and in the case of temperatures this poses various problems. The weather stations, over the years, have evolved technically, migrated from one location to another or may have been recalibrated. This is why internal recalibration algorithms and processes have been applied to this database.
Finally, from this database, it is possible to select different meteorological stations of the globe. 

\begin{figure}[h!]
\centering
\includegraphics[scale = 0.7]{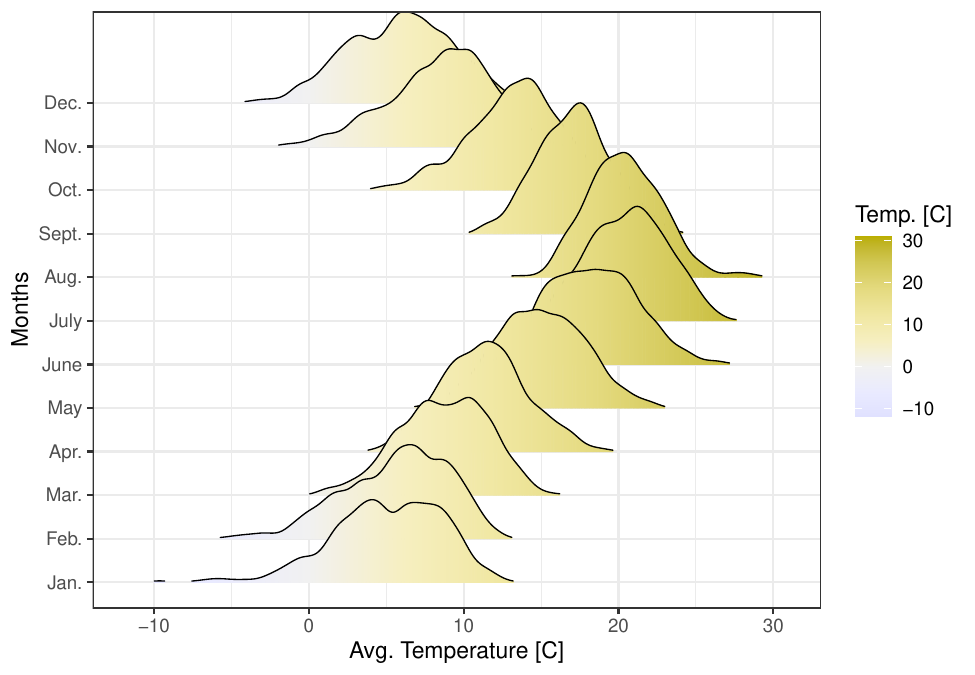}
\caption{Distribution of average daily temperatures for each month of the year. The average daily temperatures are calculated city by city, and then an average of 14 cities is considered to derive average daily temperatures for Metropolitan France. Temperatures are expressed in degrees Celsius. Negative temperatures are represented in a blue color range, while positive temperatures are represented in a yellow color range. Calibration period 1980-2019, GHCN data.}
\label{fig-dens-temp}
\end{figure}

In this paper, we select 14 stations related to major French cities: Bordeaux, Brest, Caen, Clermont-Ferrand, Dijon, Lille, Lyon, Marseille, Nantes, Paris, Perpignan, Strasbourg, Toulouse and Tours, to cover the French territory and the main part of the population over the calibration period. 
For each of them, daily records give access to the average, minimum and maximum temperature of each city. 
These three indicators are then aggregated to obtain their average at the scale of Metropolitan France, allowing for a comparison with the daily death count serie collected at an aggregated level for Metropolitan France.
\footnote{
To our knowledge, there is no existing national index of average temperatures weighted by observed mortality data. As an alternative, we have considered a series of daily mean temperatures weighted by the population size in the departments corresponding to the 14 cities studied. Since this series is nearly identical to the one obtained by calculating a simple average of the temperatures across the 14 cities, both for past and projected periods (assuming a constant population), we opted for the simple average in this study for the sake of simplicity.}
Figure~\ref{fig-dens-temp} presents the distribution of average daily temperatures over the observation period. It illustrates the temperature variations throughout the year as well as the range of values taken for each month of the year. We describe in Appendix~\ref{sec:app_hist_france} a historical overview of recent heatwaves in France.

\subsubsection{Mortality datasets\label{subsubsec:mortality}}
\paragraph{Daily data.}
Our daily mortality data come from a specific inquiry made to the Quetelet-Prodego Diffusion network \citep{insee_data}, a French network of data centers for social sciences. This portal serves as a valuable resource for researchers and students. The data is acquired through a connection with the INSEE and is provided by the \textit{Archives des Données Issues de la Statistique Publique} team.

Our request involves aggregating deaths from the civil registry. Various variables were accessible, including sex, age at death, socio-professional category, urban area, department, and region of death. The selected deaths in our inquiry are limited to metropolitan France, covering the years 1980 to 2019 and are provided on a daily basis. We select the age–period observation $\mathcal{X} = \{0, \ldots , 105\}$  $\mathcal{T}_y = \{1980, \ldots , 2019\}$ and the death counts at national level.

\begin{figure}[h!]
\centering
\includegraphics[scale = 0.5]{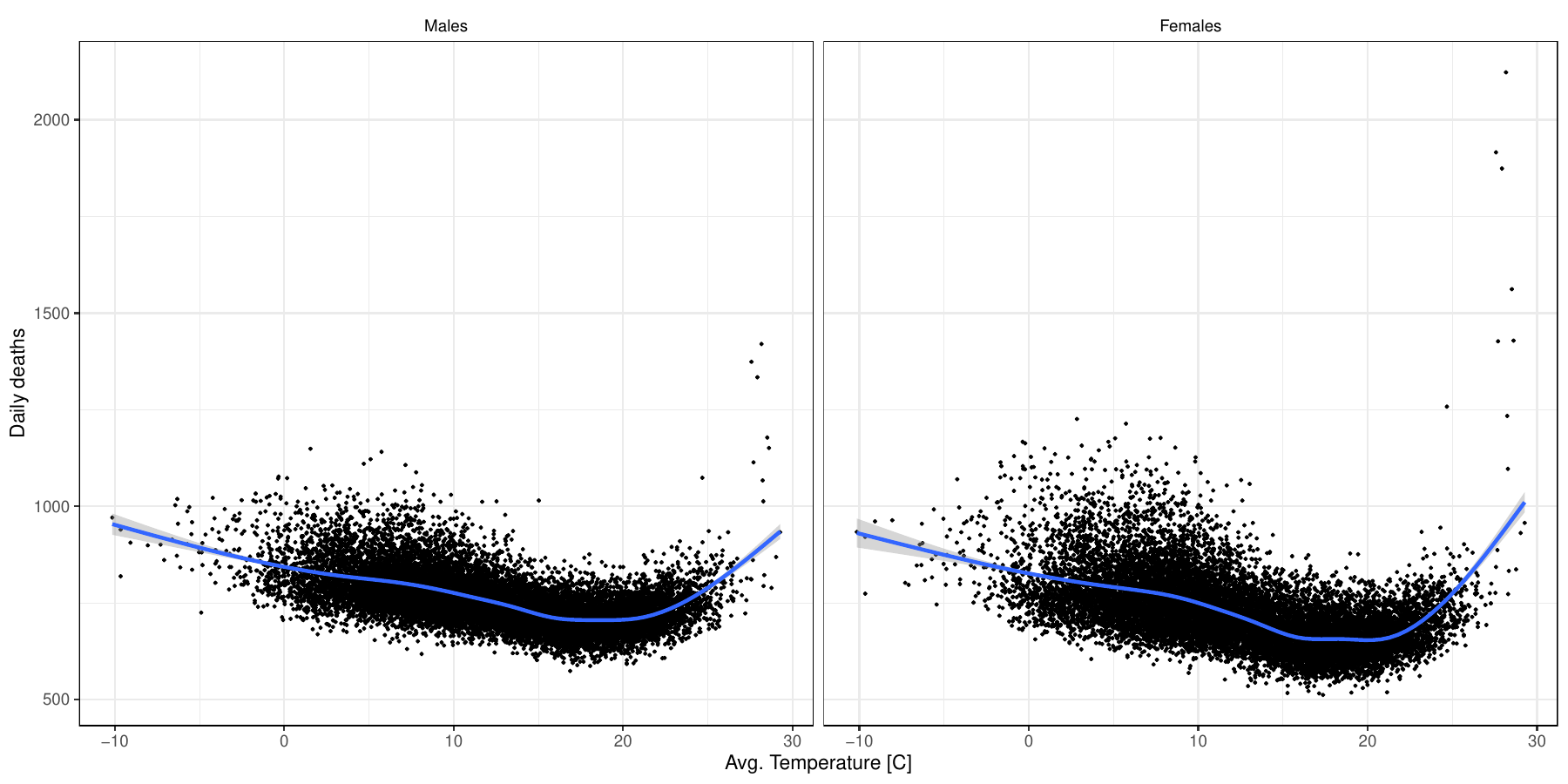}
\caption{
Representation of the daily death counts according to the average daily temperature in France for women and men.
The blue trend curve is obtained by smoothing using a generalized additive model with a confidence level of a $95 \%$.
The average daily temperatures are calculated city by city, and then an average of 14 cities is considered to derive average daily temperatures for Metropolitan France. Temperatures are expressed in degrees Celsius. Observation period: 1980-2019, based on GHCN and INSEE data.}
\label{fig-death_vs_temp}
\end{figure}

Figure~\ref{fig-death_vs_temp} depicts the evolution of daily death counts over the period and compares them with the daily average temperatures, for both women and men. This figure highlights the relationship between mortality and temperature by revealing a classical "U" shape: the daily death count tends to decrease with temperature until reaching a minimal level near 20°C. Beyond this average threshold over a day, mortality tends to increase with rising temperatures. The figure highlights a more pronounced association between heat and mortality for women.

\begin{figure}[h!]
\centering
\includegraphics[scale = 0.5]{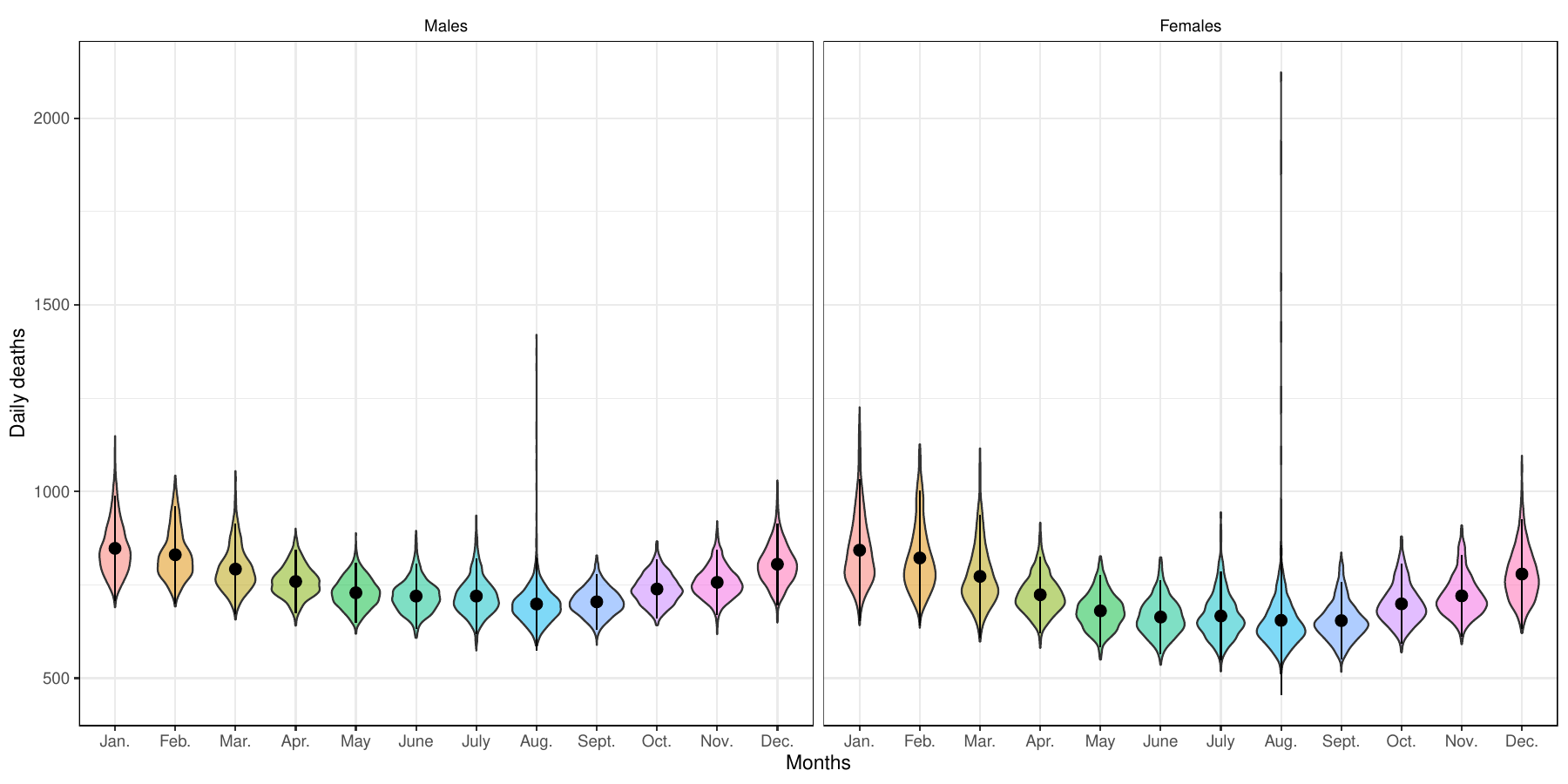}
\caption{Probability density of the number of deaths by month of the year. 
The mean plus and minus the standard deviation is displayed as a pointrange.
 Daily average temperatures are calculated for each city, and then an average of 14 cities is used to derive the daily average temperatures for Metropolitan France. Temperatures are measured in degrees Celsius. Observation period: 1980-2019, based on GHCN and INSEE data.}
\label{fig-dens_vs_month}
\end{figure}

This relationship between mortality and temperature leads to observing seasonal cycles in the number of deaths. Figure~\ref{fig-dens_vs_month} illustrates the distribution of deaths per month over the observation period for women and men. It highlights both the mortality associated with cold during the winter months, which can be marked by some extreme events, e.g., flu epidemics, and the peaks in mortality associated with heatwaves, notably an extreme event related to the summer heatwave of 2003. This figure underscores the greater variability in the number of deaths among women during the winter and in the height of summer.

\paragraph{Annual data.}
The annual mortality data considered are extracted from the Human Mortality Database \citep{human_mortality_database_university_2024}. It enables the representation of the crude central death rate $\widehat{m}_{x,t}^{(g)}$ for each age $x \in \mathcal{X}$, year $t \in \mathcal{T}_y$, and sex $g \in \mathcal{G}$. We verify the consistency of annual and daily death counts in Appendix~\ref{sec:app:data_sources}.  

\subsubsection{Climate scenario dataset\label{subsubsec:climate}}

In this paper, our daily future temperature projections are derived from twelve climate models for which we examine three climate change scenarios (Representative Concentration Pathway [RCP]2.6, RCP4.5, and RCP8.5) presented in the framework of the 5th IPCC Assessment Report \citep{ipcc2014}. These pathways depict the evolution of radiative forcing, which represents the difference between incoming solar radiation and outgoing infrared radiation. Radiative forcing is measured in $W.m^{-2}$. 

The international CORDEX project aim to coordinate global research in terms of climate projections, its European branch EUROCORDEX made various simulations available in 2014. The outputs of these simulations have been improved and corrected in 2020 and are freely available through the \citet{drias} portal. The data available for Europe have a resolution of 7 to 10 km depending on the geographical areas, with different daily variables such as mean, minimum and maximum temperature, total precipitation or wind speed. In our study, we extract temperature scenarios generated using global circulation models (GCMs) and regional circulation models (RCMs) described in Table~\ref{tab:list_climate_models}, see \citet{flato2013evaluation} for a comprehensive presentation of climate models and their evaluations. Temperature data is available in France on a daily time step over the period 2006-2100. It undergoes bias correction and statistical downscaling using the ADAMONT method \citep{adamont17} to enable projections on an 8 km resolution grid (SAFRAN).

\begin{table}[!h]

\caption{List of climate models provided by the DRIAS \label{tab:list_climate_models}}
\centering
\fontsize{8}{10}\selectfont
\begin{tabular}[t]{ccccc}
\toprule
Short Name & GCM & RCM & RCPs available & Period\\
\midrule
CNRM-CM5 / ALADIN63 & CNRM-CM5 & ALADIN63 & RCP8.5, RCP4.5, RCP2.6 & 2006-2100\\
MPI-ESM / CCLM4-8-17 & MPI-ESM & CCLM4-8-17 & RCP8.5, RCP4.5, RCP2.6 & 2006-2100\\
HadGEM2 / RegCM4-6 & HadGEM2 & RegCM4-6 & RCP8.5, RCP2.6 & 2006-2099\\
EC-EARTH / RCA4 & EC-EARTH & RCA4 & RCP8.5, RCP4.5, RCP2.6 & 2006-2100\\
IPSL-CM5A / WRF381P & IPSL-CM5A & WRF381P & RCP8.5, RCP4.5 & 2006-2100\\
\addlinespace
NorESM1 / REMO2015 & NorESM1 & REMO2015 & RCP8.5, RCP2.6 & 2006-2100\\
MPI-ESM / REMO2009 & MPI-ESM & REMO2009 & RCP8.5, RCP4.5, RCP2.6 & 2006-2100\\
HadGEM2 / CCLM4-8-17 & HadGEM2 & CCLM4-8-17 & RCP8.5, RCP4.5 & 2006-2099\\
EC-EARTH / RACMO22E & EC-EARTH & RACMO22E & RCP8.5, RCP4.5, RCP2.6 & 2006-2100\\
IPSL-CM5A / RCA4 & IPSL-CM5A & RCA4 & RCP8.5, RCP4.5 & 2006-2100\\
\addlinespace
CNRM-CM5 / RACMO22E & CNRM-CM5 & RACMO22E & RCP8.5, RCP4.5, RCP2.6 & 2006-2100\\
NorESM1 / HIRHAM5 v3 & NorESM1 & HIRHAM5 v3 & RCP8.5, RCP4.5 & 2006-2100\\
\bottomrule
\end{tabular}
\end{table}

To remain consistent with the observed mortality trends in 14 major cities described in Section~\ref{subsubsec:mortality}, we extract daily averages, minimums, and maximums of temperatures for the points of the SAFRAN grid corresponding to these major cities. Subsequently, we calculate daily averages of these temperature indicators to reconstruct national averages. This simplification is adopted to have aggregated temperature trends enabling a connection with the selected stochastic mortality model at the scale of Metropolitan France. It is important to note that this simplification results in smoothing out certain regional temperature variations.
It is conducted because we do not have access to historical daily death records by city or other areas.

To illustrate the impact of global warming on heatwaves, we introduce several descriptive indicators to characterize each heatwave period. As explained in Appendix~\ref{sec:app_hist_france}, we define a heatwave in France for a given day $d \in \mathcal{D}^\star$ as follows
$$
\bar{\vartheta}^{\max}_d = \frac{1}{3}\sum_{j=1}^3{{\vartheta}^{\max}_{d-j}}, \qquad 
\bar{\vartheta}^{\min}_d = \frac{1}{3}\sum_{j=1}^3{{\vartheta}^{\min}_{d-j}},
$$
where ${\vartheta}^{\max}_{d}$ and ${\vartheta}^{\min}_{d}$ are the aggregated maximum and minimum temperatures in Metropolitan France for day $d$. This 3-day moving average is sometimes referred to as a biometeorological indicator. Thus, a day is classified as a heatwave if the 3-day moving averages of the minimum temperature $\bar{\vartheta}^{\min}_d$ and maximum temperature $\bar{\vartheta}^{\max}_d$ exceed the respective minimum $r^{\min}$ and maximum $r^{\max}$ thresholds. These thresholds vary by geographical area. At a national scale, we consider $r^{\min}=18^{\circ}\text{C}$ and $r^{\max}=30^{\circ}\text{C}$.
For an heatwave period $\mathcal{D}_t \subset \mathcal{D}^\star$ in year $t$, we then introduce:
\begin{itemize}
	\item
		the duration corresponds to the number of consecutive days above the alert thresholds, i.e., $\vert \mathcal{D}_t\vert$,
	\item
		the cumulative severity corresponds to the sum of the daily severities $\text{sev}_d$ over the heatwave period $\mathcal{D}_t$ where
		$$
		\text{sev}_{d,t} = \vert {\vartheta}^{\min}_{d,t} - r^{\min} \vert +
		\vert {\vartheta}^{\max}_{d,t} - r^{\max} \vert,
		$$
	\item
		 the intensity of a heatwave corresponds to
		$$
		\text{int} = \max_{d \in \mathcal{D}_t} \left\lbrace 
		\vert {\vartheta}^{\min}_{d} - r^{\min} \vert
		\right\rbrace +
		\max_{d \in \mathcal{D}_t} \left\lbrace 
		\vert {\vartheta}^{\max}_{d} - r^{\max} \vert 
		\right\rbrace.
		$$	
\end{itemize}
These indicators allow for triggering the appropriate level of alert in line with the danger associated with a heatwave, and for evaluating and ranking heatwaves among themselves.

\begin{figure}[h!]
\centering
\includegraphics[scale = 0.5]{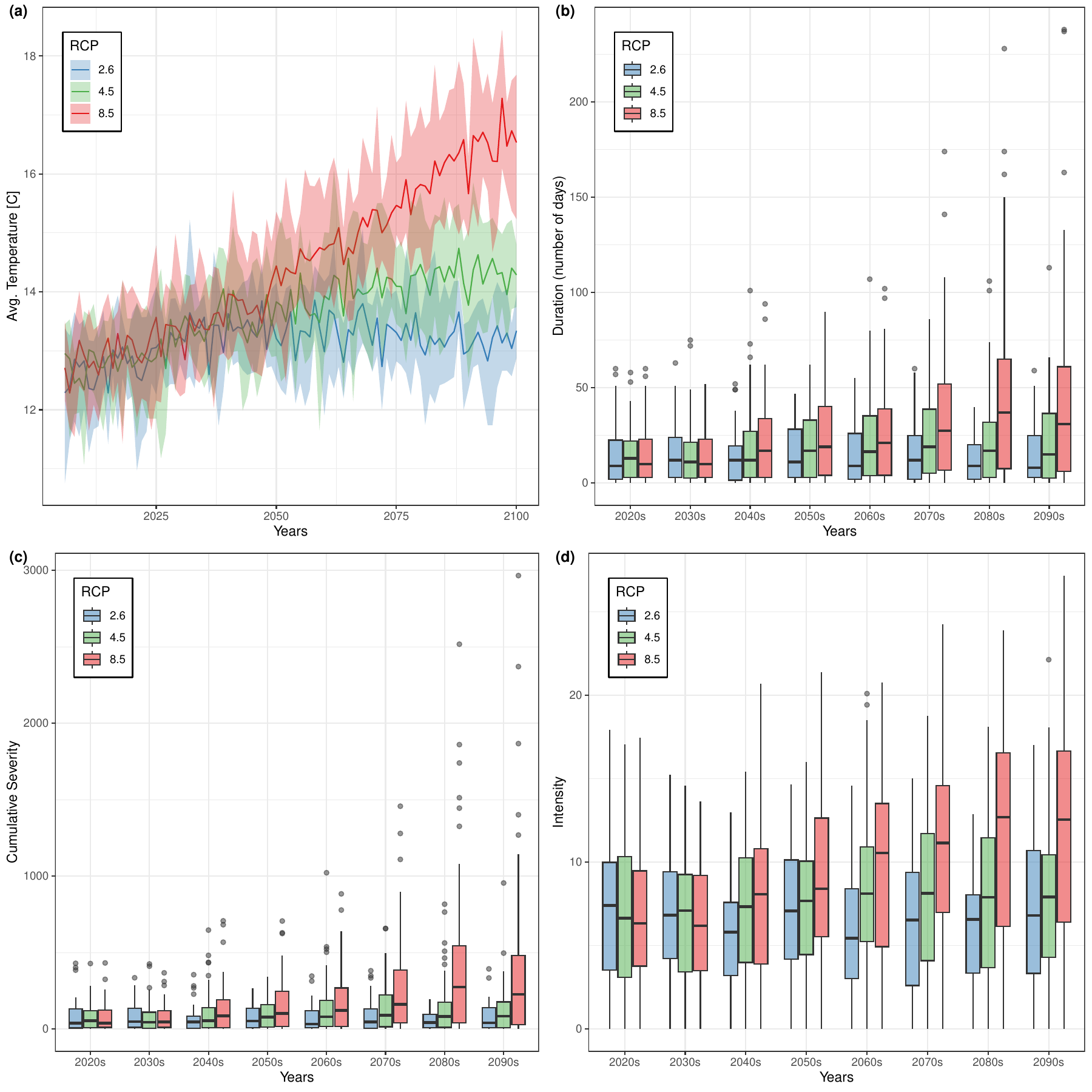}
\caption{
Projection of temperatures and heatwaves by RCP scenario in Metropolitan France over the period 2020-2100. (a) Annual average of daily temperatures by RCP scenario. Each curve is calculated as the average across available climate models. The shaded area corresponds to the $95\%$ confidence intervals across all available climate models. (b) Distribution of heatwave duration per decade and per RCP scenario. (c) Distribution of cumulative heatwave severity per decade and per RCP scenario. (d) Distribution of heatwave intensity per decade and per RCP scenario. For each RCP scenario and each decade, the boxplots represent the distribution of duration, cumulative severity, and intensity calculated over the sample consisting of all years of the decade and across all available climate models. Daily average temperatures are calculated for each city, and then an average of 14 cities is used to derive the daily average temperatures for Metropolitan France. Temperatures are measured in degrees Celsius. Based on DRIAS data.}
\label{fig_proj_temp_rcp}
\end{figure}

Figure~\ref{fig_proj_temp_rcp} depicts temperature trajectories from the three RCP2.6, RCP4.5, and RCP8.5 scenarios, as well as the distribution of indicators measuring the duration, cumulative severity, and intensity of a heatwave for each decade. Figure~\ref{fig_proj_temp_rcp}.a illustrates a gradual increase in average annual temperatures, exacerbated for the RCP8.5 scenario from the 2040s-2050s. This overall temperature rise impacts not only the number but also the characteristics of heatwaves, which gain both in duration, severity, and intensity in the RCP4.5 and RCP8.5 scenarios. Specifically, we observe a similar trajectory of these characteristics for the 2020s and 2030s for all three scenarios, followed by an increase in the danger of heatwaves starting from the 2040s in the RCP8.5 scenario. For these indicators (Figure~\ref{fig_proj_temp_rcp}.b to Figure~\ref{fig_proj_temp_rcp}.d), the increase in median values in the RCP4.5 and RCP8.5 scenarios is also accompanied by greater model volatility over time beyond the 2040s. The RCP2.6 scenario presents a relatively stable trajectory between 2020 and 2100. Additionally, we can observe a slight decrease in median values in the RCP2.6 scenario over the decades 2020 to 2040, with no clear trend thereafter. 
These observations are consistent with the findings on hot days presented by the \citet{ipcc22}. Indeed, the regions identified as the most affected by this warming are central and eastern North America, central and southern Europe, and the Mediterranean region.

Regarding the RCP2.6 scenario, the HadGEM2/RegCM4-6 model predicts the most heatwaves, with 459 heatwaves recorded over the period 2020-2100. According to this model, the most severe heatwave occurs in 2047, lasting for 30 consecutive days with a cumulative severity of 209.4. In 2097, it predicts a heatwave with an intensity of 16.3. In scenario RCP4.5, the most pessimistic model in terms of heatwaves is the HadGEM2/CCLM4-8-17 model with a total of 1,243 heatwaves. According to this model, the most severe heatwave occurs in 2062, lasting for 49 consecutive days with a cumulative severity of 572.0. In 2097, it predicts a heatwave with an intensity of 22.1.
Finally, for scenario RCP8.5, this model is again the most pessimistic and predicts the most severe heatwave in 2088 (lasting for 89 consecutive days with a cumulative severity of 1,099.5), as well as the most intense heatwave in 2094 with an intensity of 27.2. In comparison with these thresholds, the 2003 heatwave lasted 12 consecutive days, with a cumulative severity of 92 and an intensity of 9.2.

\subsection{Estimation of temperature-mortality association with the DLNM model
\label{subsec:calibrating_dlnm}}

Using the approach outlined in Section~\ref{subsec:estim_dlnm}, we calibrate model~\eqref{eq:dlnm_myspec} over the period $\mathcal{T}_y$ using daily temperature and death data with a lag of $L=21$ days. In our study, we consider $K=4$ age buckets to account for the differing sensitivities of women's and men's ages based on their specific MMTs: 0-64, 65-74, 75-84, and 85+. 

\begin{figure}[h!]
\centering
\includegraphics[scale = 0.5]{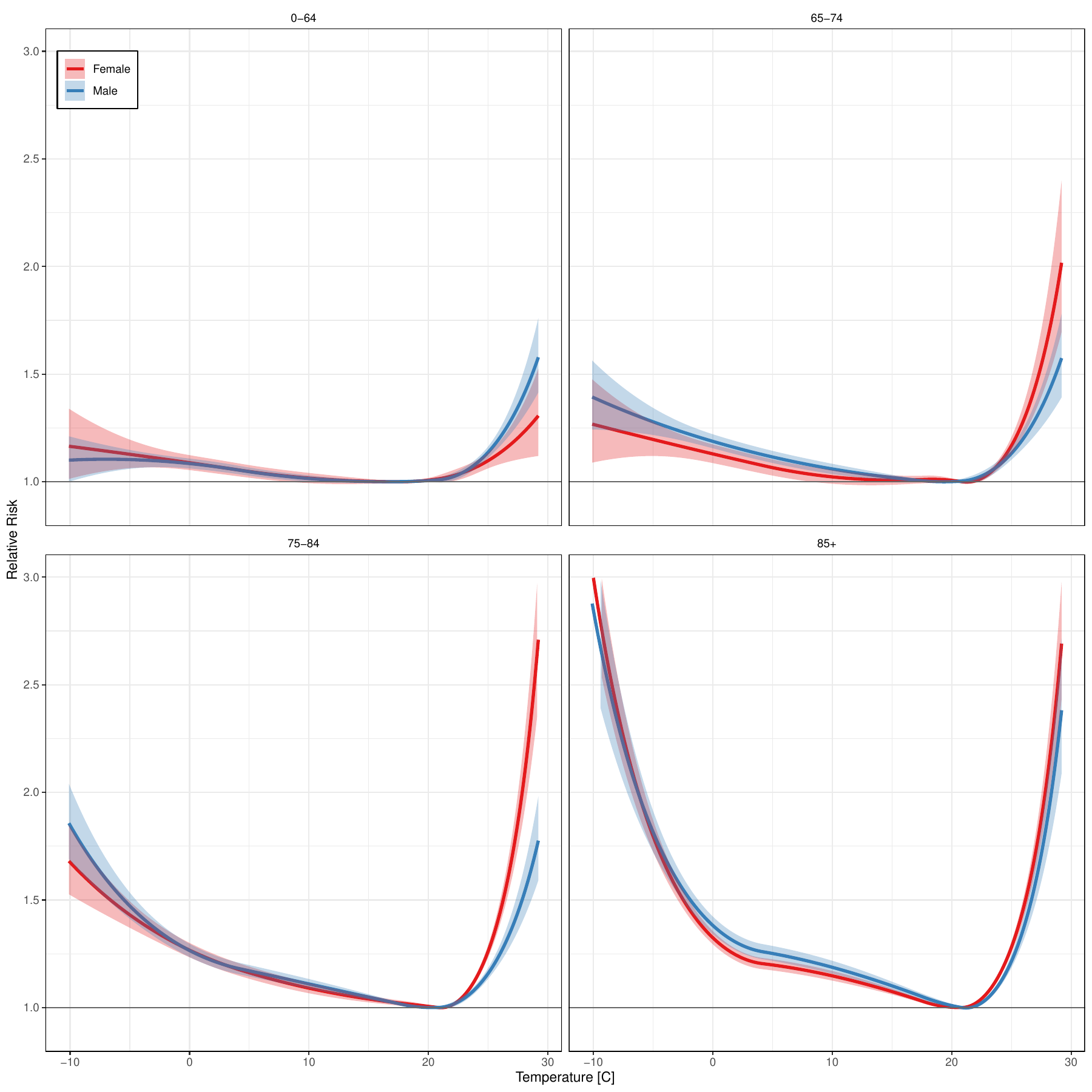}
\caption{
Cumulative relative risk of mortality over a 21-day period in Metropolitan France calculated for the years 1980-2019 for women (red) and men (blue) across age groups 0-64, 65-74, 75-84, and 85+, with their $95\%$ confidence intervals (shaded areas). Daily average temperatures are calculated for each city, and then an average of 14 cities is used to derive the daily average temperatures for Metropolitan France. Temperatures are measured in degrees Celsius. Based on GHCN and INSEE data.
}
\label{fig_dlnm_rr_curve}
\end{figure}

Figure~\ref{fig_dlnm_rr_curve} depicts the cumulative temperature-mortality association estimated with the DLNM model for both women and men, and for each age group, along with the corresponding 95\% confidence intervals (CIs). 

After centering the values of $\vartheta_{d,t}$ on the MMT threshold, it represents a cumulative RR surface that quantifies the increase or decrease in mortality in response to deviations in temperature above or below this threshold.
These curves reveal a non-linear relationship between temperature and the cumulative RR of mortality over a 21-day period. The shapes of the curves are typical, showing sensitivity to both extreme cold and heat variations, above and below an optimal temperature range, i.e. the MMT, typically between 17 and 21°C, with little variation by sex. We observe that this sensitivity increases notably with age and exhibits differences between sexes. For individuals over 85 years old, the responses to temperature variations are generally similar between women and men for both extreme cold and heat temperatures. For ages between 65 and 84 years, the RR for cold temperatures at the $1\%$ percentile of the coldest temperatures exceeds the response to heat at the $1\%$ percentile of the hottest temperatures. However, this risk significantly increases during the most extreme heatwaves. We also note a greater sensitivity of women to high temperatures and, conversely, a higher sensitivity of men to the coldest temperatures in these age groups. Regarding the age group 0-64 years, the curve for response to extreme heat is steeper for men.

\begin{figure}[h!]
\centering
\includegraphics[scale = 0.5]{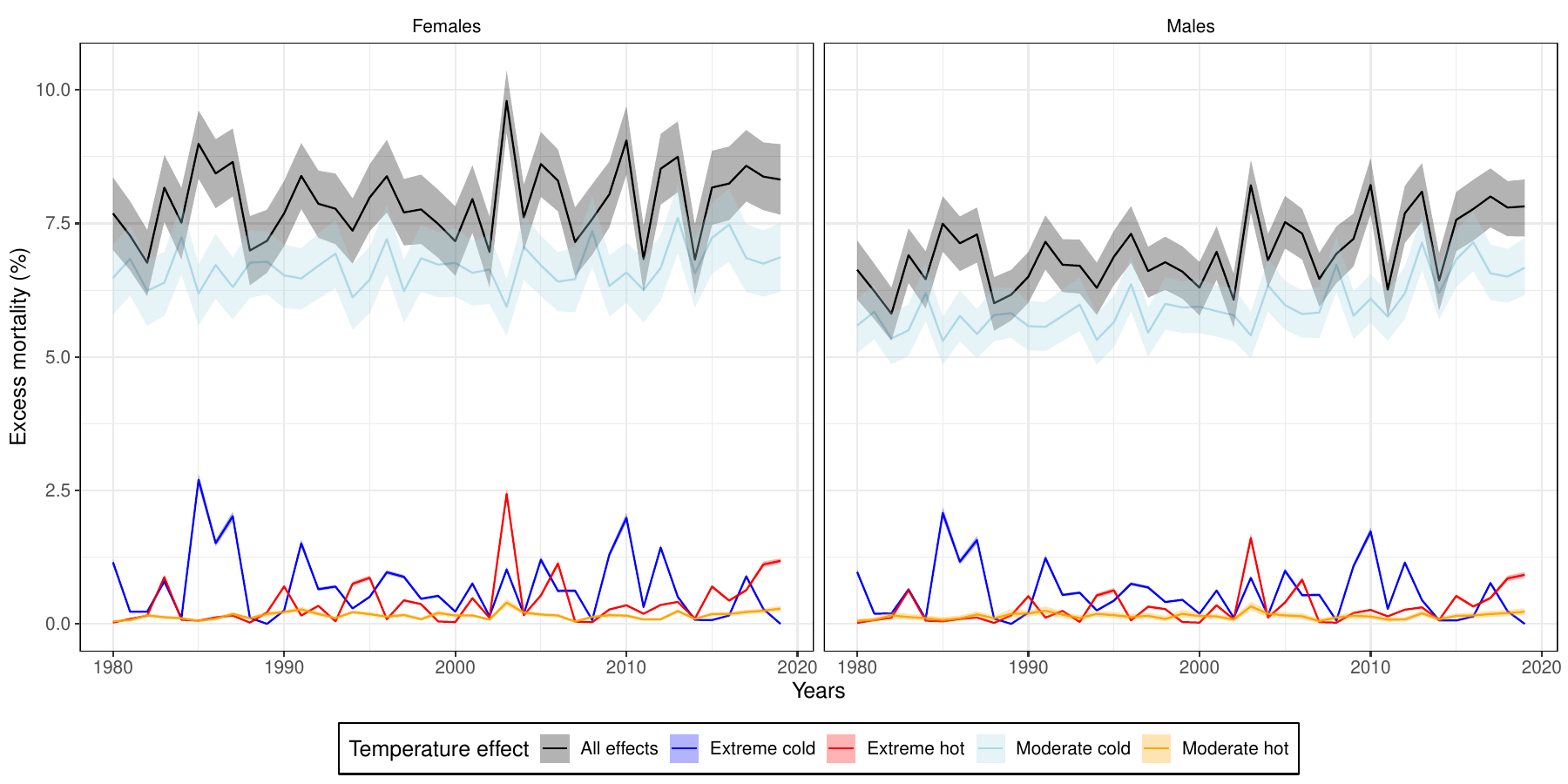}
\caption{
Temperature attributable fraction in Metropolitan France, calculated for the years 1980-2019 for both women and men across all age groups. These values, along with their $95\%$ confidence intervals (shaded areas) are calculated through 1,000 Monte Carlo simulations of the DLNM coefficients. Each line corresponds to the average of these simulations. Attributable fraction is expressed in \% for all effect, decomposed into moderate cold, moderate hot, extreme cold and extreme hot. 
}
\label{fig_temp_decomp}
\end{figure}

Over the period from 1980 to 2019, we finally represent in Figure~\ref{fig_temp_decomp} the temperature attributable fraction for both women and men, defined in Equation~\eqref{eq:estim_attrib_factor}, and aggregated over all ages to facilitate visualization. This fraction globally oscillates between $5\%$ and $9\%$ over the considered period, except for women where a peak appears for the 2003 extreme heatwave \citep{charpentier_return_2011}. This fraction is decomposed into different temperature effects, depending on the observed daily average temperatures. Hot or cold temperature days corresponds to days $\mathcal{D}_t$ with average temperature above or below the MMT. Thus, similarly to \citet{martinez-solanas_projections_2021}, we define moderate cold days (respectively moderate hot days) as cold days where the average temperature is below the $2.5\%$ quantile (respectively above the $97.5\%$ quantile)\footnote{Over the observation period, the $2.5\%$ quantile of the average temperature series is 0.86°C and the $97.5\%$ quantile is 23.44°C.} of observed average temperatures over the period. Extreme cold (respectively extreme hot) days are those where the average temperature is below the $2.5\%$ quantile (respectively above the $97.5\%$ quantile). The most significant contribution to the attributable fraction for both women and men comes from moderate cold, while the effects of moderate hot temperatures are generally constant and very low. Extreme cold and hot events produce peaks in excess mortality in certain years, such as in 1985 and 2010 for cold effect, or in 2003 for the heatwave effect. We observe from around 2015 what appears to be the beginning of an upward trend in the extreme hot effect for both women and men.

Appendix~\ref{subsec:app_fit_dlnm_desc_analysis} includes an empirical analysis presenting the predicted all-cause death counts derived from the DLNM model. We visually assess its ability to reproduce the seasonality of daily deaths for womens and mens and for each age group. This appendix also details the decomposition of seasonal mortality and illustrates that the temperature-attributable component aligns with the evolution of daily average temperatures.
As explained in Section~\ref{subsec:estim_dlnm}, we choose to adopt reference hyperparameters for the calibration of our DLNM model, as they provide a satisfactory fit to the data and facilitate comparisons with the literature. The goodness of fit of our DLNM models is assessed by comparing the distributions of observations and predictions, as well as the residuals of the deviance, see Appendix~\ref{subsec:app_fit_dlnm_model}. Sensitivity analysis to model hyperparameters is conducted in Appendix~\ref{subsec:app_dlnm_sensi} to assess its robustness. Additionally, we check that our model predicts an insignificant number of negative temperature-attributable deaths.

\subsection{Calibrating and projecting the Li-Lee model 
\label{subsec:calibrating_li_lee}}

In this section, we present the estimates obtained for the Li-Lee model described in Section~\ref{subsec:baseline_estimation}, in order to project the predicted values of the central death rate not attributable to temperatures $\widetilde{m}_{x,t}^{(g)}$. To assess the impact of adjusting exposure to risk, we compare the results of the model~\eqref{eq:2_LL}, which incorporates observed death counts and adjusted exposure to risk according to Equation~\eqref{eq:poisson_adj_expo}, with those of the original Li-Lee model for $\widehat{m}_{x,t}^{(g)}$, which is based on observed death counts and unadjusted exposure to risk, i.e. 
\begin{align}\label{eq:original_LL}
D_{x,t}^{(g)} \sim \text{Pois}\left( E_{x,t}^{(g)}  \exp{\left(  A_x + B_{x}K_{t} + \alpha_{x}^{\left( g \right)} + \beta_{x}^{\left( g \right)} \kappa_{t}^{\left( g \right)}\right)}\right).
\end{align}
In both cases, our estimates of the Li-Lee model are obtained based on annual mortality data over the period $\mathcal{T}_y$.

\begin{figure}[h!]
\centering
\includegraphics[scale = 0.5, page = 2]{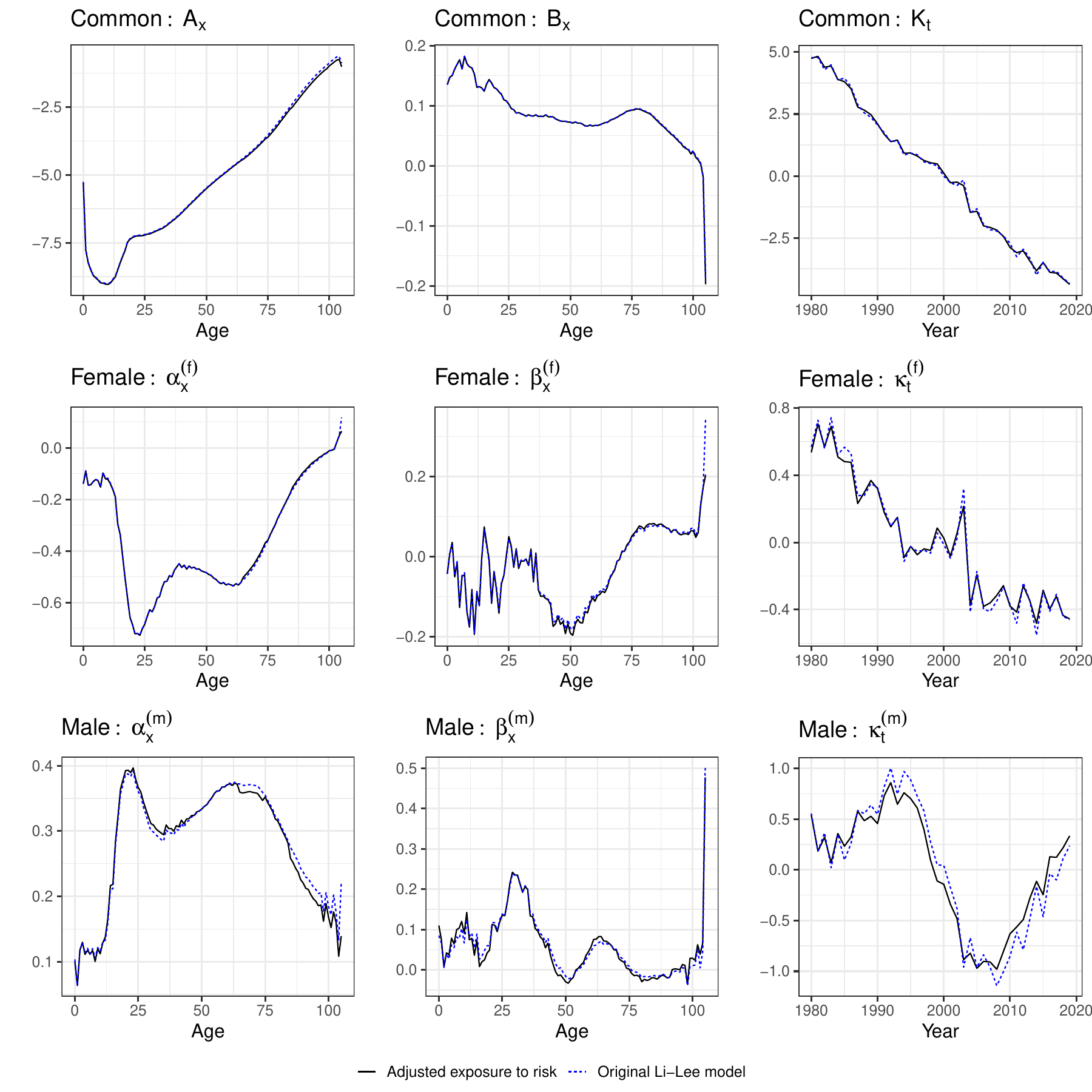}
\caption{
Estimated parameters $(\widehat{A}_x, \widehat{B}_x, \widehat{K}_{t},
\widehat{\alpha}_{x}^{(f)}, \widehat{\beta}_{x}^{(f)}, \widehat{\kappa}_{t}^{(f)},
\widehat{\alpha}_{x}^{(m)}, \widehat{\beta}_{x}^{(m)}, \widehat{\kappa}_{t}^{(m)})$ of the Li-Lee model for the calibration period 1980-2019 and ages between 0-105 for the entire population of Metropolitan France (Common), females (Female), and males (Male). We distinguish between estimated parameters of model~\eqref{eq:2_LL} represented by black solid lines, and those of model~\eqref{eq:original_LL} in blue dashed lines. 
}
\label{fig_li_lee_parameters}
\end{figure}

Figure~\ref{fig_li_lee_parameters} illustrates the dynamics of the parameters of both Li-Lee models. It first appears the close proximity between their estimated parameters. In particular, the temporal trend $\widehat{K}_t$ is not disrupted by adjusting exposure to risk. Although the impact is moderate, the most affected trend is visually that related to $\widehat{\kappa}^{(m)}$. 

\begin{figure}[h!]
\centering
\includegraphics[scale = 0.5]{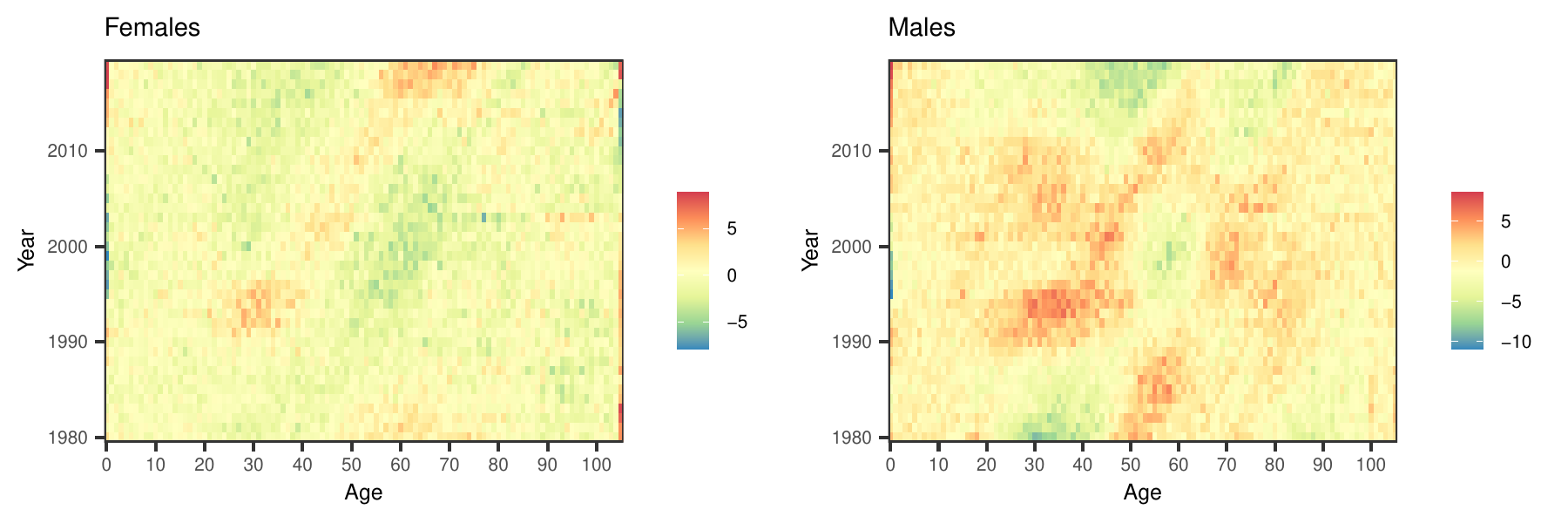}
\caption{
Pearson residuals of the Li-Lee model for the calibration period 1980-2019 and ages between 0-105 for the female and male populations of Metropolitan France. The model is fitted on observed deaths and temperature-adjusted exposures to risk.  
}
\label{fig_li_lee_residuals}
\end{figure}

Figure~\ref{fig_li_lee_residuals} examines the goodness of fit of the estimated Li-Lee model~\eqref{eq:2_LL} by presenting the Pearson residuals of the fitted Poisson model. Although we have chosen a relatively simple mortality model, it adequately captures age and period effects for both female and male populations. We observe that some cohort effects are less well taken into account, especially for the male population. This limitation could potentially be addressed by using a model with a cohort component. However, including this effect generally introduces numerous other issues, such as less robustness than an Lee-Carter model and difficulties in projecting model parameters. Additionally, cohort effect modeling should be justified by arguments supporting the existence of real cohort effects in the data \citep{hunt_structure_2021}. This situation is not straightforward concerning French data. For example, \citet{boumezoued_new_2020} shows a significant reduction in isolated cohort effects on the French population by incorporating information on monthly birth counts. An another reason is the diagonal trends in the female population appear relatively diffuse and limited. Since the issue of cohort effects is beyond the scope of this paper, we choose to retain a model capturing only age and period effects. It is also worth noting that our chosen specification for handling temperature-attributable deaths could easily be adapted to a another stochastic mortality model.

The joint dynamics of the parameters $(\widehat{K}_t$, $\widehat{\kappa}_t^{(f)}, \widehat{\kappa}_t^{(m)})$ are determined using the time series model described in Section~\ref{subsec:baseline_time_series}, employing respectively a RWD process and two AR(1) processes. The estimated parameters from this model are presented in Table~\ref{tab:li_lee_parameters}, both for the original Li-Lee model~\eqref{eq:original_LL} as well as the model with adjusted exposure to risk~\eqref{eq:2_LL}. Table~\ref{tab:li_lee_corrmatrix} describes the correlation matrix of the residuals for these two sets of parameters.

To ensure compliance with the consistency constraint specified by the Lee-Li model, the dynamics of the processes $\widehat{\kappa}_t^{(f)}$ and $\widehat{\kappa}_t^{(m)}$ must remain stationary, i.e. the absolute value of $\phi^{(g)}$, $g \in \{f,m \}$, must be strictly lower than 1. This constraint is met during the maximum likelihood estimation of the model~\eqref{eq:time_series_matrix} regarding $\kappa_t^{(m)}$ and $\kappa_t^{(f)}$ for observed deaths. However, it is observed that the values of the coefficients $\phi^{(m)}$ and $\phi^{(f)}$ are very close to 1. Regarding $\phi^{(f)}$ when adjusting exposure to risk to the temperature effect $\widehat{T}_{x,t}^{(g)}(\mathcal{D}^\star)$, the parameter values are slightly above 1. This undesirable situation would thus lead to projecting mortality rates for women diverging from the common trend $K_t$ in the long term. To overcome this issue, \citet{antonio_iabe_2020} proposes reducing the calibration period of the model or increasing the lag of the autoregressive processes. Other strategies could be proposed, such as using a mortality model based on mortality increments \citep{hunt_mortality_2023} or employing a data-driven model, as seen in \citet{guibert_forecasting_2019}. However, this topic is beyond the scope of this paper, and we choose to address it more simply by constraining the coefficient $\phi^{(g)}$ to be strictly less than 1 during maximum likelihood maximization.

\begin{table}[!h]

\caption{Calibrated parameters in the Li-Lee models~\eqref{eq:original_LL} and~\eqref{eq:2_LL} on the period 1980-2019 and age range 0-105.\label{tab:li_lee_parameters}}
\centering
\fontsize{8}{10}\selectfont
\begin{tabular}[t]{cccccc}
\toprule
Calibration data & $\delta$ & $c^{(m)}$ & $\phi^{(m)}$ & $c^{(f)}$ & $\phi^{(f)}$\\
\midrule
Original Li-Lee model & -0.2338 & -0.0083 & 0.9681 & -0.0264 & 0.9978\\
Adjusted exposure to risk & -0.2339 & -0.0055 & 0.9681 & -0.0254 & 0.9999\\
\bottomrule
\end{tabular}
\end{table}

\begin{table}[!h]

\caption{
Correlation matrix of residuals in the Li-Lee models~\eqref{eq:original_LL} and~\eqref{eq:2_LL} on the period 1980-2019 and age range 0-105.\label{tab:li_lee_corrmatrix}}
\centering
\fontsize{8}{10}\selectfont
\begin{tabular}[t]{ccccccc}
\toprule
\multicolumn{1}{c}{ } & \multicolumn{3}{c}{Original Li-Lee model} & \multicolumn{3}{c}{Adjusted exposure to risk} \\
\cmidrule(l{3pt}r{3pt}){2-4} \cmidrule(l{3pt}r{3pt}){5-7}
 & $e_t$ & $r_t^{(m)}$ & $r_t^{(f)}$ & $e_t$ & $r_t^{(m)}$ & $r_t^{(f)}$\\
\midrule
$e_t$ & 1.0000 & -0.7658 & 0.9437 & 1.0000 & -0.5826 & 0.9330\\
$r_t^{(m)}$ & -0.7658 & 1.0000 & -0.7451 & -0.5826 & 1.0000 & -0.5683\\
$r_t^{(f)}$ & 0.9437 & -0.7451 & 1.0000 & 0.9330 & -0.5683 & 1.0000\\
\bottomrule
\end{tabular}
\end{table}

\begin{figure}[h!]
\centering
\includegraphics[scale = 0.5]{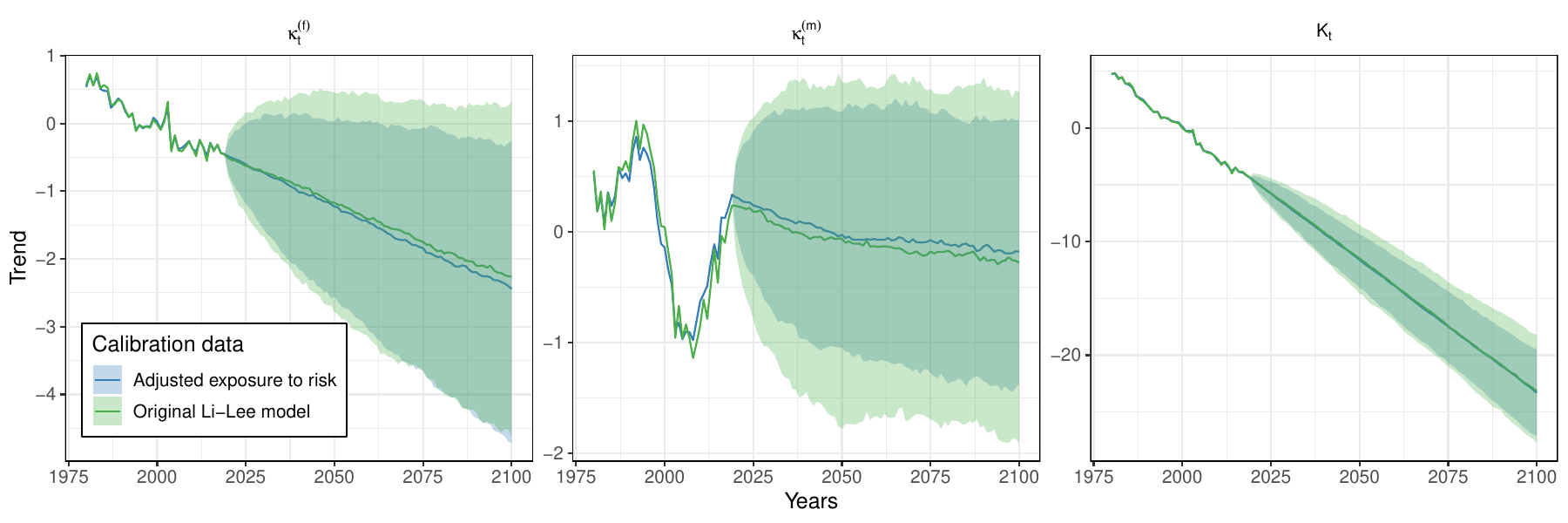}
\caption{
Projection of trend parameters $\widehat{K}_t$, $\widehat{\kappa}_t^{(f)}$, and $\widehat{\kappa}_t^{(m)}$ over the period 2020-2100 for the Li-Lee models with original exposure to risk~\eqref{eq:original_LL} and ajusted exposure to risk~\eqref{eq:2_LL} for the calibration period 1980-2019 and ages between 0-105 for the entire population of Metropolitan France. The solid lines represent from 2020 the median of the projections. The $95\%$ prediction intervals are constructed based on 1,000 Monte-Carlo simulations.
}
\label{fig_li_lee_forecast_trend}
\end{figure}

Figure~\ref{fig_li_lee_forecast_trend} finally presents the projection of parameters $\widehat{K}_t$, $\widehat{\kappa}_t^{(f)}$, and $\widehat{\kappa}_t^{(m)}$ over the period 2020-2100, along with the $95\%$ prediction intervals obtained by 1,000 Monte-Carlo simulation of innovation errors. 
These projections are shown for both models, with and without adjusting exposure to risk for temperature effects. More precisely, the trends of model~\eqref{eq:2_LL} correspond to the dynamic of central death rates not attributable to temperature effects. We first observe that the common trend $K_t$ is virtually identical in terms of median for both parameter estimations. However, as expected, the projected uncertainty for the model~\eqref{eq:original_LL} is wider than the trend for central death rates not attributable to temperatures $\widetilde{m}_{x,t}^{(g)}$. This is explained by adjustments in exposure to risk related to mortality peaks induced by temperature variations. Regarding the parameters $\widehat{\kappa}_t^{(f)}$ and $\widehat{\kappa}_t^{(m)}$, we notice the similarity between the forecasted series, and the same phenomenon in terms of uncertainty.

In Appendix \ref{subsec:app_consistency_rates}, we further verify that the projections of the Li-Lee model with original exposure to risk~\eqref{eq:original_LL} and adjusted exposure to risk~\eqref{eq:2_LL} are consistent.

\subsection{Simulating temperatures effects}
\label{subsec:simul_temp_effets}

We now focus on the simulation of total attributable fractions to temperatures $\widehat{\text{AF}}_{x,t}^{(g)} (\mathcal{D}_t)$, as defined in Equation~\eqref{eq:simul_attrib_fraction}. In this equation, we consider weights, estimated based on daily deaths not attributable to temperatures from the last five years of our historical data. \footnote{Alternatively, the use of constant weights empirically leads to a very slight increase in the temperature-attributable fractions on these data.} These attributable fractions depend on a period $\mathcal{D}_t$ for each year $t$ that allows for distinguishing the effects of heat and cold over the year. These quantities are also calculated for each age group, sex and calendar year. To simplify notation, we do not include the index corresponding to the Monte-Carlo simulation numbers.

To facilitate visual analysis, we calculate an aggregated attributable fraction to temperatures using the distribution of deaths known at the end of 2019, as follows
$$
\widehat{\text{AF}}_{t} (\mathcal{D}_t)  = \sum_{g \in \mathcal{G}}
{
\sum_{x \in \mathcal{X}}{\widehat{\text{AF}}_{x,t}^{(g)} (\mathcal{D}_t) \frac{D_{x, 2019}^{(g)}}{D_{2019}}
}},
$$
where
$
D_{2019}  = \sum_{g \in \mathcal{G}}
{\sum_{x \in \mathcal{X}}{D_{x, 2019}^{(g)}}}.
$
The details of the attributable fractions to temperatures $\widehat{\text{AF}}_{x,t}^{(g)} (\mathcal{D}_t)$ per age bucket and sex are provided in Appendix~\ref{subsec:app_attrib_fraction_sex_age}.

\begin{figure}[h!]
\centering
\includegraphics[scale = 0.55]{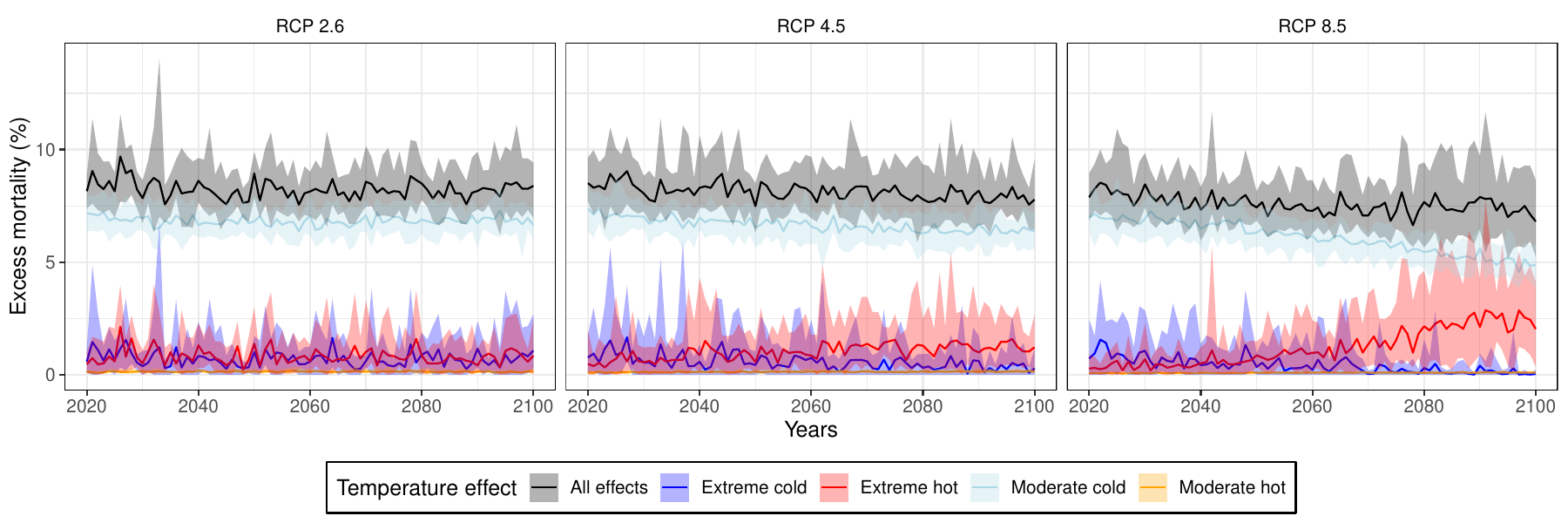}
\caption{
Temperature attributable fraction in Metropolitan France, simulated for the years 2020-2100 for both women and men across. These values, along with their $95\%$ confidence intervals (shaded areas) are calculated through 1,000 Monte Carlo simulations of the DLNM coefficients and the ensemble of climate models. Each line corresponds to the average of these simulations. Attributable fraction is expressed in \% for all effect, decomposed into moderate cold, moderate hot, extreme cold and extreme hot. 
}
\label{fig_attrib_global}
\end{figure}

Figure~\ref{fig_attrib_global} breaks down the evolution of attributable fractions related to hot and cold temperatures according to the RCP scenarios for the period 2020-2100 and all population. The presented $95\%$ confidence intervals combine, for each RCP scenario, the estimation error associated with the coefficients $\thetavec_{k}^{(g)}$, measured by 1,000 bootstrap replications, and the uncertainty related to different climate models listed in Table~\ref{tab:list_climate_models}. Similar to Figure~\ref{fig_temp_decomp}, we examine the evolution of components associated with moderate cold and hot days and extreme cold and hot days. Firstly, we observe a relatively stable overall effect of mean temperatures until the end of the century in the RCP2.6 scenario and a slightly decreasing effect in the RCP4.5 scenario for women and men. The RCP4.5 scenario is also characterized by a decrease in the effect of moderate cold and a progressive increase in the effect of extreme heat. It is noteworthy that there is an increase in the uncertainty of the effect of extreme heat over time.

The RCP8.5 scenario exhibits a different behavior. Specifically, it shows a more pronounced decrease over time than the RCP4.5 scenario in the effect of moderate cold temperatures, which is the main component of temperature-attributable mortality for both women and men. Concurrently, a marked increase in the effect of extreme heat is observed starting from the 2050s. However, as highlighted in Appendix~\ref{subsec:app_attrib_fraction_sex_age}, this increase is not sufficient to reverse the overall declining trend in the effect of temperatures over the century for the males, except for those aged between 0 and 64 years. For the female population, we observe a slight resurgence in the effect of temperatures towards the end of the century, as well as greater uncertainty in temperature-attributable mortality.

It is also noteworthy that the effects of temperatures vary according to geographical locations \citep{martinez-solanas_projections_2021}. To illustrate this pronounced effect between the north and south of France, we simulate the trajectory of attributable fractions for different French cities across the country, namely Brest, Marseille, Paris, Perpignan, and Strasbourg, using data from DRIAS. Detailed results are presented in Appendix~\ref{subsec:app_attrib_fraction_city}. For this purpose, we employ the DLNM model estimated on the daily mortality data of Metropolitan France as presented in Section~\ref{subsec:calibrating_dlnm}. This is an approximation that could be improved by having access to time-series of death counts per city. Indeed, each of the sub-populations residing there may have specific adaptation capacities to hot and cold temperatures, e.g. air conditioning or thermal insulation of buildings, which we do not account for. 

These simulations reveal different sensitivities of each of these cities to extreme hot temperatures and moderate cold temperatures. For the city of Brest for example, the contribution of extreme hot temperatures to mortality remains very limited in the latter part of the century, even under the RCP8.5 scenario. On the other hand, the contribution of cold temperatures decreases gradually, which could tend to decrease the contribution of temperatures to mortality in this region. In contrast, the city of Perpignan appears significantly more exposed to extreme hot temperatures in the RCP4.5 scenario and in the RCP8.5 scenario.

\subsection{Impacts on life-years lost due to temperature\label{subsec:life_expect_lost}}

\begin{figure}[h!]
\centering
\includegraphics[scale = 0.35]{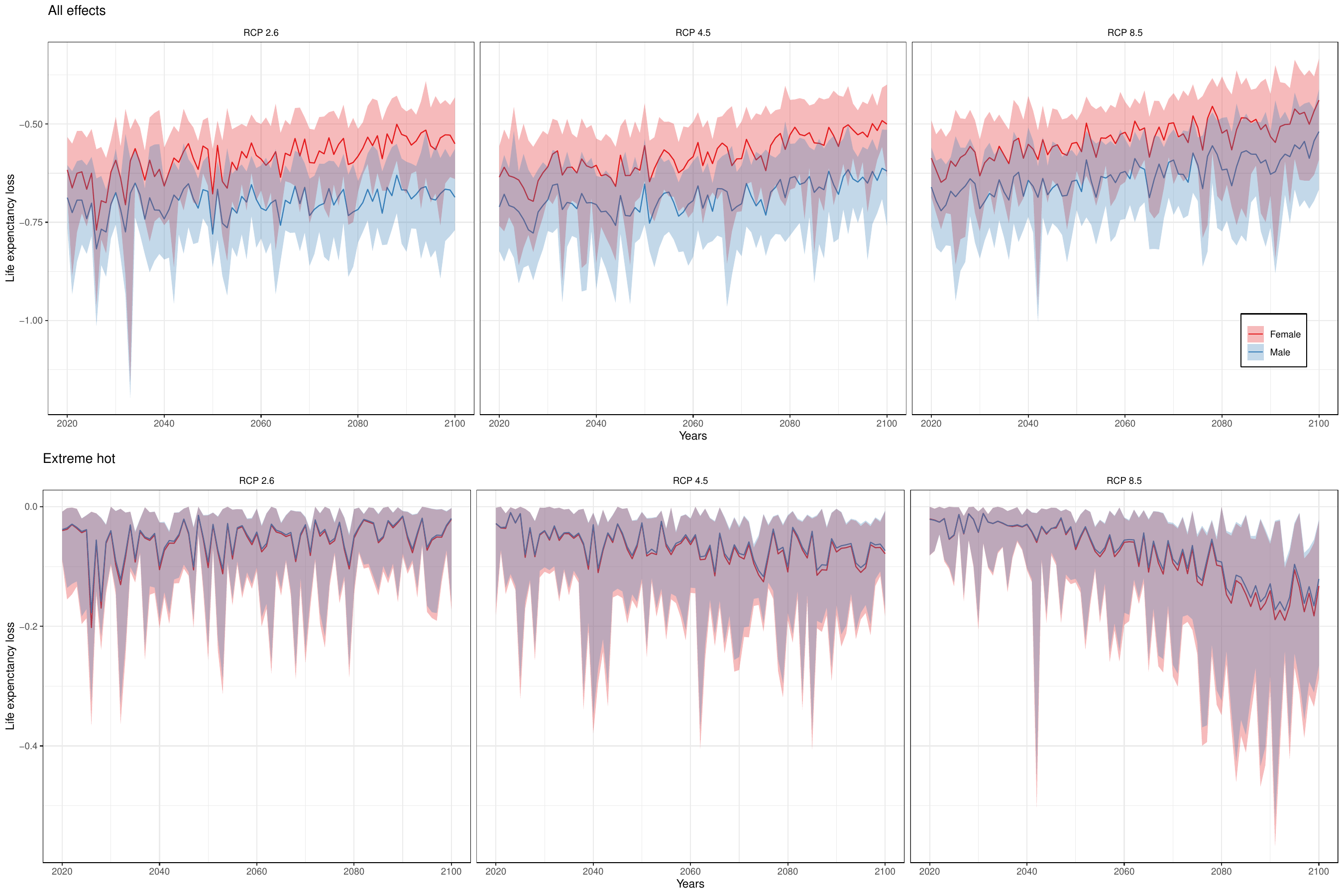}
\caption{
Life expectancy at birth lost in Metropolitan France, simulated for the years 2020-2100 for both women and men. We present both the loss related to all temperature effects and extreme hot effects only. These values, along with their $95\%$ confidence intervals (shaded areas), are calculated through 1,000 Monte Carlo simulations of the DLNM coefficients, the Li-Lee model and the ensemble of climate models. Each line corresponds to the median of these simulations. 
}
\label{fig_ev_gap}
\end{figure}

Finally, we examine the effect of temperatures on the loss of life expectancy at birth. Figure~\ref{fig_ev_gap} shows the loss of life expectancy at birth, as defined in Equation~\eqref{eq:loss_le}, for women and men between 2020 and 2100 due to temperatures in Metropolitan France. We specifically distinguish the overall temperature effects, which account for both hot and cold temperature-related mortality, and the sole effects of extreme hot temperatures. The different trajectories presented correspond to RCP2.6, RCP4.5 and RCP8.5 scenarios. These assessments are conducted under the assumption that populations and healthcare systems do not adapt to temperature changes throughout the century and evolve similarly at national level. Hence, the potential evolution of adaptation to both heat and cold, and the differences between regions are not analyzed in this study. This is a strong hypothesis but difficult to address in a prospective approach as it may evolve according to numerous factors (physiological, technological, immunological, etc.). It is noteworthy that recent literature suggest that populations can adapt to their local environment \citep{wu_temperature_2024}. 

For the overall temperature effects, we first note that the loss of life expectancy for men is greater (in absolute value) than that for women. This difference is explained by men's higher sensitivity to cold temperatures compared to women (Figure~\ref{fig_dlnm_rr_curve}), especially in the age groups between 65 and 84 years. In the three RCP scenarios presented, rising temperatures would limit the contribution of cold to mortality. This would conduct to a general reduction of temperature-related life expectancy loss. We also observe in the RCP8.5 scenario a small reduction of the gap between women and men from 2050 onward due to increasing temperatures.

At the same time, attention must be paid to the evolution of life expectancy loss related to extreme hot temperatures. These excesses lead to reductions in life expectancy loss that generally do not exceed 0.1 years, except in a few exceptional years, under the RCP2.6 scenario. This loss does not show any particular trend throughout the century. The RCP4.5 scenario shows a slight but steady trend of reduction (in absolute value) in life expectancy loss for both women and men with the gradual increase in temperatures. The $95\%$ uncertainty associated with our life expectancy loss predictions becomes wider from 2050 onwards in this scenario. On the other hand, the RCP8.5 scenario comprises an even greater worsening of life expectancy loss due to extreme heat from the second half of the century, potentially reaching nearly 0.2 years by 2100. The uncertainty associated with these predictions increases significantly over time due to the longer duration, severity, and intensity of heatwave episodes (Figure~\ref{fig_proj_temp_rcp}).

While many studies quantify the impact of temperature in terms of excess mortality, fewer examine its effect on life expectancy to our knowledge. Among these rare studies, \citet{hauer_inaction_2021} estimates the impact of climate change under a business-as-usual scenario, equivalent to our RCP8.5 scenario. Specifically for France, they find a reduction in life expectancy due to climate change of approximately 0.65 years by 2080, which is more pessimistic than our results. However, it is important to note that their study only accounts for the extreme effects of temperature (heatwaves and cold spells) and ignores the compensatory mechanisms between the effects of cold and heat, particularly the gain in life expectancy resulting from the reduction in the attributable effect of moderate cold. Therefore, their estimates should be more closely compared to the approximate 0.2-year loss in life expectancy that we observe for heatwaves by the end of the century, with a large uncertainty depending on the climate model used.

This situation at the scale of Metropolitan France varies according to geographical location. Thus, the same figures for temperature-related life expectancy loss are reproduced for Brest, Marseille, Paris, Perpignan and Strasbourg. For that purpose, we use the projections of central death rates not attributable to temperature effects, as described in Section~\ref{subsec:add_deaths_temp}, and then apply the attributable fractions simulated from the temperature trajectories specific to each city using data from DRIAS. Detailed results are presented in Appendix~\ref{subsec:app_loss_le_city}. For these different cities, the trajectory of temperature-related life expectancy loss clearly depends on the evolution of hot and cold contributions in each area of France. In the northern part of France, including Paris, Brest and Strasbourg, the life expectancy loss related to all temperature effects tends to decrease (in absolute value) in all RCP scenarios due to the reduced contribution of cold to mortality. In the RCP8.5 scenario where average temperatures tend to rise rapidly, the life expectancy loss seems to become lower more quickly. In the southern part of France (Marseille and Perpignan), the trajectory of life expectancy loss reduction is less clear in RCP2.6 and RCP4.5. In the RCP8.5 scenario for Perpignan, the temperature-related life expectancy loss increases (in absolute value) in the second half of the century. For this city, where extreme heat will be more frequent at the end of the century, the median life expectancy loss due to heatwaves alone reaches between 0.3 and 0.4 years.

\section{Conclusion} \label{sec:conclusion}

Temperature changes linked to climate change create disturbances that will affect the frequency of observed deaths throughout the year. In this paper, we design a multi-population mortality model incorporating the effect of temperature changes on mortality. This framework includes a classic climate epidemiology model, known as the Distributed Lag Non-Linear Model (DLNM), which measure deaths attributable to hot and cold temperatures over different periods of the year. By coupling these two modeling frameworks and integrating temperature trajectories from various climate models, we project central death rates impacted by temperature effects and measure these impacts in terms of gains or losses in life expectancy.

Our results describe a situation dependent on both climate scenarios and the geographical location of populations. Specifically, the simulated gains or losses in life expectancy until the end of the century, particularly from 2050 in Metropolitan France, show a downward trend in mortality related to cold and an upward trend in mortality related to extreme heat. These two effects evolve in opposite directions and reveal a source of divergence between certain regions of France, especially under the RCP8.5 scenario. For all results, we incorporate multiple sources of uncertainty, including climate scenarios, prediction errors of the mortality model, and estimation errors of the DLNM model. The modeling framework established here and illustrated with French data offers an adaptable approach for other countries or regions, whether for demographic needs or for insurers, actuaries, or supervisory authorities interested in designing climate scenarios. Further research perspectives are opened by extending the multi-population mortality model to other countries or by refining the geographical resolution of the considered territory. For instance, some regions will be more affected than others by climate change, and more pronounced impacts of heat-related mortality are expected in Southern Europe or the MENA region \citep{hajat_current_2023}. Otherwise, the study of daily mortality or on finer time scales may encompass other environmental components such as air pollution or humidity, which we do not consider in this study. These factors introduce additional sources of uncertainty that should be addressed in future research. In this regard, the DLNM model could be calibrated on multiple sites and additional environmental variables. Thus, a possible extension would be to model the simultaneous occurrence of mortality shocks related to temperature extremes and other environmental variables.

In this paper, we make the strong assumption that populations do not adapt to their local environment, which is debated in recent literature \citep{wu_temperature_2024}. Indeed, the number of deaths related to temperatures could decrease thanks acclimatization. This can result both from technology, e.g. improvement of house insulation, development of air conditioning, or by a physiological process, see \citet{anderheiden20} among others. The vulnerability of populations to low temperatures, influenced in particular by respiratory virus infections, also needs to be questioned in light of changes in acquired immunity and the duration of infection periods across different regions \citep{walkowiak_exploring_2024}. Furthermore, prevention and protection plans could be set up to alert and guide the population on how to behave in the face of heatwaves. Concrete actions like raising awareness of population about heatwaves, changing working hours, opening air-conditioned reception areas, distributing water in public transport, monitoring the most vulnerable at home, and limiting or prohibiting outdoor sports activities could also affect heat-related mortality. Considering these acclimatization factors offers research perspectives to be explored in the construction of prospective scenarios.

\paragraph{Supplementary material:}
The results in this paper is obtained using $\mathtt{R}$. Supplementary material related to this paper can be found at \url{https://github.com/qguibert/mortalityheatwaves}.

\paragraph{Author Contributions:}
Q.G., G.P. and F.P. design the study. Q.G. and G.P. collect, cure and verify the data. Q.G. designs the models and the methodology. Q.G. implements methods, produces results and visualizations. Q.G. and G.P. write the manuscript. Q.G., G.P and F.P. edit the manuscript. 

\paragraph{Funding:} This research received no external funding.

\paragraph{Acknowledgments:}
The authors are very grateful for the useful suggestions of the two anonymous referees, which led to significant improvements of this article. This paper also benefits from fruitful discussions with members of the \textit{Réseau thématique MATRISK} during the Inaugural Conference of this group at Le Mans (France) in June 2024. For the purpose of Open Access, a CC-BY public copyright licence has been applied by the authors to the present document and will be applied to all subsequent versions up to the Author Accepted Manuscript arising from this submission.

\newpage

\appendix

\section{Heatwaves in France}
\label{sec:app_hist_france}

Heatwaves represent a particular risk of acute mortality and their importance has increased over the past 20 years. To understand this phenomenon for the studied population, this appendix provides a synthetic overview of historical recent heatwaves in metropolitan France since 1950. Additional details are provided by \citet{pincemin_risques_2021}.

Heatwaves do not have a single definition. Depending on the temperature distribution and the adaptation capacity of populations in the region concerned, the definitions and tolerance thresholds adopted may vary. In particular, the identification of a heatwave may depend on the climate, urbanization of the area concerned, as well as indicators measuring its intensity (average temperature, daytime or nighttime temperature, duration of the reference period, humidity rate, etc.). In the context of this study on French population, we define a heatwave according to the criteria used for heatwave alerts by Météo France, the French national meteorological service. These criteria are also used by Santé Publique France \citep{santepubliquefrance19}. Météo France defines a heatwave period as a period where the average of daily temperatures during the day and at night exceeds the $99.5 \%$ percentile of observations made between 1981 and 2010.

Gradually, Western European countries such as France have experienced a decrease in the intensity and frequency of cold spells \citep{meteofrance22}. Conversely, heatwaves have become more frequent, and summers are increasingly hotter. In France, the major heatwaves of 1976 and 2003 resulted in an excess of 4,500 and 15,000 deaths, respectively. In recent years, heatwaves have occurred almost every single year with varying intensities. Some of them have had temperatures similar to the historical heatwave of 2003. In 2019 and 2020, they even reached the highest level of the alert plan for the first time in history. 

Their consequences in terms of mortality are listed in Table~\ref{tab:hist_canicule}, which contains data compiled from \textit{Santé Publique France} and the Centre for Research on the Epidemiology of Disasters (CRED) Database. It is noteworthy that the dates and duration of each heatwave are calculated on average by department and are extracted from the annual reports of \textit{Santé Publique France}. Considering the average for the entirety of France would result in a shorter duration of heatwaves. For comparative purposes, the CRED data is also included in this table. Significant differences in excess mortality attributed to these various events are observed depending on the data source, due to variations in the methodologies employed.

\begin{table}[!h]
\centering
\footnotesize
\caption{
Main heatwaves in Metropolitan France since 1976. This table presents for each year the cumulative number of heatwave days and the attributed number of deaths, calculated on average per department by \textit{Santé Publique France}. These data are extracted from the \textit{Bulletins de Santé Publique France} (reports analyzing the departmental, regional, and national situation) on the relevant summers \citep{spf18, stf19, stf20} and from the publication of the \citet{InVS15}. Data from the earlier years (1976, 1983, 2003, 2006) are sourced from a study on heatwaves since 1970 \citep{santepubliquefrancecanicule19}. The CRED data is drawn from the EM'DAT database \citep{cred_2021}. This open and free database is a project of the Integrated Research on Disaster Risk. It is one of the three largest databases for disasters, along with MunichRe's NatCatSERVICE and SwissRe's Sigma. It is fed by numerous governmental and non-governmental sources, insurers and reinsurers, press agencies, specialized entities and various global organizations (United Nations, SwissRe, MunichRe, NOAA, etc.).
}
\begin{tabular}[t]{>{ \arraybackslash}p{30em}>{\centering\arraybackslash}p{5em}>{\centering\arraybackslash}p{5em}}
\toprule
\multicolumn{1}{c}{ } & \multicolumn{2}{c}{\textbf{Excess Death}} \\
\cmidrule(l{3pt}l{3pt}){2-3}
\textbf{Key information} & \textit{Santé Publique France} & \textit{Cred Em'Dat}\\
\rowcolor{gray!6}
\midrule
\addlinespace[0.3em]
\multicolumn{3}{l}{\textbf{1976}}\\
\rowcolor{gray!6}
\hspace{1em}Late June – mid-July & 4,540 & -\\
\addlinespace[0.3em]

\multicolumn{3}{l}{\textbf{1983}}\\
\hspace{1em}July 9 – July 31 (22 days) & 2,900 & -\\
\addlinespace[0.3em]

\rowcolor{gray!6}
\multicolumn{3}{l}{\textbf{2003}}\\
\rowcolor{gray!6}
\hspace{1em}The summer of 2003 was the hottest in France and Europe in over 50 years, with exceptional intensity during the first half of August. This episode was associated with significant ozone pollution. & 14,800 & 19,490\\
\addlinespace[0.3em]

\multicolumn{3}{l}{\textbf{2006}}\\
\hspace{1em}July 2006 witnessed a very intense heatwave, although less severe than in 2003, but with a longer duration. & 1,442 & 1,388\\

\rowcolor{gray!6}
\addlinespace[0.3em]
\multicolumn{3}{l}{\textbf{2015}}\\
\rowcolor{gray!6}
\hspace{1em}Three waves unevenly distributed over the territory. The first was quite intense, affecting the north for 10 days. The second, lasting 11 days, was less intense and concentrated in the south-east. The last one, shorter, was very localized in some eastern and southern departments. & 2,040 & 3,275\\
\addlinespace[0.3em]

\multicolumn{3}{l}{\textbf{2018}}\\
\hspace{1em}A two-week heatwave peaking in early August, exposing 70\% of the French population. & 1,480 & -\\
\addlinespace[0.3em]

\rowcolor{gray!6}
\multicolumn{3}{l}{\textbf{2019}}\\
\rowcolor{gray!6}
\hspace{1em}Two very extensive and intense heatwaves. For the first time, several departments were under red alert during both waves. & 2,034 & 1,435\\

\addlinespace[0.3em]
\multicolumn{3}{l}{\textbf{2020}}\\
\hspace{1em}Three heatwaves, one particularly severe in the north, resulting in a red alert for the second consecutive year, over a cumulative period of around ten days. & 1,924 & 1,924\\

\bottomrule
\end{tabular}
\label{tab:hist_canicule}
\end{table}

In 2022, we observed an excess mortality of 73 deaths per million (26, 124), which corresponds to 4,807 (1,739, 8,123)\footnote{The values in parentheses represent the $95\%$ confidence intervals.} deaths \citep{ballester_heat-related_2023}. 

\section{Monthly and annual mortality data \label{sec:app:data_sources}}

\begin{figure}[h!]
\centering
\includegraphics[scale=0.35]{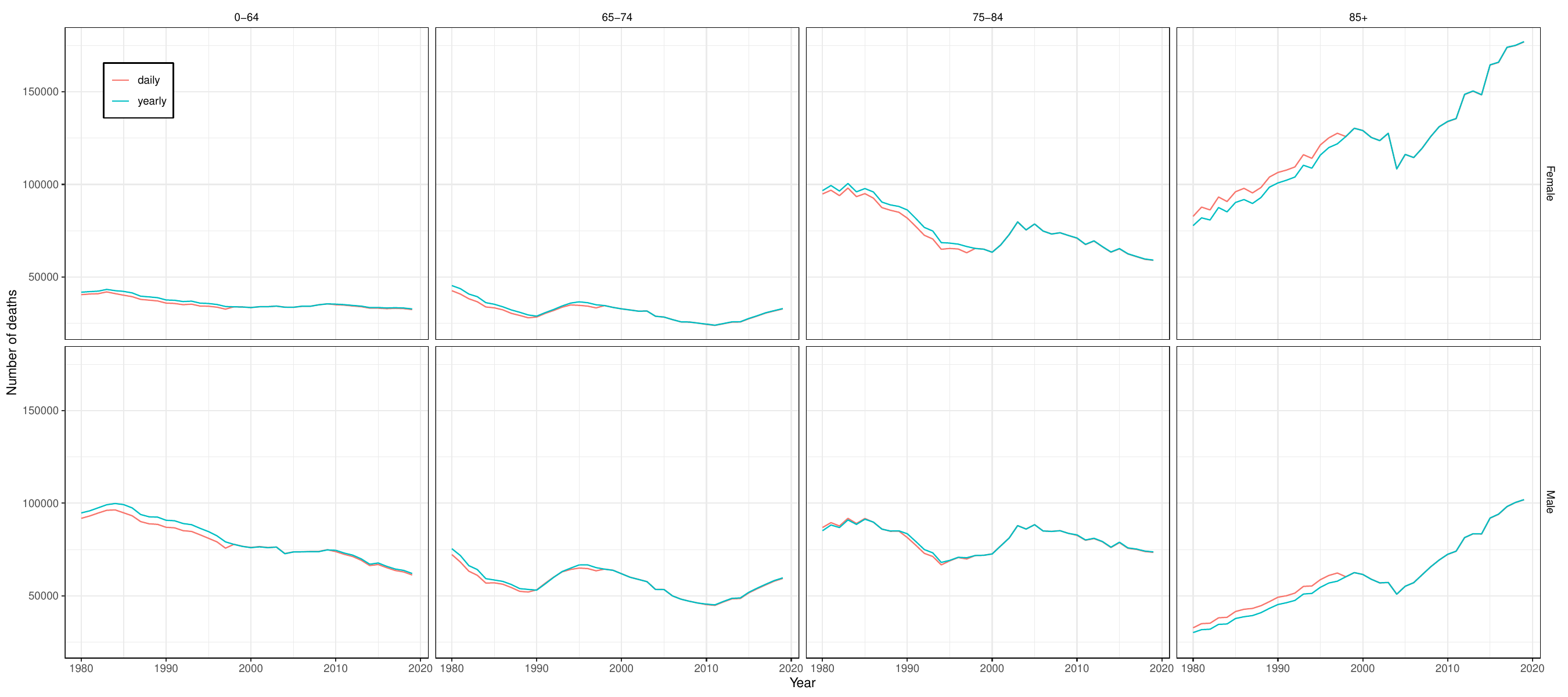}
\caption{
Comparison of annual death counts from daily data \citep{insee_data} and annual data \citep{human_mortality_database_university_2024} for both women and men with age groups 0-64, 65-74, 75-84, and 85+. For daily data, deaths are aggregated annually. Observation period: 1980-2019, based on INSEE and HMD data.
}
\label{fig_compar_data_source}
\end{figure}

In this appendix, we reconcile monthly and annual mortality data. These data originate from different sources and we compare in Figure~\ref{fig_compar_data_source} annual discrepancies in death counts by age group and sex. For the period from 1980 to 1998, we observe slight differences between the two data sources. Beyond this period, the two sources provide equivalent death counts.

Regarding the DLNM model, these minor gaps are not expected to influence the estimation conducted in Section~\ref{subsec:estim_dlnm}, as the differences are not specifically causes by temperature or seasonal mortality variations. Given the small magnitude of the observed differences, it is reasonable to proportionally adjust the daily temperature-attributable death counts in Equation~\eqref{eq:pred_death_temp} using the annual death counts provided by the HMD database. This adjustment does not affect the estimation of the total attributable fraction given in Equation~\eqref{eq:estim_attrib_factor}, which is used to adjust the exposure to risk for the Li-Lee model estimation in Section~\ref{subsec:baseline_estimation}. For this estimation, we choose to rely on the annual HMD data rather than the aggregated daily data, since the former is publicly accessible which facilitates the reproducibility of our results.

\section{The DLNM model}
\label{sec:app_dlnm_model}

\subsection{Visual examination of predicted death counts}
\label{subsec:app_fit_dlnm_desc_analysis}

\begin{figure}[h!] 
\centering
\includegraphics[scale=0.35]{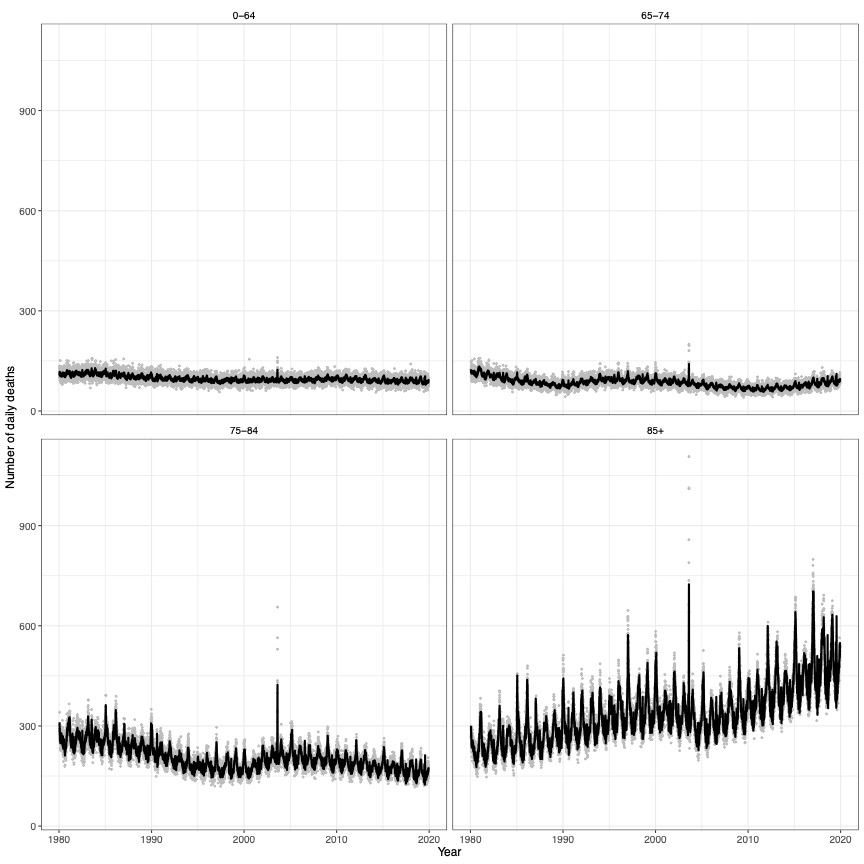}
\caption{Comparison of observed daily deaths (grey points) and predicted deaths (black line) for women. Observation period: 1980-2019, based on INSEE data}
\label{fig_daily_death_f}
\end{figure}

\begin{figure}[h!]
\centering
\includegraphics[scale=0.35]{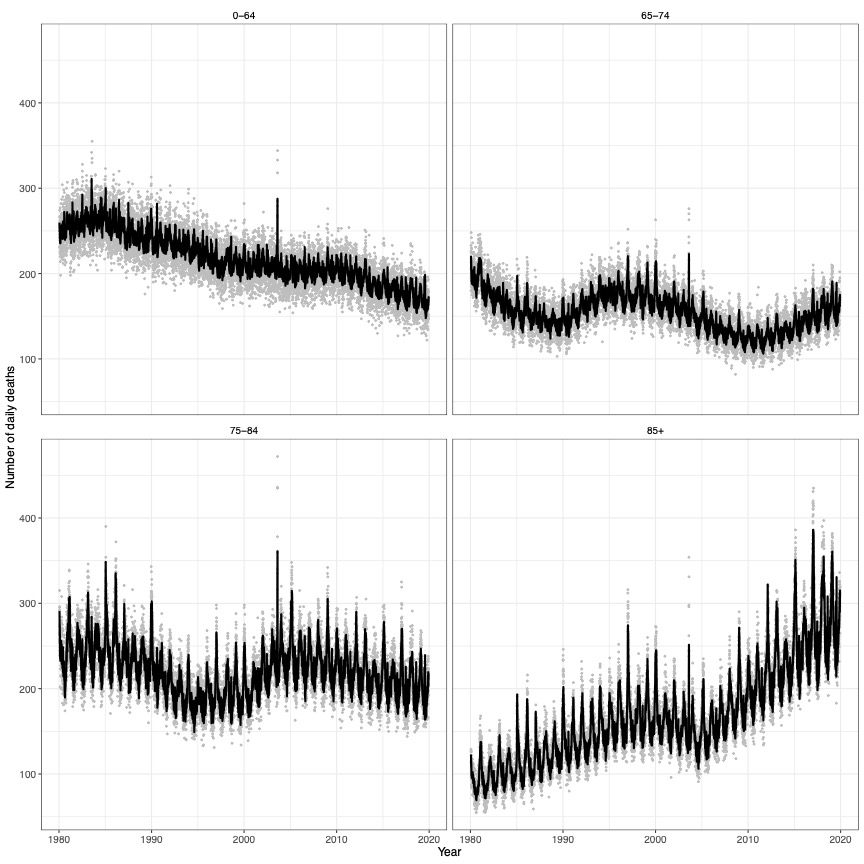}
\caption{Comparison of observed daily deaths (grey points) and predicted deaths (black line) for men. Observation period: 1980-2019, based on INSEE data}
\label{fig_daily_death_m}
\end{figure}

In this appendix, we present the predicted all-cause death counts obtained from the DLNM model, as given by Equation~\eqref{eq:pred_all_cause_death}, and visually assess its ability to reproduce the seasonality of daily deaths across age groups and genders. Figures~\ref{fig_daily_death_f} and~\ref{fig_daily_death_m} respectively display the observed daily deaths (grey points) and those predicted by the DLNM model (black line) for women and men by age group. The results show the model's ability to reproduce the daily death patterns across all-cause mortality for each age group. Notably, the model effectively captures seasonality, which intensifies with age, and reproduces long-term trends driven primarily by changes in population age structure and aging.

\begin{figure}[h!]
\centering
\includegraphics[scale=0.35]{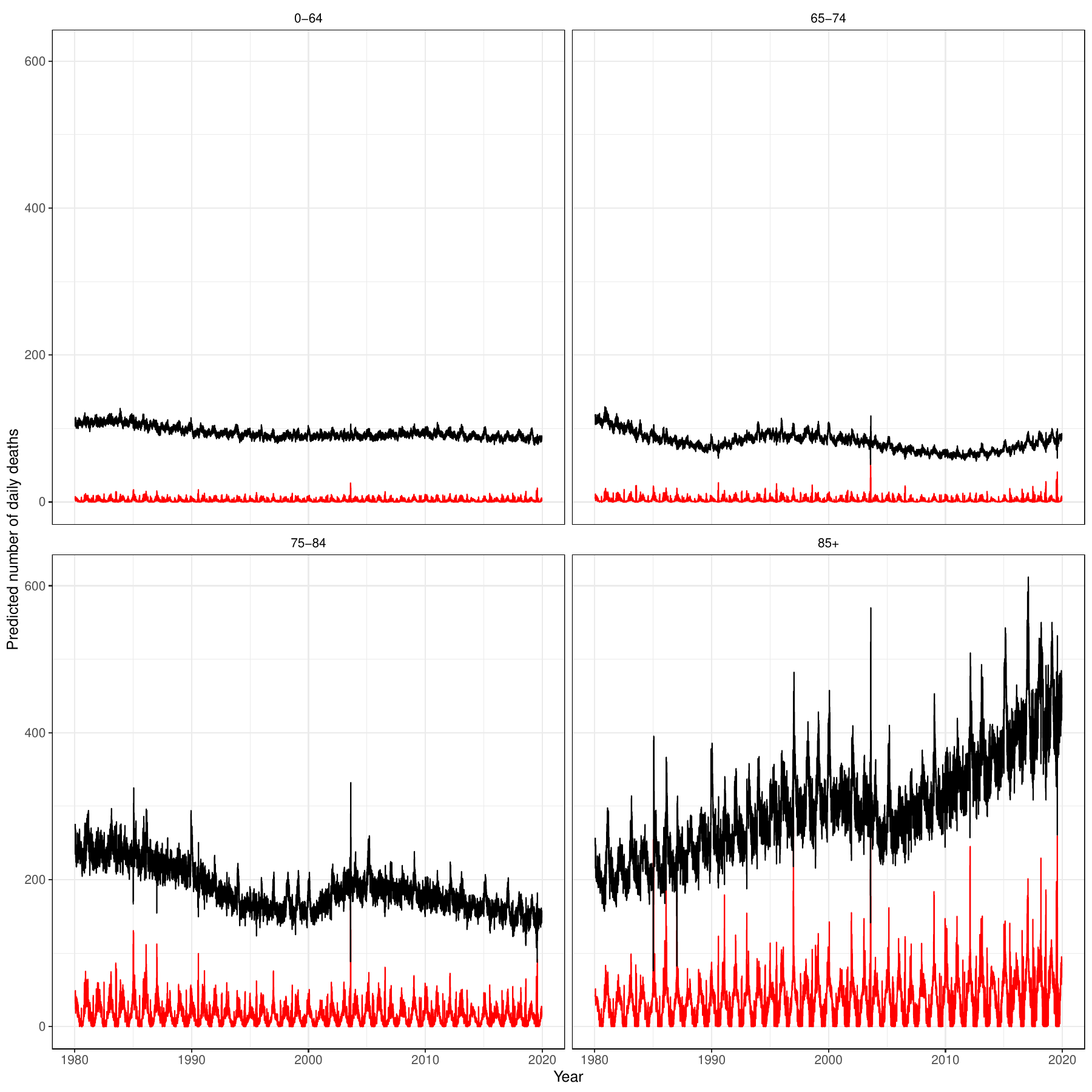}
\caption{
Decomposition of daily death counts for women predicted by the DLNM model into those attributable to temperatures (in red) and those not attributable to temperatures (in black).
Observation period: 1980-2019, based on INSEE data 
}
\label{fig_decomp_daily_death_f}
\end{figure}

\begin{figure}[h!]
\centering
\includegraphics[scale=0.35]{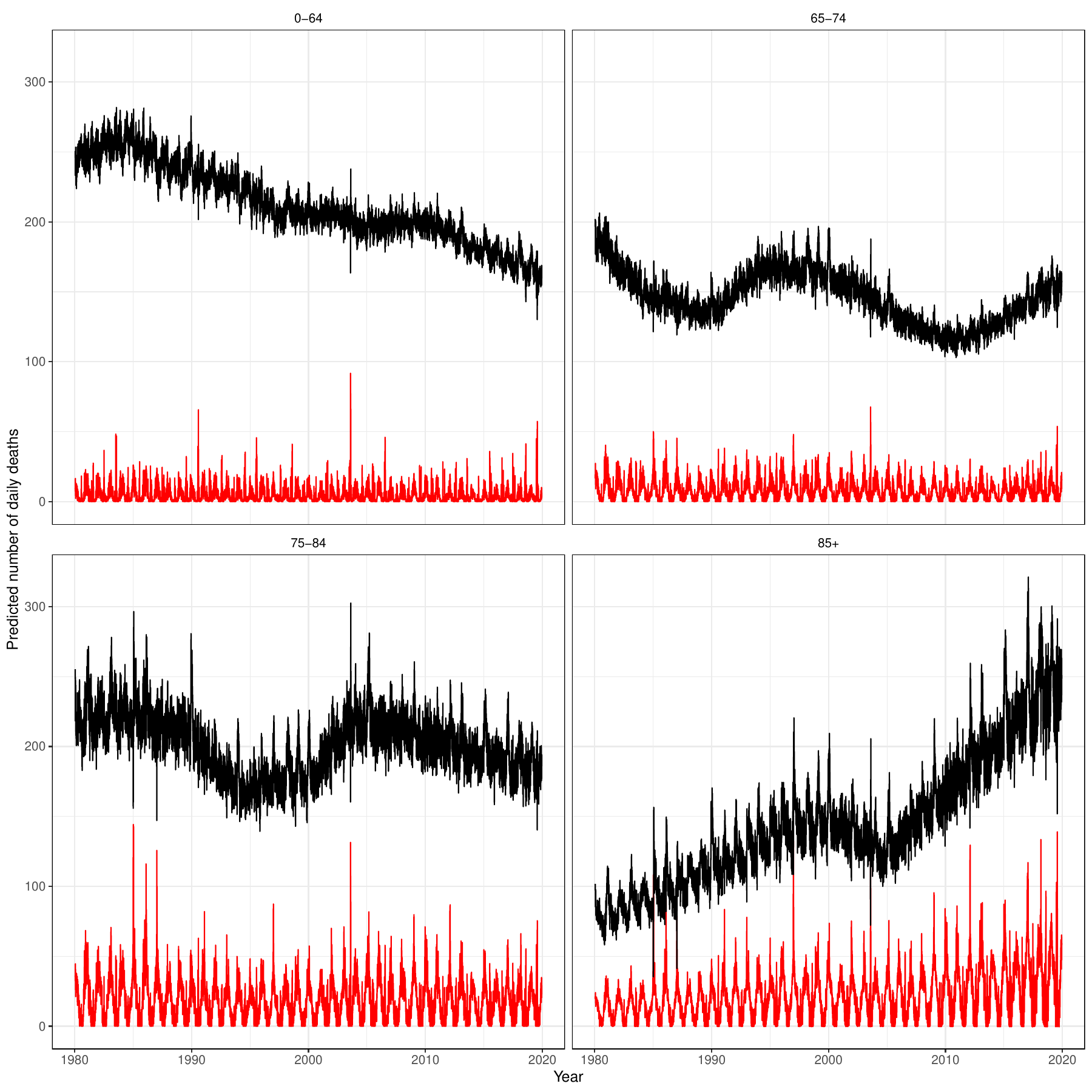}
\caption{
Decomposition of daily death counts for men predicted by the DLNM model into those attributable to temperatures (in red) and those not attributable to temperatures (in black).
Observation period: 1980-2019, based on INSEE data. 
}
\label{fig_decomp_daily_death_m}
\end{figure}

\begin{figure}[h!]
\centering
\includegraphics[scale=0.35]{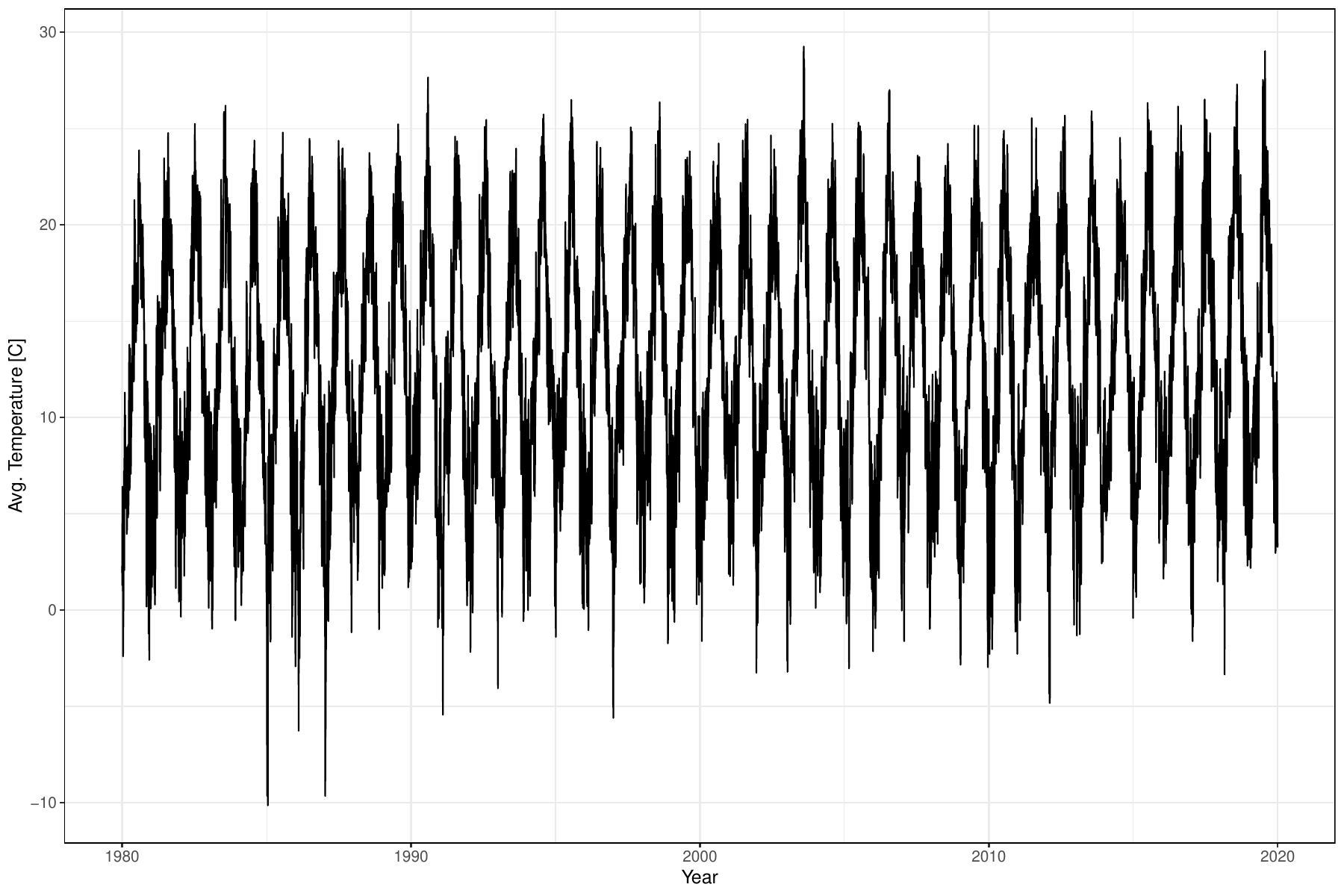}
\caption{
Evolution of daily average temperatures. Daily average temperatures are calculated for each city, and then an average of 14 cities is used to derive the daily average temperatures for Metropolitan France. Temperatures are measured in degrees Celsius. Observation period: 1980-2019, based on GHCN data.
}
\label{fig_pattern_temperature}
\end{figure}

Figure~\ref{fig_decomp_daily_death_f} decomposes daily death counts for women into temperature-attributable deaths, as given by Equation~\eqref{eq:pred_death_temp}, and non-temperature-attributable deaths, obtained by difference with the all-cause predicted death counts in Equation~\eqref{eq:pred_all_cause_death}. These visualizations align with our model specifications in Equation~\eqref{eq:dlnm_myspec}, where temperature-attributable deaths exhibit seasonal dynamics linked to temperature patterns (Figure~\ref{fig_pattern_temperature}), and non-temperature-attributable deaths display residual seasonality and distinct trends. Consistent with Figure~\ref{fig_daily_death_f}, the seasonal effect attributable to temperatures becomes more pronounced with age. Figure~\ref{fig_decomp_daily_death_m} illustrates a similar decomposition for the male population.

\clearpage

\subsection{Goodness of fit analysis and model checking}
\label{subsec:app_fit_dlnm_model}

In this appendix, we conduct a statistical analysis to evaluate the goodness of fit of the DLNM model. Since, the attributable and non-attributable death counts related to temperature are unobservable, they cannot be used to evaluate the model's performance. Therefore, we compare the observed daily death counts  $D_{k,t,d}^{(g)}$ with those predicted by the DLNM model. To ensure that the model accurately replicates both the seasonality of daily deaths and long-term trends, we examine the monthly distribution of deaths over the four decades comprising the period 1980–2019.

\begin{figure}[h!]
\centering
\includegraphics[scale = 0.4]{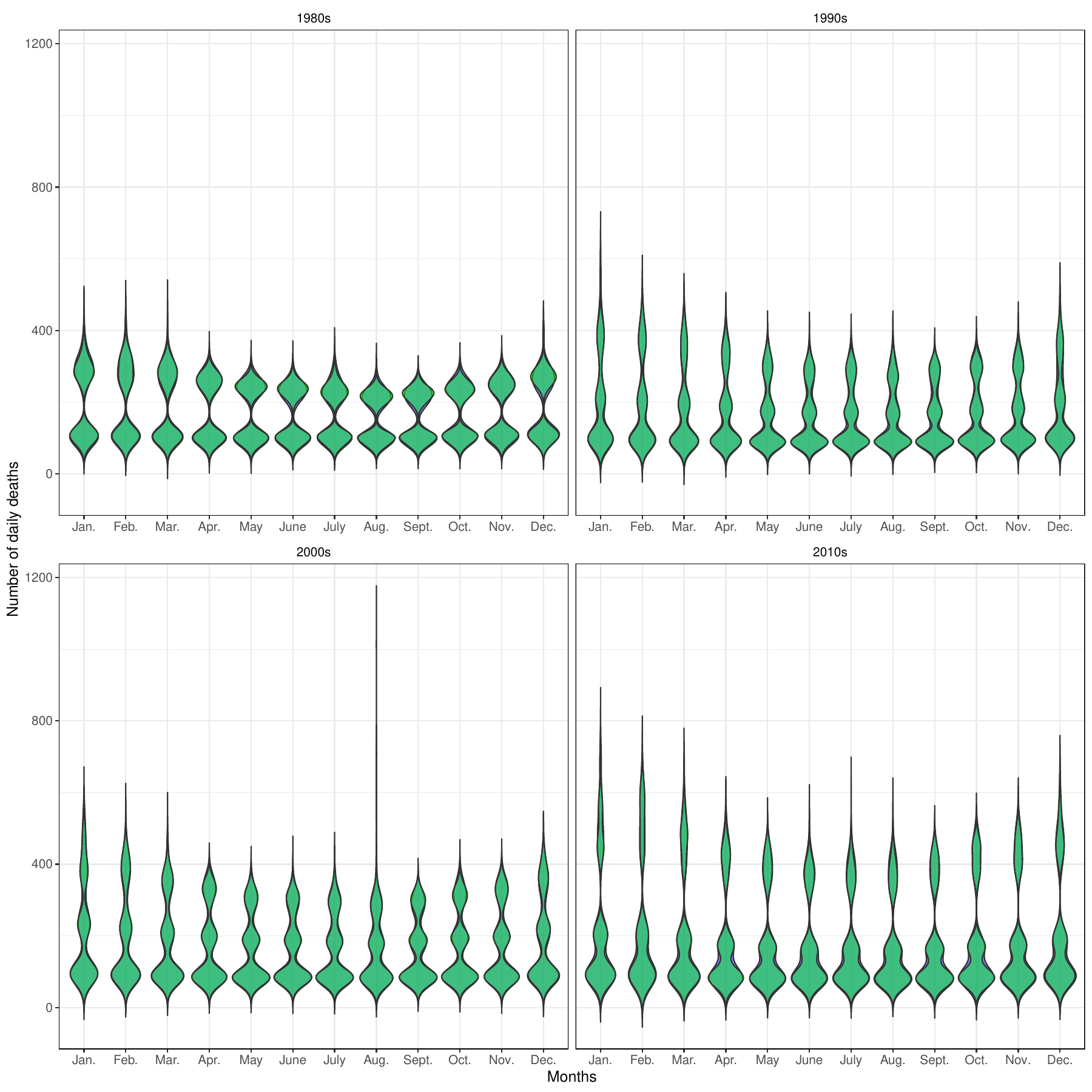}
\caption{Monthly distribution of observed (blue) and predicted (green) numbers of deaths based on the DLNM model~\eqref{eq:dlnm_myspec} per year for women in Metropolitan France for the years between 1980 and 2019. The distributions are grouped by decade.}
\label{fig_dlnm_obs_pred_f}
\end{figure}

\begin{figure}[h!]
\centering
\includegraphics[scale = 0.4]{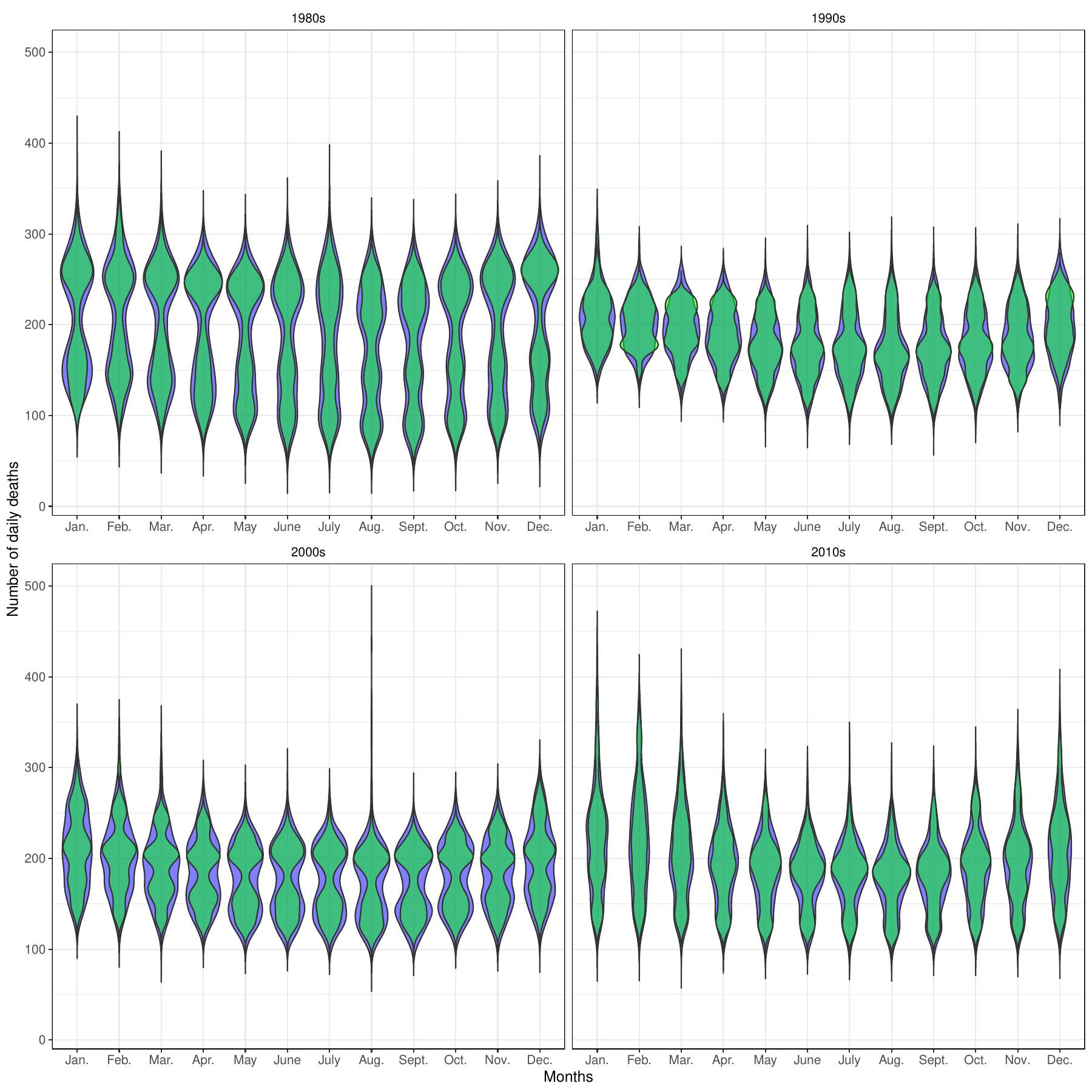}
\caption{
Monthly distribution of observed (blue) and predicted (green) numbers of deaths based on the DLNM model~\eqref{eq:dlnm_myspec} per year for men in Metropolitan France for the years between 1980 and 2019. The distributions are grouped by decade.}
\label{fig_dlnm_obs_pred_m}
\end{figure}

Figures~\ref{fig_dlnm_obs_pred_f} and~\ref{fig_dlnm_obs_pred_m} compare the monthly distributions of observed and modeled death counts over the observation period for females and males. We observe a close similarity in the monthly distributions. For females, the model replicates the distribution of deaths accurately, except for some extreme values. For males, the model fits less well in the central part of the distribution, but it still provides a good overall fit, especially in the last decade.

\begin{figure}[h!]
\centering
\includegraphics[scale = 0.7]{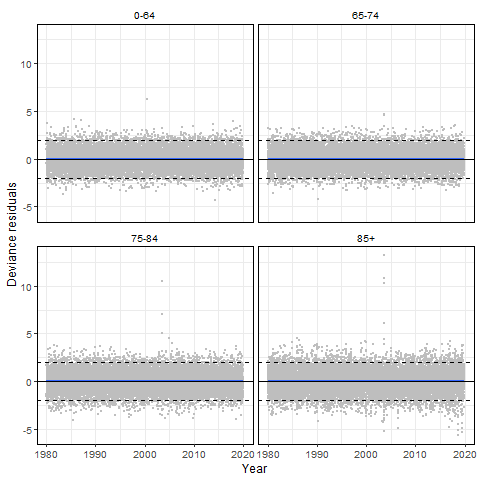}
\caption{Representation of deviance residuals for DLNM models associated with age groups 0-64, 65-74, 75-84, and 85+ for women in Metropolitan France from 1980 to 2019. The blue curve represents the smoothed residuals, while the dashed lines indicate the +2 and -2 thresholds.}
\label{fig_dlnm_resid_f}
\end{figure}

\begin{figure}[h!]
\centering
\includegraphics[scale = 0.7]{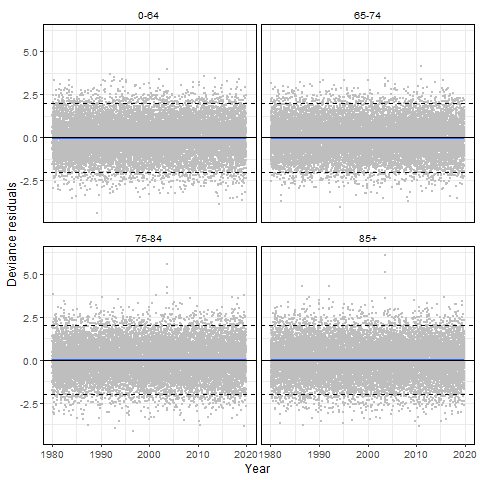}
\caption{
Representation of deviance residuals for DLNM models associated with age groups 0-64, 65-74, 75-84, and 85+ for men in Metropolitan France from 1980 to 2019. The blue curve represents the smoothed residuals, while the dashed lines indicate the +2 and -2 thresholds.}
\label{fig_dlnm_resid_m}
\end{figure}

As suggested by \citet{bhaskaran_time_2013}, we examine the deviance residuals to check for any residual trends not captured by the DLNM model. Figures~\ref{fig_dlnm_resid_f} and~\ref{fig_dlnm_resid_m} present the analysis of deviance residuals by age group for women and men. Since seasonality and long-term trends are controlled in Equation~\eqref{eq:dlnm_myspec}, the results indicate a relatively good fit to the data for both women and men, as well as the absence of observable trends in deviance residuals. It is worth noting a few extreme points that the model did not capture well, such as the 2003 heatwave for the 85+ age group, which was an event of extreme magnitude. Extreme variations in the number of deaths could potentially be explained by other factors not included in our model, such as air pollution, flu epidemics, etc.

\clearpage

\subsection{Sensitivity analysis}
\label{subsec:app_dlnm_sensi}

As explained in Section~\ref{subsec:excess_mortality}, the proportion of deaths attributable to temperatures is not directly observable but is instead predicted by a model. Following the recommendations of \citet{bhaskaran_time_2013}, we compare in this appendix various specifications of the DLNM model to assess the robustness of our estimates.

\subsubsection{Degrees of freedom and knots hyperparameters}
\label{subsec:app_dlnm_hyper_knots}

First, we compare different specifications of the DLNM model and assess the model's goodness of fit across varying hyperparameters used in Equation~\eqref{eq:dlnm_myspec}. There is no definitive consensus in the empirical literature regarding these choices. For instance, \citet{bhaskaran_time_2013} recommend using 7 knots per year for the natural cubic B-spline in the lag-response curve $w(\cdot)$. They strikes that this is "a balance between providing adequate control for seasonality and other confounding by trends in time, while leaving sufficient information from which to estimate exposure effects." Other studies, such as \citet{gasparrinietal15, gasparrini_projections_2017, vicedo-cabrera_multi-country_2018, lee_projections_2020, martinez-solanas_projections_2021, ballester_heat-related_2023}, use eight knots, while \citet{lee_mortality_2018} employ seven and nine knots. For modeling warm periods only, \citet{guo_quantifying_2018,vicedo-cabrera_burden_2021} use four knots. We thus investigate the robustness of the choices made in Section~\ref{sec:estimation}.

We examine the sensitivity of the DLNM model to the number of internal knots for the natural cubic B-spline in the lag-response curve $w(\cdot)$, as well as to the degrees of freedom for $h_p(\cdot)$ which controls the residual seasonality. We have also tested the impact of the placement of internal knots for the natural cubic B-spline of the exposure–response curve $f(\cdot)$. Different specifications were considered, such as three internal knots placed at the 25th, 50th, and 75th percentiles of the daily average temperature distribution, or four internal knots placed at the 20th, 40th, 60th, and 80th percentiles. Although these specifications can sometimes slightly improve model fit, we ultimately retain the initial configuration with three internal knots placed at the 10th, 75th, and 90th percentiles of the daily average temperature distribution. Indeed, this choice minimizes overfitting and facilitates the extrapolation of results to extreme temperatures. Therefore, we only present the results of the sensitivity analysis for the first two hyperparameters.

Here, we vary the degrees of freedom per year between 4 and 12 for $h_p(\cdot)$ functions, corresponding to one degree of freedom per quarter or per month. Additionally, we evaluate the impact of the number of internal knots between 1 and 10 used for the natural cubic B-spline in the lag-response curve $w(\cdot)$. The combination of these two hyperparameters leads us to compare 90 specifications of the DLNM model. Following the notation from Section~\ref{subsec:dlnm_mortality}, we compute the modified Akaike and Bayesian information criteria for GLMs with overdispersed responses, defined as follows for $k \in \{1, \ldots, K\}$ and $g \in \mathcal{G}$
\begin{align*}
& \text{QAIC}_k^{(g)} = - 2 \mathcal{L} \left( \widehat{\eta}_k^{(g)}, \widehat{\thetavec}_{k}^{(g)}, \widehat{\zetavec}_{k,1}^{(g)}, \widehat{\zetavec}_{k,y_{\min}}^{(g)}, \ldots, \widehat{\zetavec}_{k,y_{\max}}^{(g)}\right) + 2 \widehat{\varphi}_k^{(g)} \xi_k^{(g)},\\
& \text{QBIC}_k^{(g)} = - 2 \mathcal{L} \left( \widehat{\eta}_k^{(g)}, \widehat{\thetavec}_{k}^{(g)}, \widehat{\zetavec}_{k,1}^{(g)}, \widehat{\zetavec}_{k,y_{\min}}^{(g)}, \ldots, \widehat{\zetavec}_{k,y_{\max}}^{(g)} \right) + \ln\left(n_k^{(g)}\right) \widehat{\varphi}_k^{(g)} \xi_k^{(g)},
\end{align*}
where $\xi_k^{(g)}$ is the number of parameters, $n_k^{(g)}$ is the number of observations, $\mathcal{L}$ represents the log-likelihood of the DLNM fitted with parameters $\widehat{\eta}_k^{(g)}, \widehat{\thetavec}_{k}^{(g)}, \widehat{\zetavec}_{k,1}^{(g)}, \widehat{\zetavec}_{k,y_{\min}}^{(g)} \ldots, \widehat{\zetavec}_{k,y_{\max}}^{(g)}$, and $\widehat{\varphi}_k^{(g)}$ is the estimated dispersion parameter. The QAIC and QBIC values are then summed across each subgroup $k$ and for each sex $g$.

Table \ref{tab:sensi_dlnm} compares the QAIC and QBIC values of the ten best models, ranked by QBIC, to the model selected in Section \ref{subsec:calibrating_dlnm}. The model with 6 degrees of freedom and 2 knots achieves the lowest QBIC statistic. Our model ranks 8th, but its QBIC is close to those of the top models. However, it outperforms them according to the QAIC statistic. 

\begin{table}[!h]
\centering\centering
\fontsize{8}{10}\selectfont
\begin{tabular}[t]{ccccc}
\toprule
Model & Number of knots (lag-response) & Seanonality degree & QAIC & QBIC\\
\midrule
Model 1 & 2 & 6 & 952354.8 & 971552.8\\
Model 2 & 2 & 5 & 954926.7 & 971594.7\\
Model 3 & 3 & 6 & 952213.0 & 971743.6\\
Model 4 & 2 & 8 & 947781.3 & 971793.2\\
Model 5 & 3 & 5 & 954789.9 & 971802.5\\
Model 6 & 4 & 6 & 951960.9 & 971807.6\\
Model 7 & 4 & 5 & 954542.9 & 971885.8\\
\textbf{Model 8} & \textbf{3} & \textbf{8} & \textbf{947653.4} & \textbf{971979.6}\\
Model 9 & 2 & 7 & 950322.1 & 972019.5\\
Model 10 & 4 & 8 & 947410.4 & 972028.9\\
\bottomrule
\end{tabular}
\caption{Comparison of QAIC and QBIC statistics for the 10 best DLNM models estimated over the period 1980-2019, varying the number of knots for the lag-response curve and the degrees of freedom associated with residual seasonality. These top ten models are ranked according to the QBIC statistic. The results of the model used in Section \ref{subsec:calibrating_dlnm} are presented in bold.}
\label{tab:sensi_dlnm}
\end{table}

\begin{figure}[h!]
\centering
\includegraphics[scale = 0.5]{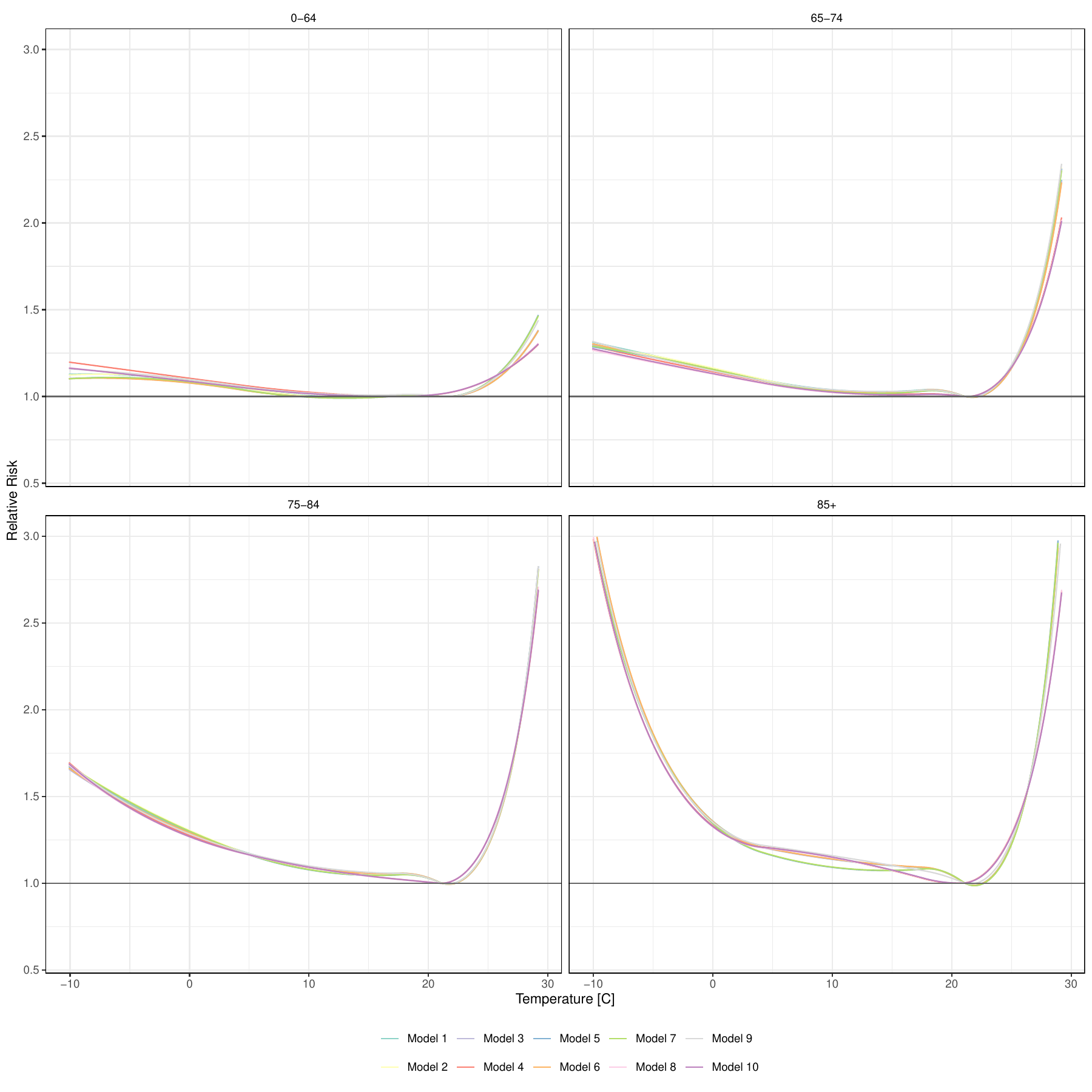}
\caption{
Cumulative relative risk of mortality over a 21-day period in Metropolitan France calculated for the years 1980-2019 for women across age groups 0-64, 65-74, 75-84, and 85+ for the ten best DLNM models. Daily average temperatures are calculated for each city, and then an average of 14 cities is used to derive the daily average temperatures for Metropolitan France. Temperatures are measured in degrees Celsius. Based on GHCN and INSEE data.
}
\label{fig_sensi_rr_f}
\end{figure}

\begin{figure}[h!]
\centering
\includegraphics[scale = 0.5]{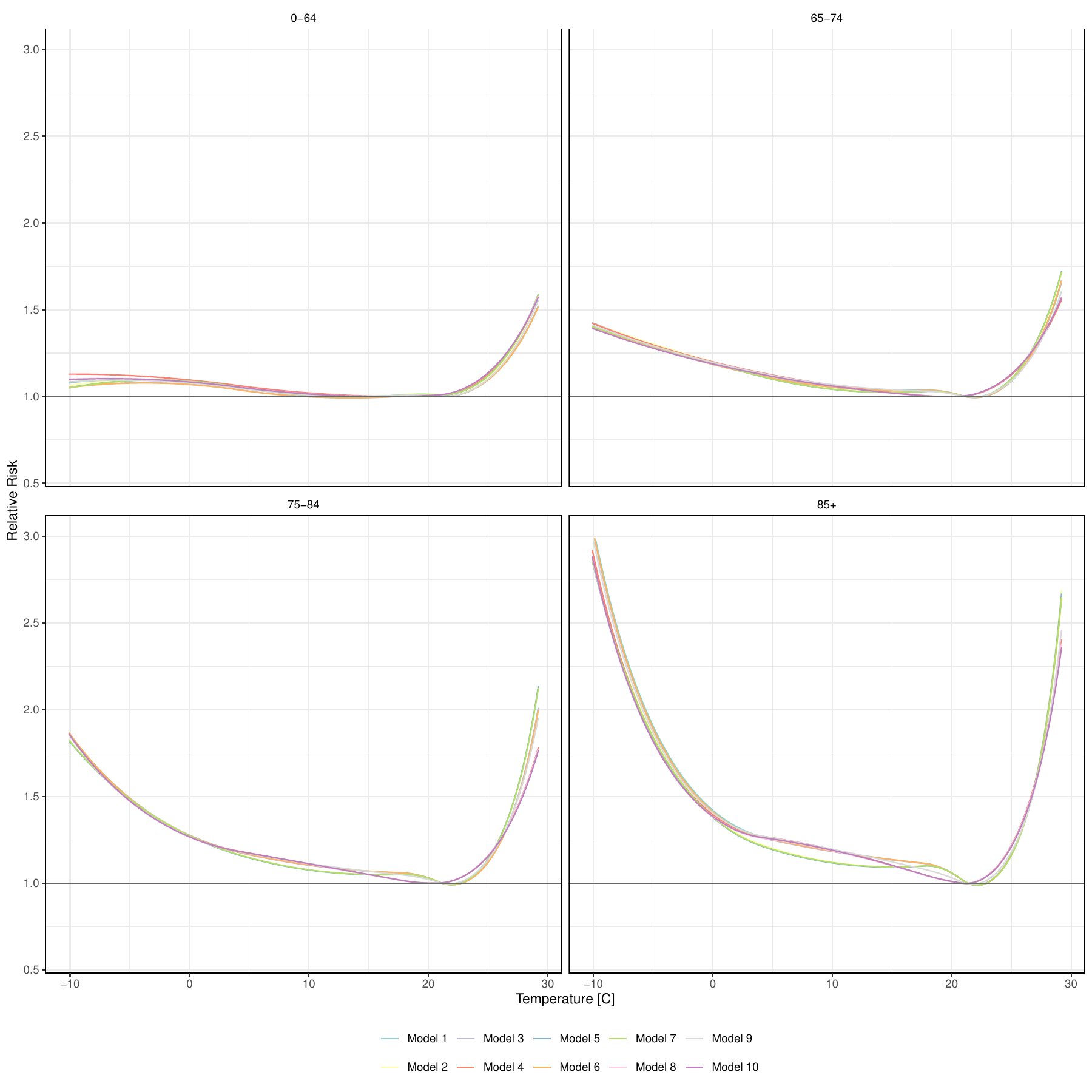}
\caption{
Cumulative relative risk of mortality over a 21-day period in Metropolitan France calculated for the years 1980-2019 for men across age groups 0-64, 65-74, 75-84, and 85+ for the ten best DLNM models. Daily average temperatures are calculated for each city, and then an average of 14 cities is used to derive the daily average temperatures for Metropolitan France. Temperatures are measured in degrees Celsius. Based on GHCN and INSEE data.
}
\label{fig_sensi_rr_m}
\end{figure}

Finally, we examine the sensitivity of the curve representing the cumulative temperature-mortality association estimated for each of these 10 DLNM models. Figures~\ref{fig_sensi_rr_f} and~\ref{fig_sensi_rr_m} respectively compare the cumulative relative risk of mortality over a 21-day period for women and men. These figures demonstrate the robustness of our model (Model 8) to variations in the hyperparameters listed in Table~\ref{tab:sensi_dlnm}.

\clearpage

\subsubsection{Lag hyperparameter}
\label{subsec:app_dlnm_sensi_lag}

The choice of lag for the function $s(\cdot,\cdot)$ in Equation~\eqref{eq:bidim_sline} influences the short-term seasonality attributable to temperature. In studies focusing exclusively on the effects of heatwaves \citep{guo_quantifying_2018,vicedo-cabrera_burden_2021}, the impact of temperature on mortality typically dissipates after approximately ten days. However, when accounting for the effects of cold, a longer lag period is necessary. For instance, \citet{gasparrini_projections_2017} use 21 days. If the lag is set too long, the $s(\cdot,\cdot)$ component of the DLNM model \eqref{eq:dlnm_myspec} might inadvertently capture seasonal mortality effects not related to temperature. To remain consistent with the literature, we limit the lag to 21 days and we analyze in this appendix the impact of shorter lags on the cumulative effects of the temperature-mortality association.

\begin{figure}[h!]
\centering
\includegraphics[scale = 0.5]{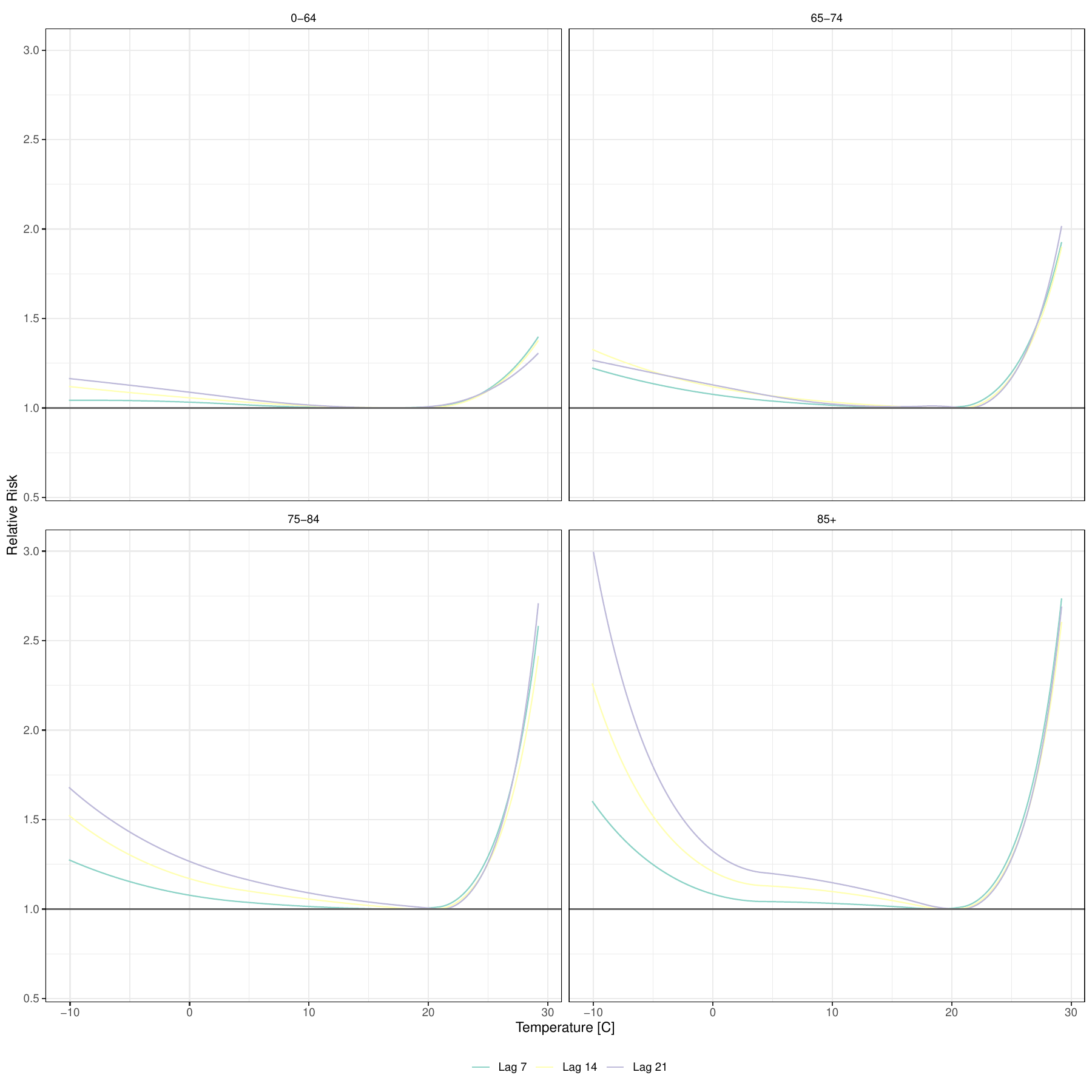}
\caption{
Cumulative relative risk of mortality over a 7, a 14 or 21-day period in Metropolitan France calculated for the years 1980-2019 for women across age groups 0-64, 65-74, 75-84, and 85+. Daily average temperatures are calculated for each city, and then an average of 14 cities is used to derive the daily average temperatures for Metropolitan France. Temperatures are measured in degrees Celsius. Based on GHCN and INSEE data.
}
\label{fig_sensi_lag_rr_f}
\end{figure}

\begin{figure}[h!]
\centering
\includegraphics[scale = 0.5]{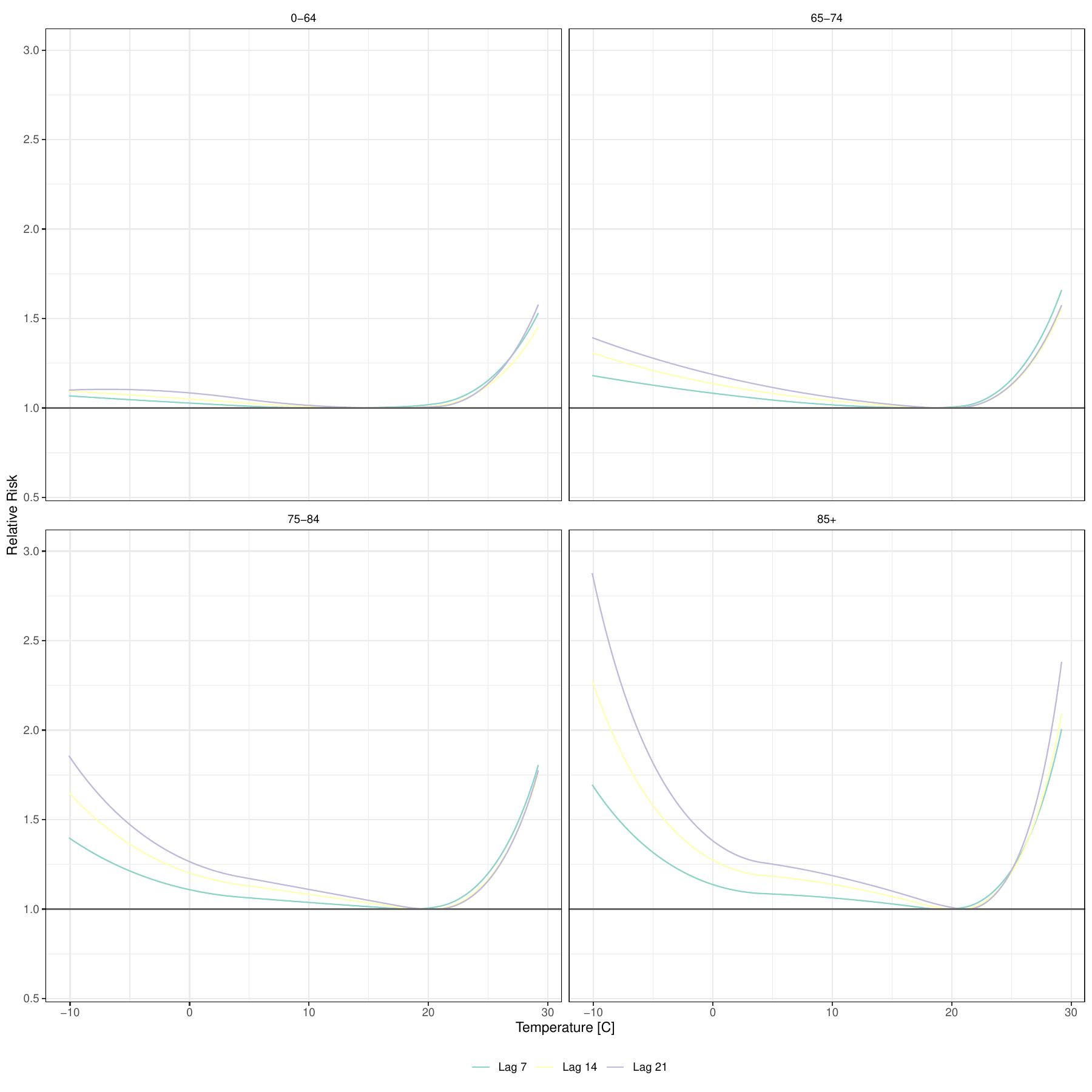}
\caption{
Cumulative relative risk of mortality over a 7, a 14 or 21-day period in Metropolitan France calculated for the years 1980-2019 for men across age groups 0-64, 65-74, 75-84, and 85+. Daily average temperatures are calculated for each city, and then an average of 14 cities is used to derive the daily average temperatures for Metropolitan France. Temperatures are measured in degrees Celsius. Based on GHCN and INSEE data.
}
\label{fig_sensi_lag_rr_m}
\end{figure}

To illustrate the sensitivity to the number of lagged days, Figures~\ref{fig_sensi_lag_rr_f} and~\ref{fig_sensi_lag_rr_m} display the response curves in terms of cumulative relative risk (RR) for women and men, respectively, using lags of 7, 14, and 21 days. Differences in lag selection primarily affect the excess mortality attributable to cold, while the impact on heat-related excess mortality is less pronounced. Specifically, the RR of mortality for 7- and 14-day periods fails to capture the longer-term effects of cold and to a lesser extent the delayed effects of heat, such as the harvesting effect associated with certain extreme heatwaves.

\clearpage

\section{Consistency of projected central death rates}
\label{subsec:app_consistency_rates}

In this appendix, we verify the consistency of the projections obtained from the original Li-Lee model~\eqref{eq:original_LL} and the Li-Lee model with adjusted exposures to risk~\eqref{eq:2_LL}. This verification is carried out for the different methods of calculating the temperature-attributable fraction used in this paper. Indeed, the adjustment of death rates can be performed in two ways:

\begin{enumerate}  
\item For historical data, the annual attributable fractions $\text{AF}_{x,t}^{(g)}$ are estimated using Equation~\eqref{eq:estim_attrib_factor}, which relies on daily deaths.  The central death rates are then obtained via Equation~\eqref{eq:multi_deaths_rates}.
\item During the projection phase, daily deaths are unknown. In this case, we use Equation~\eqref{eq:forecast_all2}, using both constant weights $\omega_{x,t,d}^{(g)}$ and weights calibrated on historical data.  
\end{enumerate}

\begin{figure}[h!]
\centering
\includegraphics[scale = 0.35]{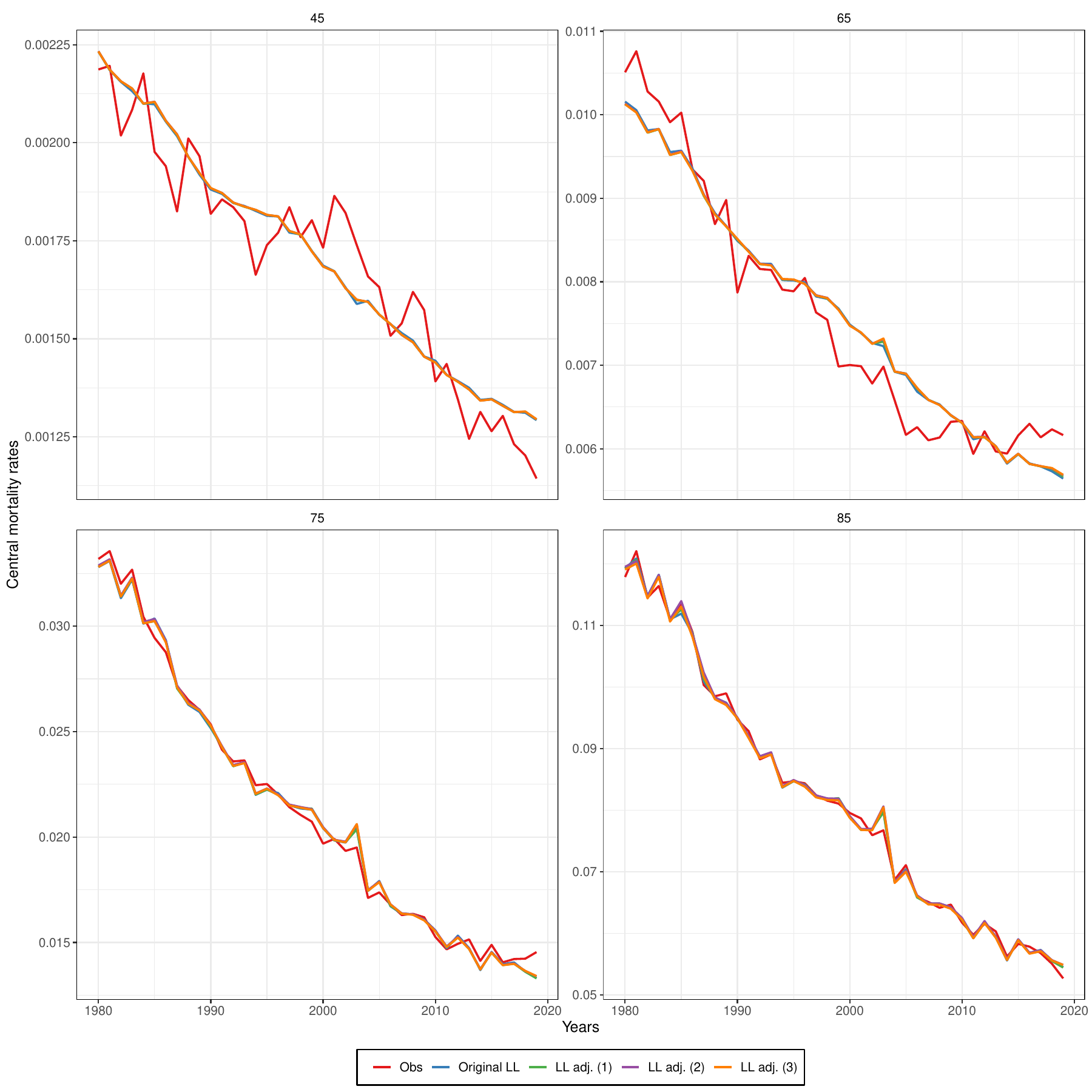}
\caption{
Comparison of observed and predicted central death rates at ages 45, 65, 75, and 85 for females over the observation period. The red curves represent the crude death rates. The blue curves correspond to the central death rates predicted by the original Li-Lee model. The green curves represent the central death rates predicted using Equation~\eqref{eq:attributable_fract}, after estimating the total attributable fractions via Equation~\eqref{eq:estim_attrib_factor}. The purple curves utilize Equation~\eqref{eq:forecast_all2} with constant weights, while the orange curves use weights estimated based on daily deaths unaffected by temperature from the last five years of our historical data.
}
\label{fig_check_inconsistency_f}
\end{figure}

\begin{figure}[h!]
\centering
\includegraphics[scale = 0.35]{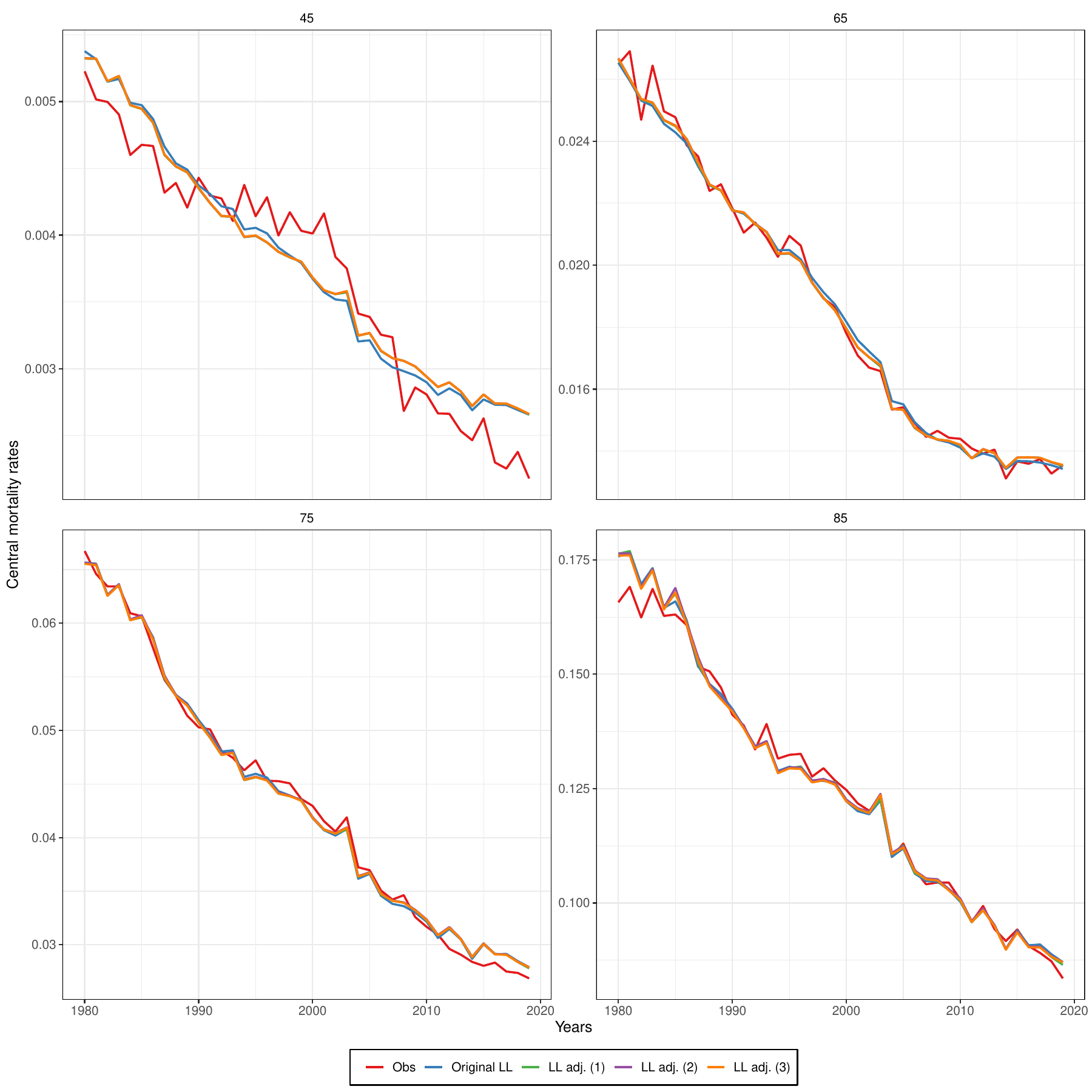}
\caption{
Comparison of observed and predicted central death rates at ages 45, 65, 75, and 85 for males over the observation period. The red curves represent the crude death rates. The blue curves correspond to the central death rates predicted by the original Li-Lee model. The green curves represent the central death rates predicted using Equation~\eqref{eq:attributable_fract}, after estimating the total attributable fractions via Equation~\eqref{eq:estim_attrib_factor}. The purple curves utilize Equation~\eqref{eq:forecast_all2} with constant weights, while the orange curves use weights estimated based on daily deaths unaffected by temperature from the last five years of our historical data.
}
\label{fig_check_inconsistency_m}
\end{figure}

Figures~\ref{fig_check_inconsistency_f} and~\ref{fig_check_inconsistency_m} illustrate the trajectories of the crude central death rates at ages 45, 65, 75, and 85 for females and males, comparing them with the central death rates predicted by the two approaches described above. We specifically distinguish two sub-cases when central death rates are calculated using Equation~\eqref{eq:forecast_all2}: one considering constant daily weights and the other using weights estimated based on daily deaths unaffected by temperature from the last five years of our historical data. First, we note that the death rates obtained from the original Li-Lee model and those predicted using the attributable fractions from Equation~\eqref{eq:estim_attrib_factor} are strictly equivalent. This is expected, as we apply Equation~\eqref{eq:multi_deaths_rates}, while the Li-Lee model is estimated based on adjusted exposure to risk. When death rates are derived from Equation~\eqref{eq:forecast_all2}, a slight bias may appear compared to predictions from the original Li-Lee model. Empirically, we observe however that this bias is small and decreases when non-constant weights $\omega_{x,t,d}^{(g)}$ are applied in Equation~\eqref{eq:forecast_all2}.

\clearpage

\section{Additional figures on projected temperature-attributable fractions}
\label{subsec:app_attrib_fraction}

\subsection{Attributable fraction per age bucket and sex}\label{subsec:app_attrib_fraction_sex_age}

\begin{figure}[h!]
\centering
\includegraphics[scale = 0.35]{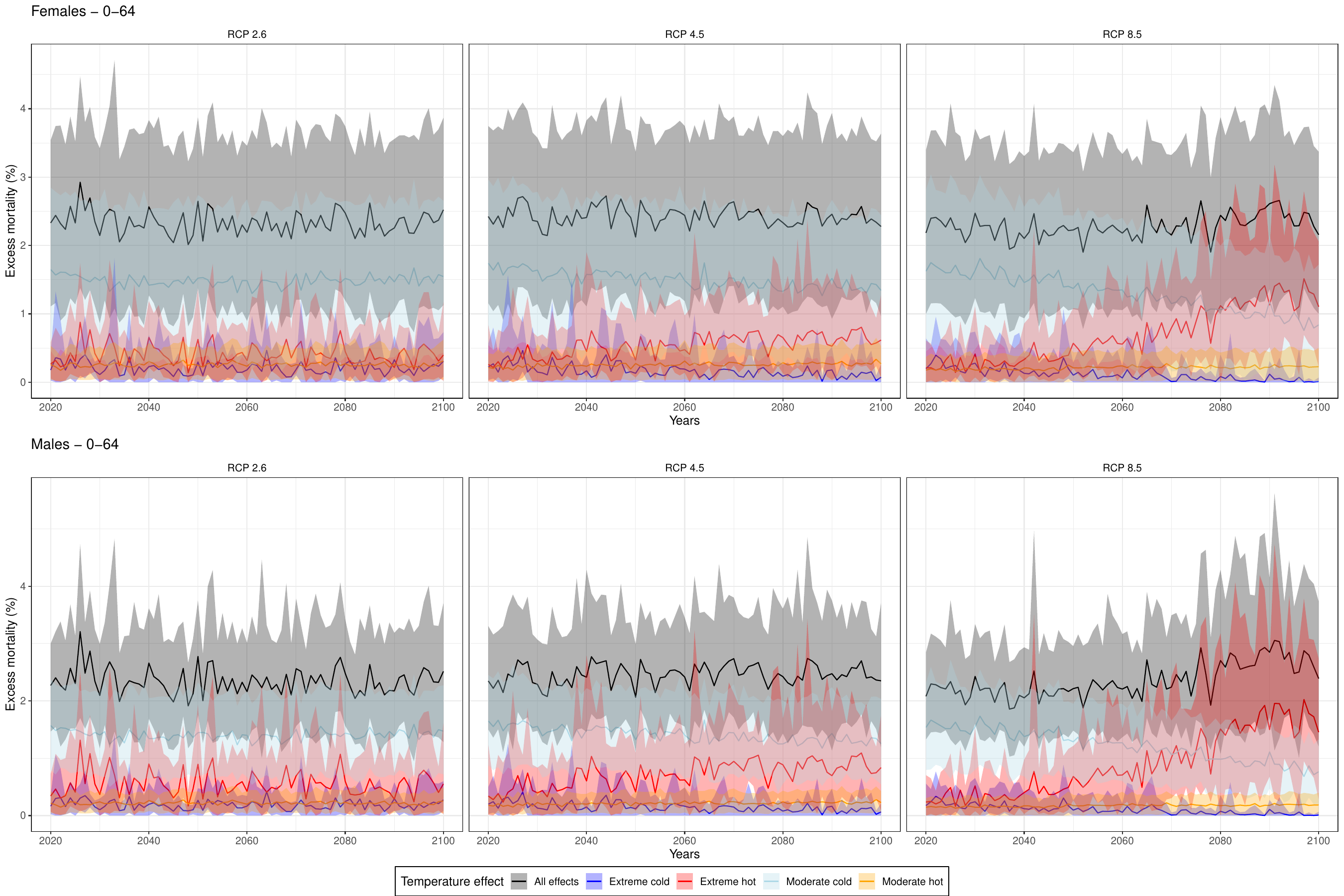}
\caption{
Temperature attributable fraction in Metropolitan France, simulated for the years 2020-2100 for both women and men across for the age group 0-64. These values, along with their $95\%$ confidence intervals (shaded areas) are calculated through 1,000 Monte Carlo simulations of the DLNM coefficients and the ensemble of climate models. Each line corresponds to the average of these simulations. Attributable fraction is expressed in \% for all effect, decomposed into moderate cold, moderate hot, extreme cold and extreme hot. 
}
\label{fig_attrib_0-64}
\end{figure}

\begin{figure}[h!]
\centering
\includegraphics[scale = 0.35]{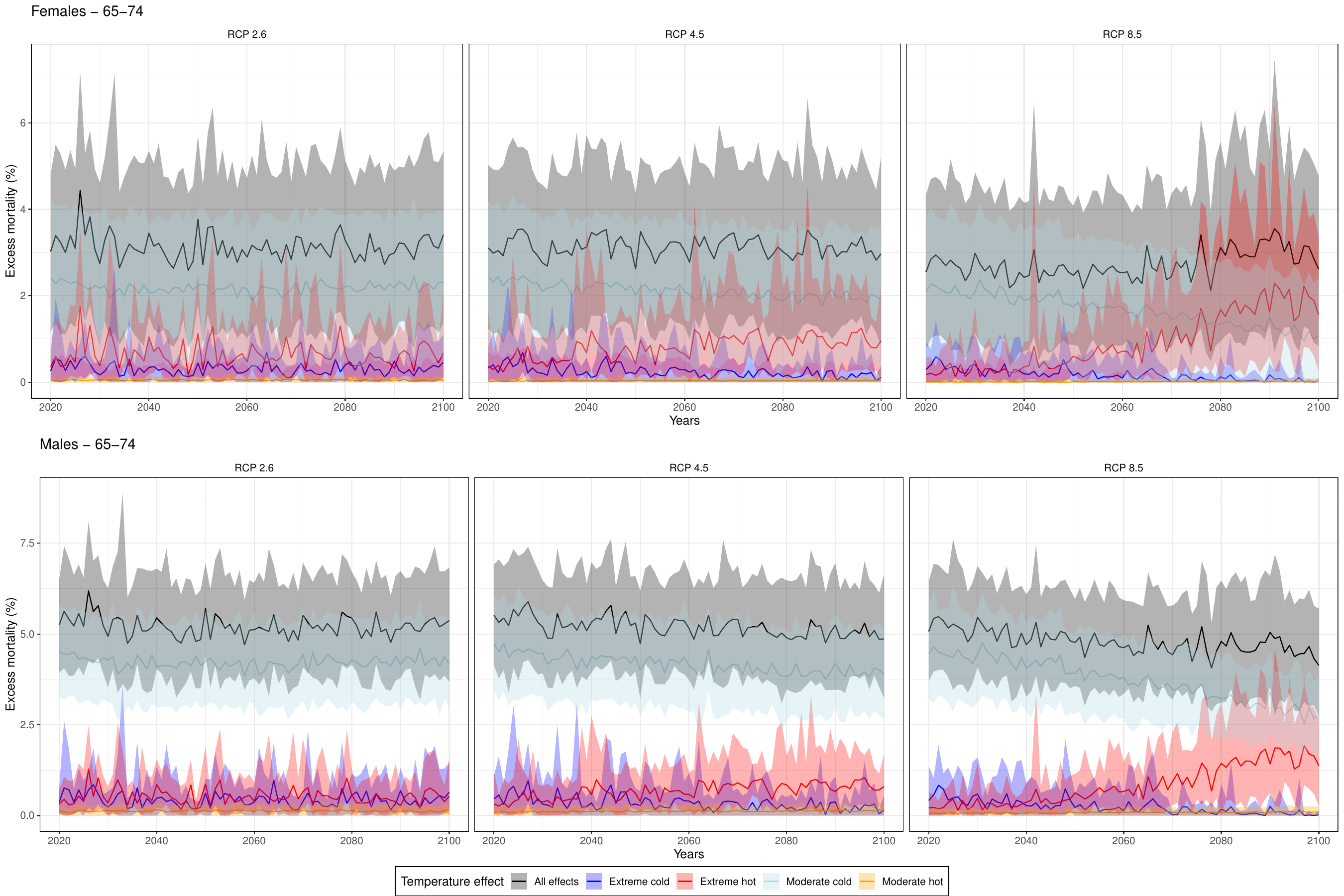}
\caption{
Temperature attributable fraction in Metropolitan France, simulated for the years 2020-2100 for both women and men across for the age group 65-74. These values, along with their $95\%$ confidence intervals (shaded areas) are calculated through 1,000 Monte Carlo simulations of the DLNM coefficients and the ensemble of climate models. Each line corresponds to the average of these simulations. Attributable fraction is expressed in \% for all effect, decomposed into moderate cold, moderate hot, extreme cold and extreme hot. 
}
\label{fig_attrib_65-74}
\end{figure}

\begin{figure}[h!]
\centering
\includegraphics[scale = 0.35]{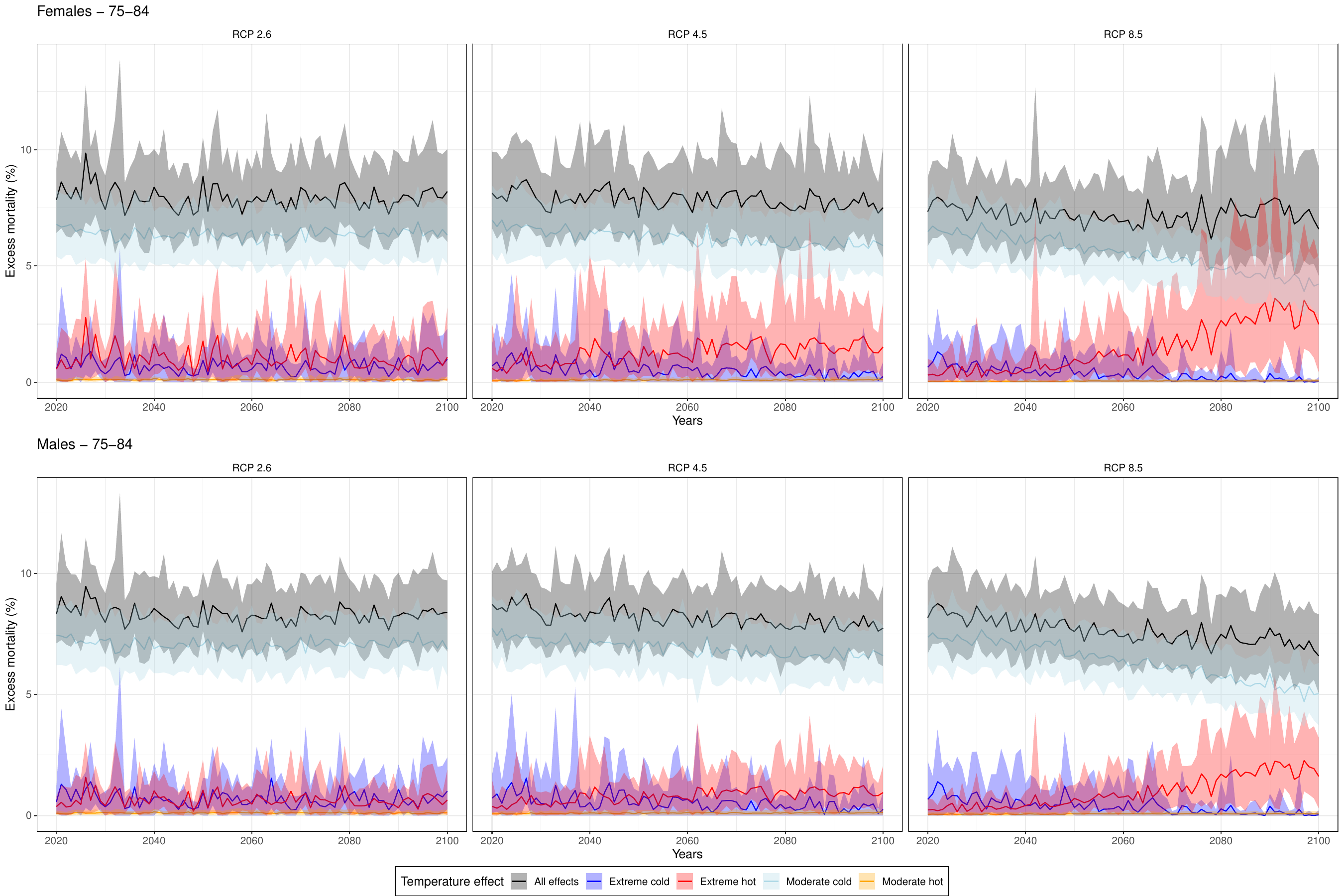}
\caption{
Temperature attributable fraction in Metropolitan France, simulated for the years 2020-2100 for both women and men across for the age group 75-84. These values, along with their $95\%$ confidence intervals (shaded areas) are calculated through 1,000 Monte Carlo simulations of the DLNM coefficients and the ensemble of climate models. Each line corresponds to the average of these simulations. Attributable fraction is expressed in \% for all effect, decomposed into moderate cold, moderate hot, extreme cold and extreme hot. 
}
\label{fig_attrib_75-84}
\end{figure}

\begin{figure}[h!]
\centering
\includegraphics[scale = 0.35]{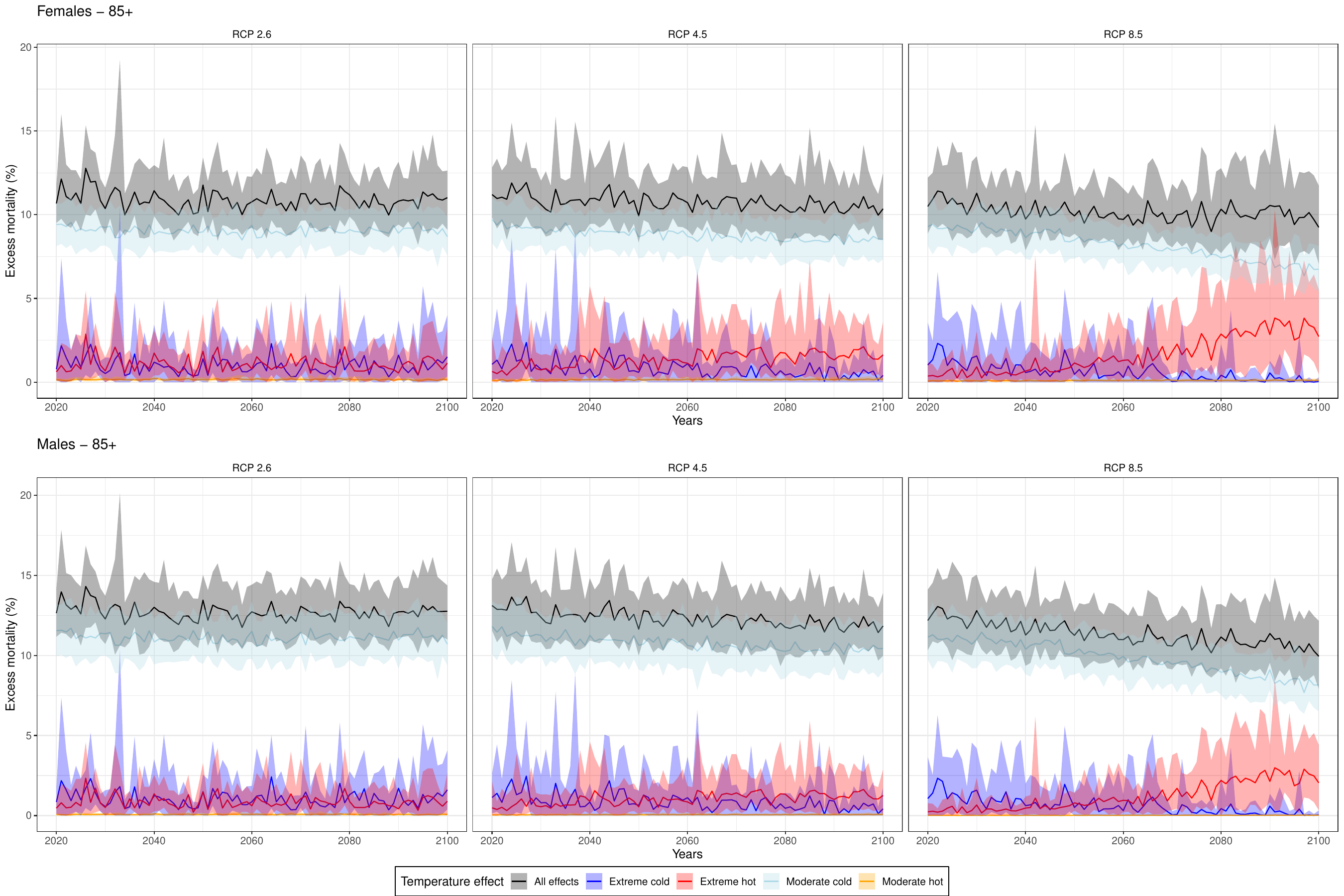}
\caption{
Temperature attributable fraction in Metropolitan France, simulated for the years 2020-2100 for both women and men across for the age group 85+. These values, along with their $95\%$ confidence intervals (shaded areas) are calculated through 1,000 Monte Carlo simulations of the DLNM coefficients and the ensemble of climate models. Each line corresponds to the average of these simulations. Attributable fraction is expressed in \% for all effect, decomposed into moderate cold, moderate hot, extreme cold and extreme hot. 
}
\label{fig_attrib_85+}
\end{figure}

Figures~\ref{fig_attrib_0-64},~\ref{fig_attrib_65-74},~\ref{fig_attrib_75-84} and~\ref{fig_attrib_85+} describe the projected temperature-attributable fractions simulated by sex and age group.

\clearpage

\subsection{Attributable fraction per city}\label{subsec:app_attrib_fraction_city}

\begin{figure}[h!]
\centering
\includegraphics[scale = 0.55]{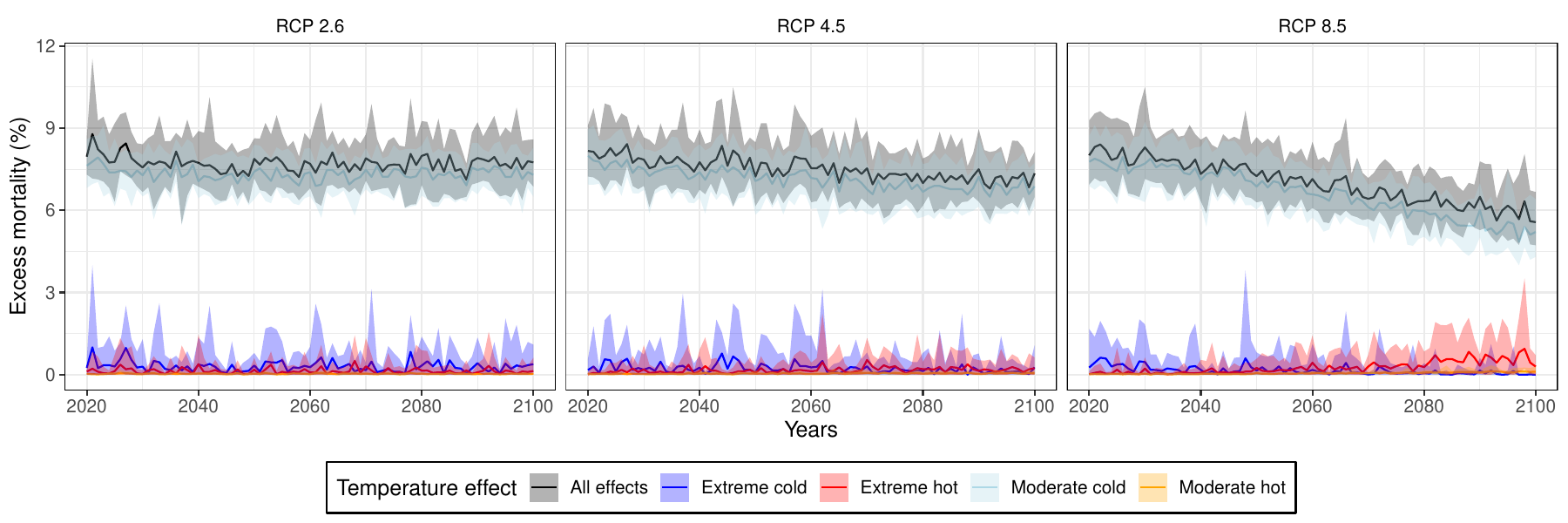}
\caption{
Temperature attributable fraction in Brest, simulated for the years 2020-2100 for both women and men across. These values, along with their $95\%$ confidence intervals (shaded areas) are calculated through 1,000 Monte Carlo simulations of the DLNM coefficients and the ensemble of climate models. Each line corresponds to the average of these simulations. Attributable fraction is expressed in \% for all effect, decomposed into moderate cold, moderate hot, extreme cold and extreme hot. 
}
\label{fig_attrib_global_brest}
\end{figure}

\begin{figure}[h!]
\centering
\includegraphics[scale = 0.55]{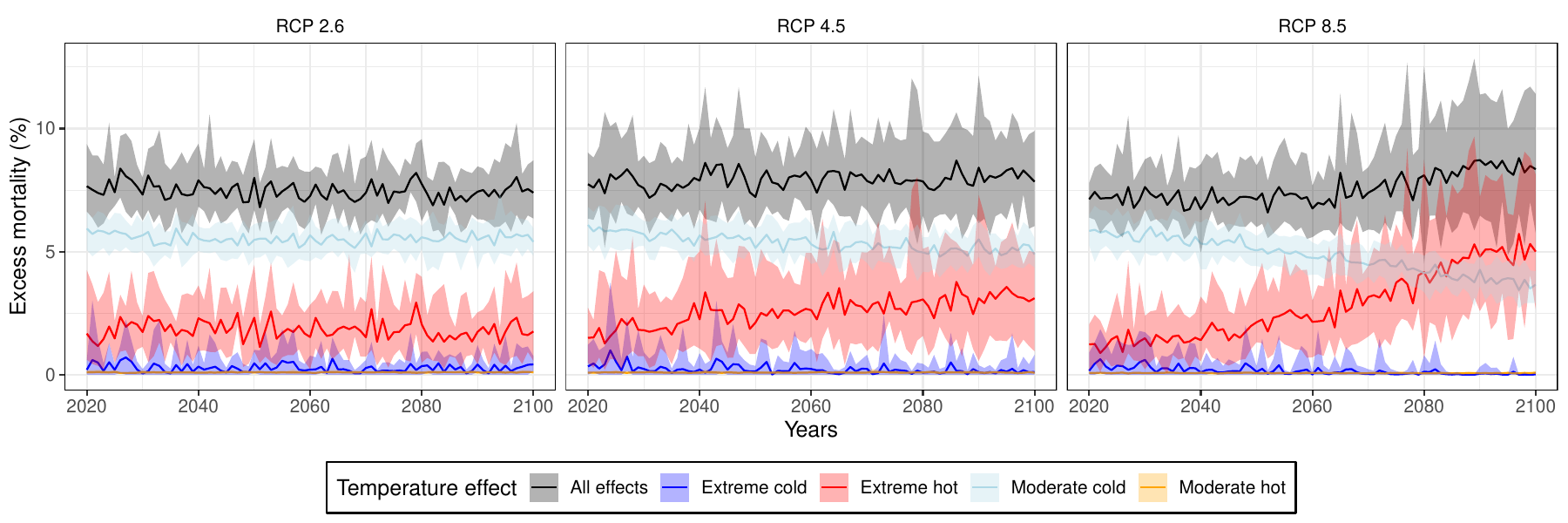}
\caption{
Temperature attributable fraction in Marseille, simulated for the years 2020-2100 for both women and men across. These values, along with their $95\%$ confidence intervals (shaded areas) are calculated through 1,000 Monte Carlo simulations of the DLNM coefficients and the ensemble of climate models. Each line corresponds to the average of these simulations. Attributable fraction is expressed in \% for all effect, decomposed into moderate cold, moderate hot, extreme cold and extreme hot. 
}
\label{fig_attrib_global_marseille}
\end{figure}

\begin{figure}[h!]
\centering
\includegraphics[scale = 0.55]{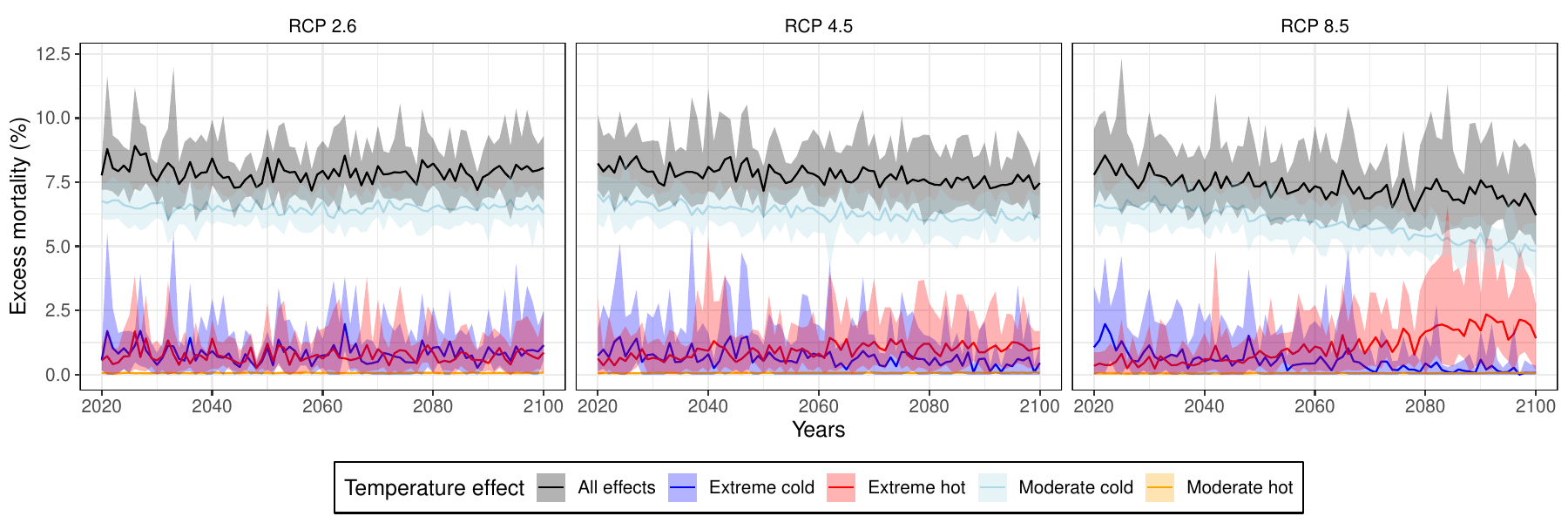}
\caption{
Temperature attributable fraction in Paris, simulated for the years 2020-2100 for both women and men across. These values, along with their $95\%$ confidence intervals (shaded areas) are calculated through 1,000 Monte Carlo simulations of the DLNM coefficients and the ensemble of climate models. Each line corresponds to the average of these simulations. Attributable fraction is expressed in \% for all effect, decomposed into moderate cold, moderate hot, extreme cold and extreme hot. 
}
\label{fig_attrib_global_paris}
\end{figure}

\begin{figure}[h!]
\centering
\includegraphics[scale = 0.55]{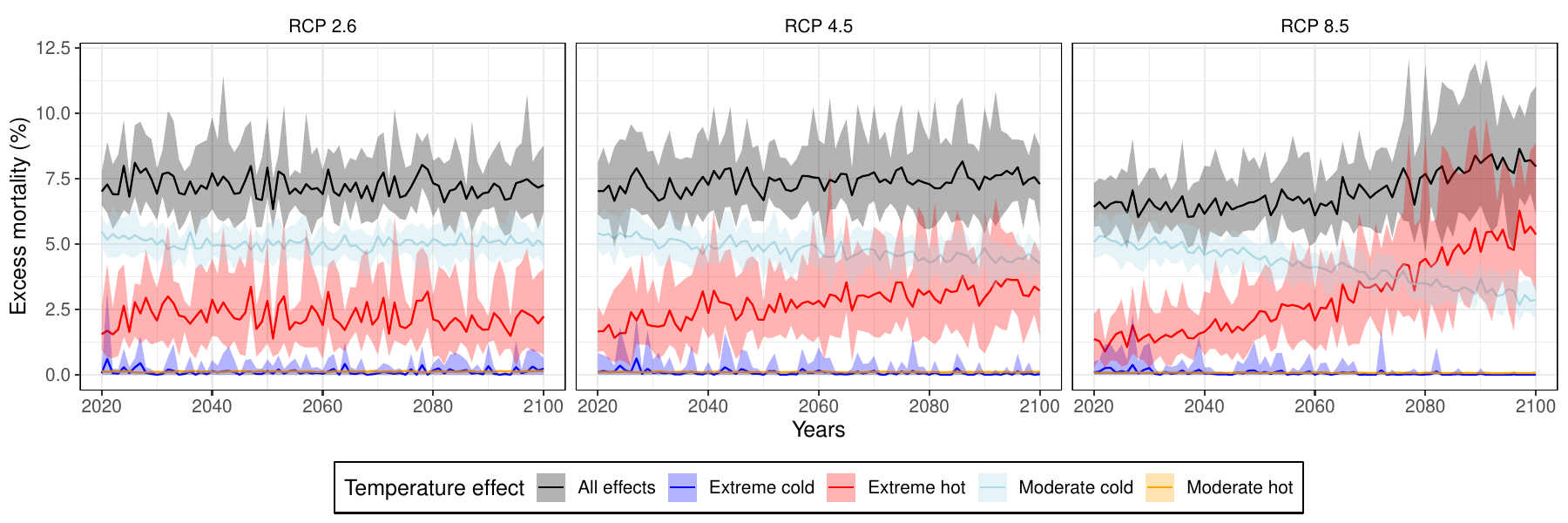}
\caption{
Temperature attributable fraction in Perpignan, simulated for the years 2020-2100 for both women and men across. These values, along with their $95\%$ confidence intervals (shaded areas) are calculated through 1,000 Monte Carlo simulations of the DLNM coefficients and the ensemble of climate models. Each line corresponds to the average of these simulations. Attributable fraction is expressed in \% for all effect, decomposed into moderate cold, moderate hot, extreme cold and extreme hot. 
}
\label{fig_attrib_global_perpignan}
\end{figure}

\begin{figure}[h!]
\centering
\includegraphics[scale = 0.55]{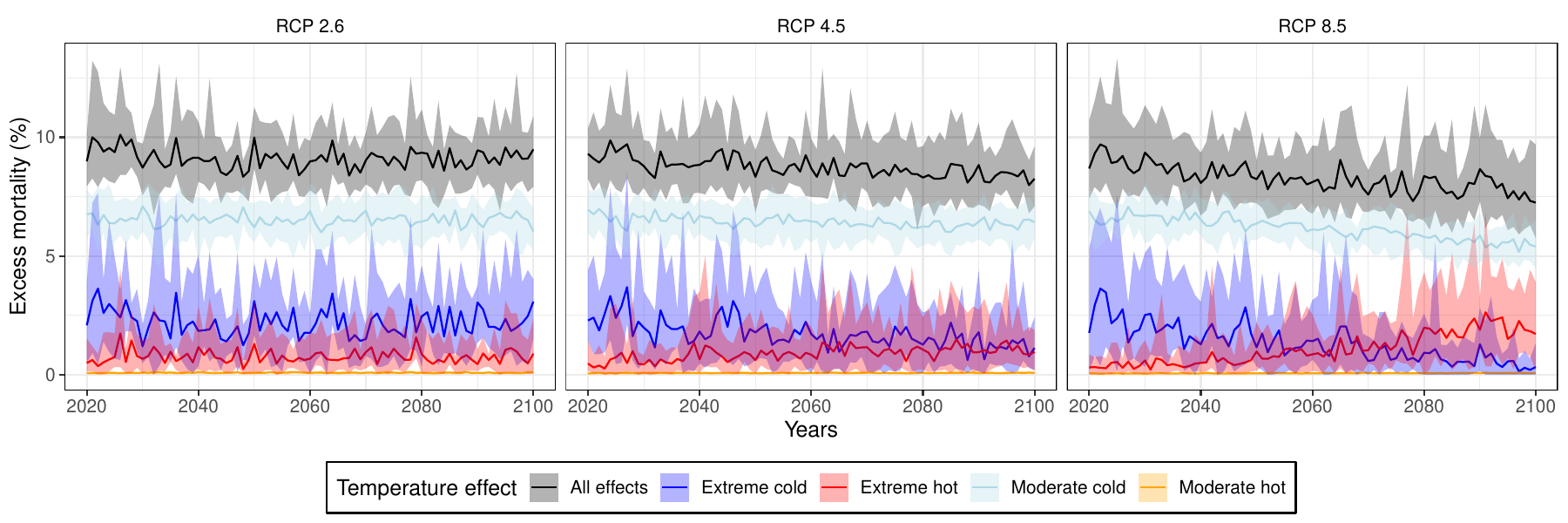}
\caption{
Temperature attributable fraction in Strasbourg, simulated for the years 2020-2100 for both women and men across. These values, along with their $95\%$ confidence intervals (shaded areas) are calculated through 1,000 Monte Carlo simulations of the DLNM coefficients and the ensemble of climate models. Each line corresponds to the average of these simulations. Attributable fraction is expressed in \% for all effect, decomposed into moderate cold, moderate hot, extreme cold and extreme hot. 
}
\label{fig_attrib_global_strasbourg}
\end{figure}

In this appendix, Figures~\ref{fig_attrib_global_brest}, \ref{fig_attrib_global_marseille}, \ref{fig_attrib_global_paris}, \ref{fig_attrib_global_perpignan} and~\ref{fig_attrib_global_strasbourg} respectively present the projected fractions of deaths attributable to temperatures for the cities of Brest, Marseille, Paris, Perpignan and Strasbourg.

\clearpage

\section{Additional figures on projected life expectancy losses by city}\label{subsec:app_loss_le_city}

\begin{figure}[h!]
\centering
\includegraphics[scale = 0.35]{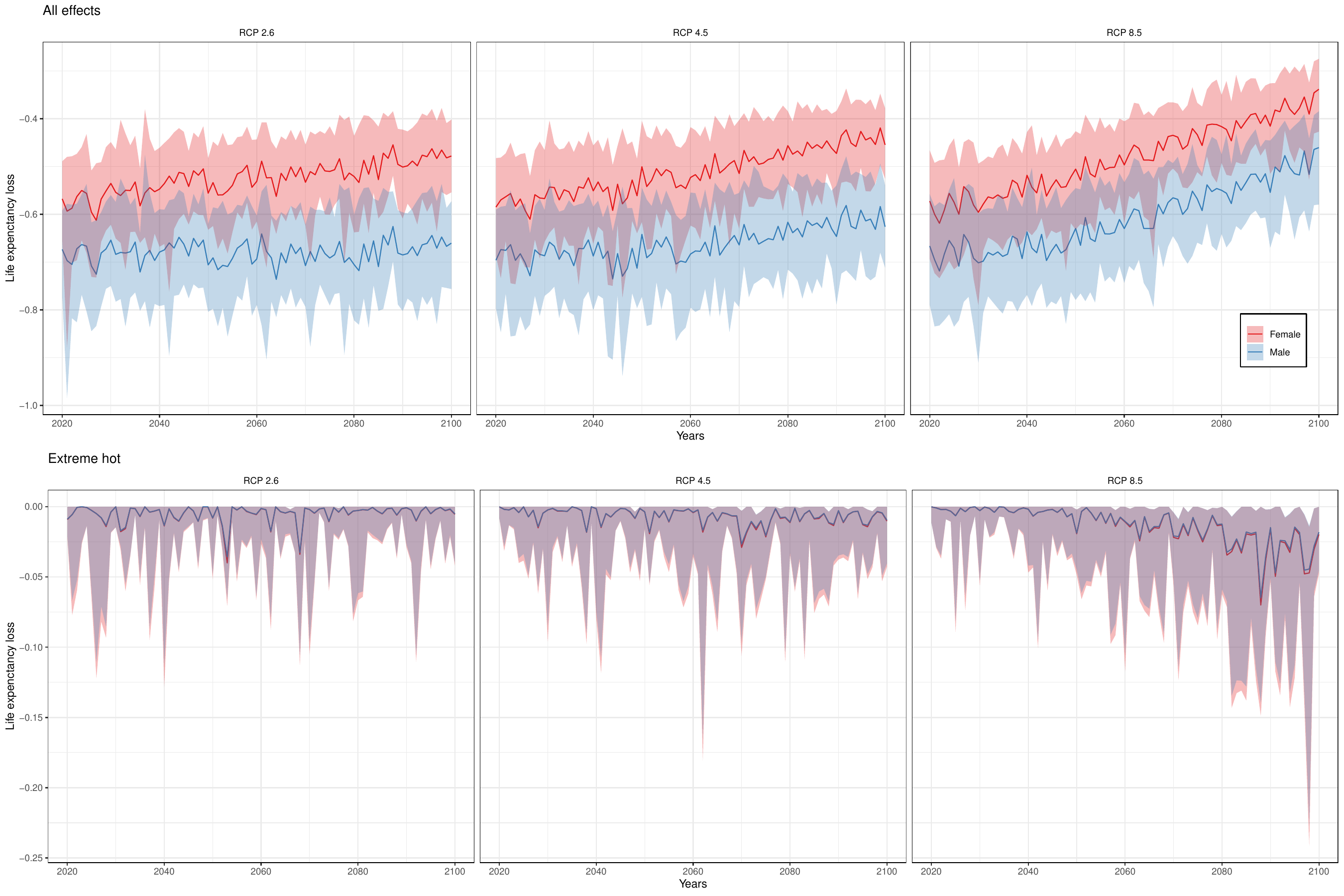}
\caption{
Life expectancy at birth lost in Brest, simulated for the years 2020-2100 for both women and men. We present both the loss related to all temperature effects and extreme hot effects only. These values, along with their $95\%$ confidence intervals (shaded areas), are calculated through 1,000 Monte Carlo simulations of the DLNM coefficients, the Li-Lee model and the ensemble of climate models. Each line corresponds to the median of these simulations. 
}
\label{fig_ev_gap_brest}
\end{figure}

\begin{figure}[h!]
\centering
\includegraphics[scale = 0.35]{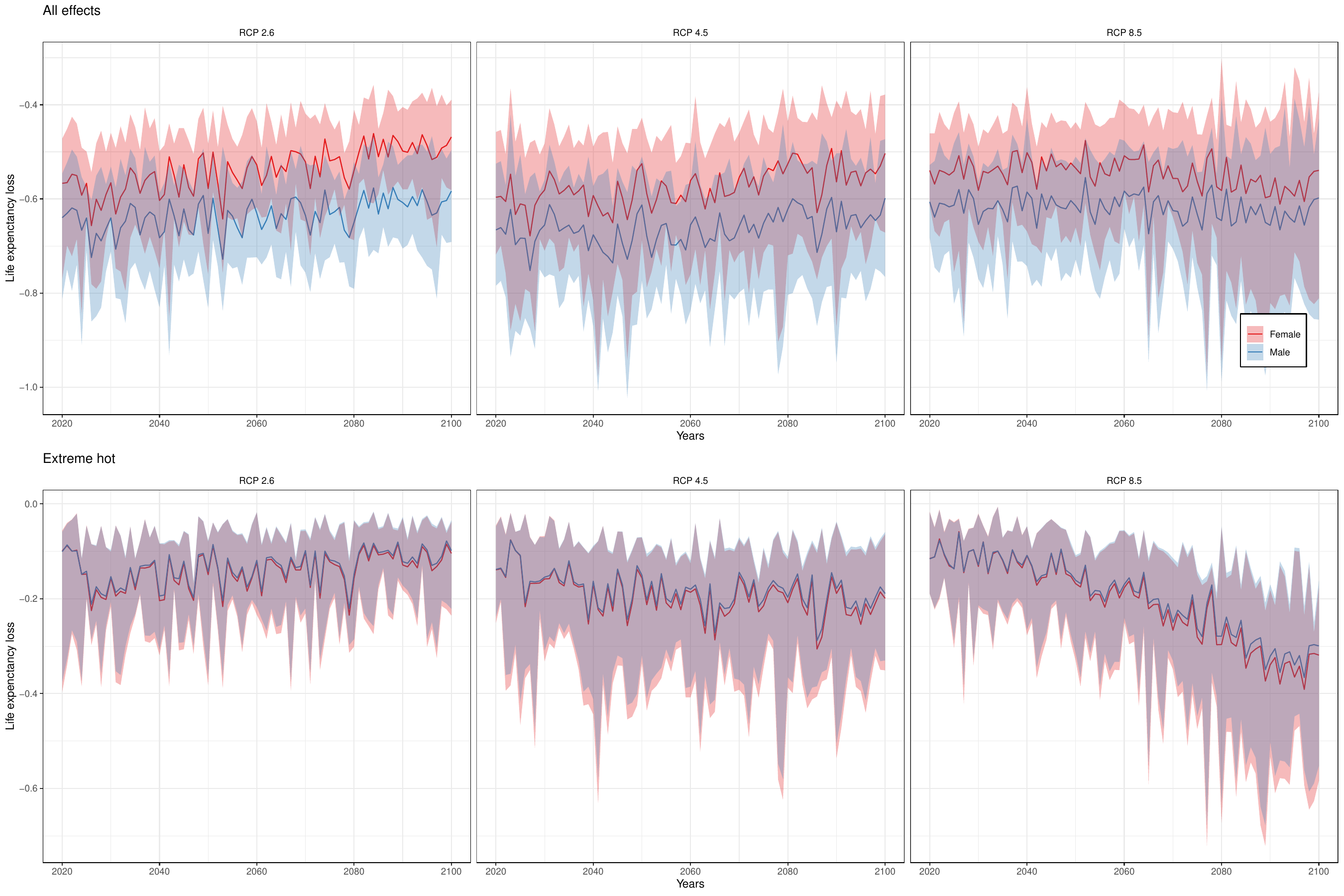}
\caption{
Life expectancy at birth lost in Marseille, simulated for the years 2020-2100 for both women and men. We present both the loss related to all temperature effects and extreme hot effects only. These values, along with their $95\%$ confidence intervals (shaded areas), are calculated through 1,000 Monte Carlo simulations of the DLNM coefficients, the Li-Lee model and the ensemble of climate models. Each line corresponds to the median of these simulations. 
}
\label{fig_ev_gap_marseille}
\end{figure}

\begin{figure}[h!]
\centering
\includegraphics[scale = 0.35]{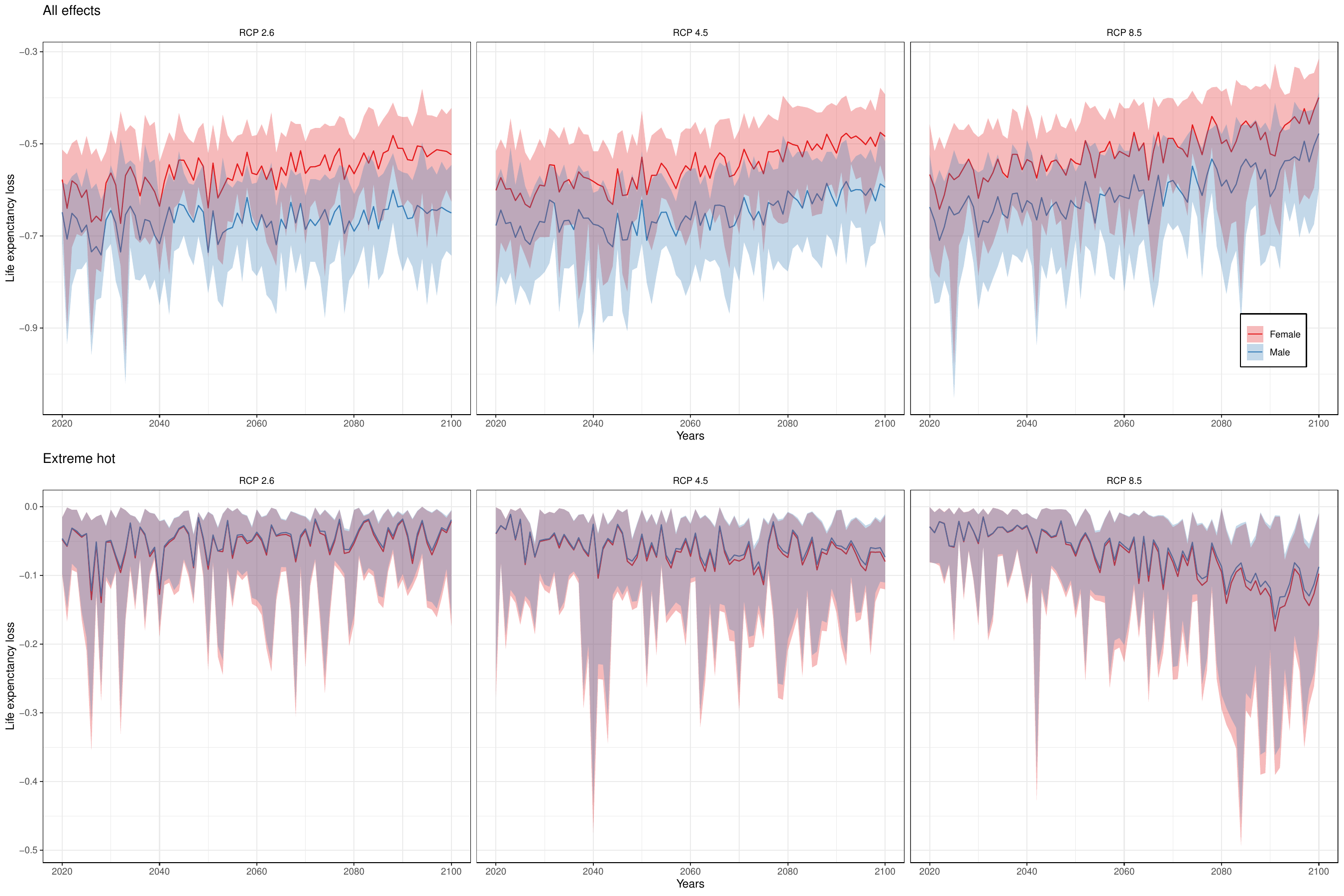}
\caption{
Life expectancy at birth lost in Paris, simulated for the years 2020-2100 for both women and men. We present both the loss related to all temperature effects and extreme hot effects only. These values, along with their $95\%$ confidence intervals (shaded areas), are calculated through 1,000 Monte Carlo simulations of the DLNM coefficients, the Li-Lee model and the ensemble of climate models. Each line corresponds to the median of these simulations. 
}
\label{fig_ev_gap_paris}
\end{figure}

\begin{figure}[h!]
\centering
\includegraphics[scale = 0.35]{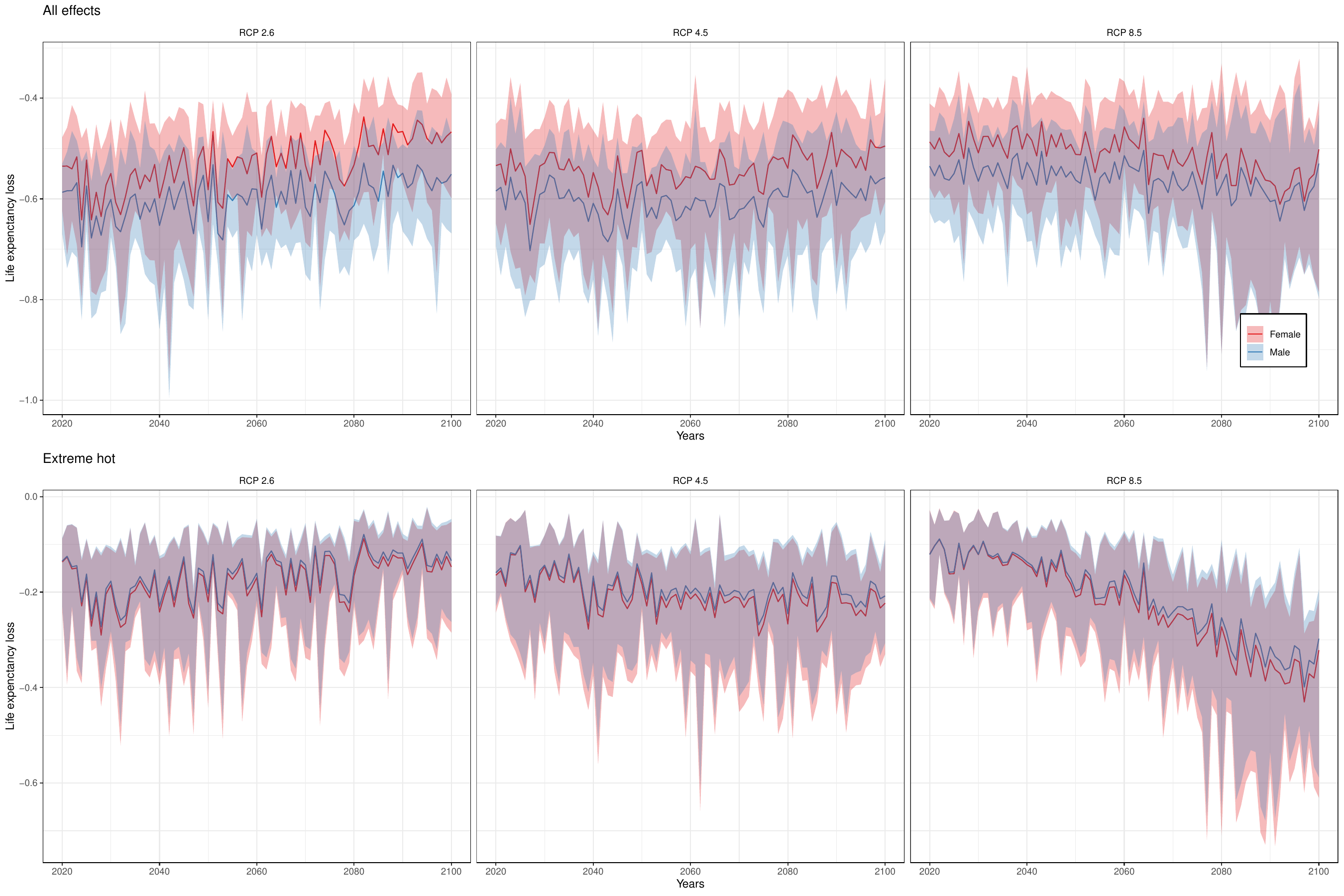}
\caption{
Life expectancy at birth lost in Perpignan, simulated for the years 2020-2100 for both women and men. We present both the loss related to all temperature effects and extreme hot effects only. These values, along with their $95\%$ confidence intervals (shaded areas), are calculated through 1,000 Monte Carlo simulations of the DLNM coefficients, the Li-Lee model and the ensemble of climate models. Each line corresponds to the median of these simulations. 
}
\label{fig_ev_gap_perpignan}
\end{figure}

\begin{figure}[h!]
\centering
\includegraphics[scale = 0.35]{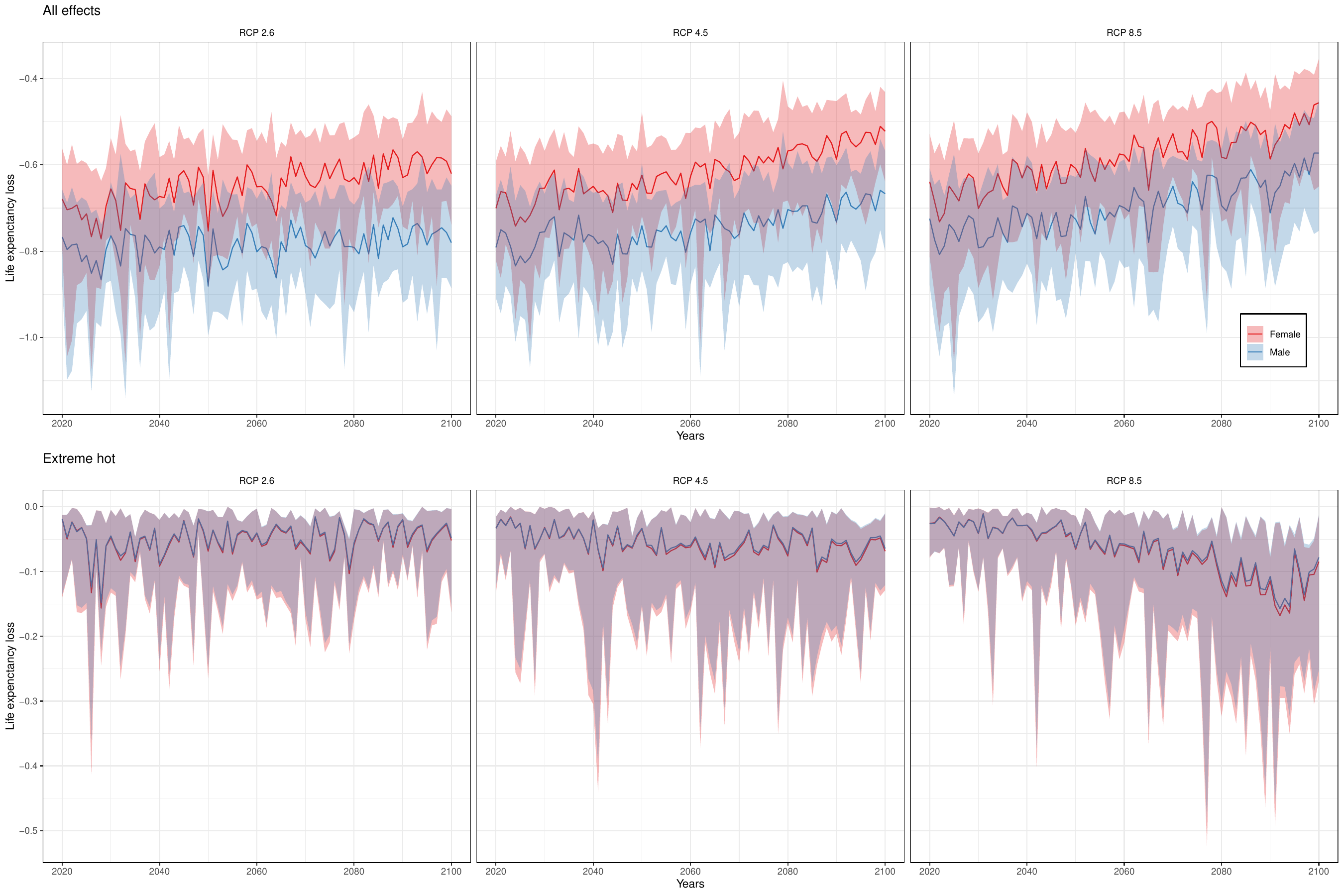}
\caption{
Life expectancy at birth lost in Strasbourg, simulated for the years 2020-2100 for both women and men. We present both the loss related to all temperature effects and extreme hot effects only. These values, along with their $95\%$ confidence intervals (shaded areas), are calculated through 1,000 Monte Carlo simulations of the DLNM coefficients, the Li-Lee model and the ensemble of climate models. Each line corresponds to the median of these simulations.
}
\label{fig_ev_gap_strasbourg}
\end{figure}

In this appendix, Figures~\ref{fig_ev_gap_brest}, \ref{fig_ev_gap_marseille}, \ref{fig_ev_gap_paris}, \ref{fig_ev_gap_perpignan} and~\ref{fig_ev_gap_strasbourg} respectively present the projected life expectancy at birth losses attributable to overall temperatures and extreme hot temperatures for the cities of Brest, Marseille, Paris, Perpignan and Strasbourg.

\clearpage


\addcontentsline{toc}{section}{References}

\printbibliography


\end{document}